\documentclass[namedreferences]{solarphysics}
\usepackage[hyperref,optionalrh]{spr-sola-addons} 
\usepackage{graphicx}        
\usepackage{color}           
\usepackage{breakurl}        
\usepackage{lscape}

\chardef\us=`\_

\begin{document}

\begin{article}
\begin{opening}

\title{Solar Energetic Particle-Associated Coronal Mass Ejections Observed by the Mauna Loa Solar Observatory Mk3 and Mk4 Coronameters}

\author[addressref={aff1,aff2},corref,email={ian.g.richardson@nasa.gov}]{\inits{I.~G.}~\fnm{I.~G.}~\lnm{Richardson}}
\author[addressref={aff1,aff3},email={ocstcyr2@gmail.com}]{\inits{O.~C.}~\fnm{O.~C.}~\lnm{St. Cyr}}
\author[addressref=aff4,email={iguana@ucar.edu}]{\inits{J.~T.}\fnm{J.~T.}~\lnm{Burkepile}}
\author[addressref={aff1,aff5},email={hong.xie-1@nasa.gov}]{\inits{H.}\fnm{H.}~\lnm{Xie}}
\author[addressref=aff1,email={barbara.j.thompson@nasa.gov}]{\inits{B.~J.}\fnm{B.~J.}~\lnm{Thompson}} 

\address[id=aff1]{Heliophysics Division, NASA Goddard Space Flight Center, Greenbelt, MD 20771, USA}
\address[id=aff2]{Department of Astronomy, University of Maryland, College Park, MD 20742}
\address[id=aff3]{Retired}
\address[id=aff4]{HAO/NCAR, Boulder, Colorado, USA}
\address[id=aff5]{Catholic University of America}

\runningauthor{I.~G.~Richardson et al.}
\runningtitle{SEPs and Mauna Loa CMEs}

\begin{abstract}
We report on the first comprehensive study of the coronal mass ejections (CMEs) associated with $\sim25$~MeV solar energetic proton (SEP) events in 1980-2013 observed in the low/inner corona by the Mauna Loa Solar Observatory (MLSO) Mk3 and Mk4 coronameters.  Where possible, these observations are combined with spacebased observations from the Solar Maximum Mission C/P, P78-1 SOLWIND or SOHO/LASCO coronagraphs. The aim of the study is to understand directly-measured (rather than inferred from proxies) CME motions in the low to middle corona and their association with SEP acceleration, and hence attempt to identify early signatures that are characteristic of SEP acceleration in ground-based CME observations that may be used to warn of impending SEP events. Although we find that SEP events are associated with CMEs that are on average faster and wider than typical CMEs observed by MLSO, a major challenge turns out to be determining reliable estimates of the CME dynamics in the low corona from the 3-minute cadence Mk3/4 observations since different analysis techniques can produce inconsistent results.  This complicates the assessment of what early information on a possible SEP event is available from these low coronal observations

\end{abstract}
\end{opening}

\section{Introduction}
     \label{S-Introduction} 

Since the discovery of coronal mass ejections (CMEs) in the early-1970s and their association with solar energetic particle (SEP) events later that decade \citep{kahler1978}, researchers have sought to expand our understanding of the connection between these phenomena.  The low solar corona ($<2.5$~R$_S$, where $R_S$ is in units of solar radii and is measured from Sun-center) is the region in which the initial maximum CME acceleration occurs  \citep[e.g.,][]{Zhang2001}and where direct observations and type II radio observations suggest that the formation of CME-driven shocks occurs \citep[e.g.,][]{wild1950, Raymond2000, Ciaravella2005}.  Furthermore, several lines of reasoning suggest that the SEPs with the hardest spectra arise from this low coronal region \citep[e.g.,][]{kahler1994, reames2009, gopalswamy2012,gopalswamy2013}.  It is therefore a key location for investigations of SEP acceleration by CME-driven shocks.  

Observations very low in the corona are necessary to detect the rapid CME accelerations that lead to shock formation and to assess the speeds of CMEs before they reach the middle corona. However, the necessary direct CME measurements of the initial acceleration are difficult to make with most spacebased coronagraphs.  Externally-occulted coronagraphs such as SOHO-LASCO C2 \citep{brueckner1995} have an inner field-of-view beginning at $\sim2.5$~R$_S$.  CME trajectories measured by this and similar instruments are usually well characterized by a single (constant) speed: e.g., $\sim80$\% of the speeds for the Solar Maximum Mission (SMM) CMEs and SOHO LASCO CMEs discussed by \cite{hundhausen1994} and \cite{stcyr2000}, respectively. It is also well-known that the extrapolation of space borne coronagraph measurements back to the low corona is subject to large errors \citep[e.g.,][]{macqueen1985}.  Thus, there is a need to combine low coronal measurements with those from space borne coronagraphs to cover the range of altitudes necessary to fully characterize the motion of CMEs from their initiation to the outer reaches of the fields of view of spacebased coronagraphs.  

This paper summarizes an observational investigation comparing the characteristics of SEP events with spacebased and groundbased observations of their associated coronal mass ejections (CMEs).  In particular, it exploits Mauna Loa Solar Observatory (MLSO) Mark~3 and Mark~4 coronameter observations that are able to image directly the early evolution and motion of CMEs close to the Sun, below the field of view of spacebased coronagraphs; such low-altitude CME dynamics have previously generally been inferred using proxy observations. (The distinction between a ``coronameter" and ``coronagraph" is that a coronameter scans the corona using a 1-D sensor to build up an image (see section~2.1 for further details) whereas a coronagraph takes an image using a 2-D sensor.) Since the most energetic SEPs appear to be accelerated close to the Sun (in the region imaged by MLSO), combining direct observations of CMEs from MLSO with spacebased coronagraphs allows us to determine the development of CMEs from their initiation and may lead to new insights and understanding into the production of energetic particles.

As a motivation for this study, \cite{stcyr2017} discussed an SEP event on January 1-2, 2016. A fast CME associated with this eruption was first detected by the SOHO LASCO C2 coronagraph when the leading edge of the CME was at a height of 2.7~R$_S$; the CME could be tracked out to 29~R$_S$ in the C3 coronagraph field-of-view.  The CME was also observed closer to the Sun by the K-Cor coronagraph at the Mauna Loa Solar Observatory. These MLSO observations indicated that a high initial acceleration of $\sim1500$ m/s$^2$ was measured in the first few minutes of CME formation, and the combined K-Cor and LASCO C2 height-time measurements indicated an average acceleration ($\sim370$ m/s$^2$) that was far larger than that inferred from C2 alone, with a final speed in C2 of $\sim1700$~km/s at 6.5~R$_S$.  This example illustrates how combining direct measurements of CMEs from MLSO with a space borne coronagraph may significantly enhance our understanding of their motion in the low corona. 

Another aim of \cite{stcyr2017} was to draw attention to both the utility and the availability of near-real-time coronagraph observations of fast CMEs, particularly in the low corona, as warnings for possible SEP events. Furthermore, \cite{thompson2017} have described an automated CME detection scheme based on archival data that has been implemented at MLSO to monitor real-time coronal images. It is well known that SEP intensities tend to be correlated with the CME speed as observed by space-based coronagraphs \citep[e.g.,][]{kahler1978, reames1999, richardson2014}, and this relationship has been exploited by using the observed properties of CMEs as a ``predictor" of SEP event intensity and spectra (e.g., \cite{richardson2018, bruno2021}; see \cite{Whitman2022} for a recent review of SEP prediction methods).  However, the association between the properties of SEP events and CME motions {\it in the low corona as measured directly by coronagraphs} remains largely unexplored. As \cite{stcyr2017} note, MLSO observations may be used to infer CME motions before the arrival of SEPs at Earth, whereas SEPs may already be arriving by the time the CME enters the field of view of a spacebased coronagraph.  

The importance of exploring the use of groundbased CME observations for SEP prediction (and other space weather uses) is underscored by the need to maintain CME observations if the ability to monitor CMEs from space is ever lost. This situation is hardly hypothetical: The aging SOHO spacecraft carries the workhorse LASCO coronagraphs \citep{brueckner1995}, which have revolutionized space weather forecasting, primarily by proving their utility in forecasting the timing of major geomagnetic storms. As reported by \cite{Fleck2014}, the field of space weather forecasting was substantially transformed by the SOHO mission.  But SOHO was launched more than 26 years ago, and the LASCO C1 coronagraph viewing the low corona did not survive the temporary loss of the spacecraft in 1998.  The twin STEREO spacecraft carrying coronagraphs designed to observe CMEs from viewpoints well-separated from Earth were launched in 2006 \citep{howard2008}, but contact with STEREO~B was lost in 2014. The Compact Coronagraphs to be carried on the NOAA Space Weather Follow on mission, due to be launched in 2025, and on GOES-U, scheduled for 2024, will provide future capabilities to observe CMEs above 3.0 and 3.7~R$_s$, respectively.  Nevertheless, a potential loss of continuous near-real-time spacebased coronagraph data is a real threat both to many fields in heliophysics \citep[e.g.,][]{stcyr2014} and to NASA’s plans to expand human presence beyond low-Earth-orbit in the near future. As an example, we note that 21 out of 38 SEP prediction models in Table~10 of \cite{Whitman2022} use spacebased coronagraph observations as an input.

The work reported here is the first comprehensive study of the association between SEP events and CMEs observed by MLSO. Several generations of white-light coronal instruments have operated at MLSO beginning in 1980. In this study we consider observations made by the Mark~3 and Mark~4 coronameters covering the period 1980 to 2013 and with an observing cadence of 3 minutes. Since K-Cor, installed in 2013 and still in operation, used by \cite{stcyr2017}, is of a different design with a higher (15~s) cadence, SEP-associated CMEs observed by K-Cor will be discussed in a future study.  In Section~2, we first describe the MLSO and spacebased coronagraphs and the SEP data sets used in this study.  We then discuss the identification of SEP events during this period and their association with CMEs observed by the MLSO Mk3/4 coronameters, provide a table of these events, and summarize the properties of these SEPs and the related solar events. In Section~3, we summarize the relationships between CME dynamics in the low corona and SEP properties, including details of methods to determine the CME motions. Section~4 summarizes the conclusions of this study. The results are discussed in Section~5, in particular focusing on how well SEP properties may be related to CME motions in the low corona and whether the results support the idea of establishing a world-wide network of coronagraphs for real-time monitoring of the low corona, as well as directions for further work.

\section{Observations}
\label{S-observations}

\subsection{Coronal Observations}
\label{SS-corobs}
This work makes use of the archive of observations from coronal imaging instruments installed at MLSO (\url{https://www2.hao.ucar.edu/mlso}).   
Groundbased imaging of the white-light corona was realized at MLSO with the installation of the Mark 3 K-coronameter (hereafter referred to as Mk3) in 1979.  First science images were made on January 2, 1980 and nominal observing began on February 4, 1980 \citep{fisher1981}. A brief history of the development of the MLSO coronal imaging instruments can be found in \cite{stcyr2015}. Mk3 used novel internal-occultation techniques to image the low corona at 1.12-2.44 R$_S$ in linearly polarized light in order to detect the faint coronal signal above the brightness of the sky. The Mk3 coronameter utilized a $1\times2048$ diode array to acquire a scan of the corona every 0.5$^\circ$. Scans alternated between clockwise and counterclockwise directions, with each rotation taking three minutes. A rotating 1/2 wave plate and dual beam splitter were used to simultaneously capture two different linear polarization states on different halves of the diode array. Mk3 was deployed in time to complement the middle corona field of view ($\sim2$-5 R$_S$) of the externally-occulted coronagraph/polarimeter on NASA’s Solar Maximum Mission \citep[SMM C/P;][]{macqueen1980}. Mk3 operated more-or-less continuously through the 1980s and 1990s until it was upgraded to Mk4 in September 1999 \citep{elmore2003} with a field of view of 1.12--2.8~R$_s$.  Mk4 was retired and replaced in late 2013 with K-Cor \citep{dewijn2012}, the currently operating instrument. K-Cor (not used in the present study) has significant improvements in spatial resolution (including uniformity over the field of view) and temporal cadence (15~s, compared to 3 minutes for earlier coronameters) and includes an operational automated detection scheme for CMEs (Thompson et al., 2017). The MLSO coronal instruments typically operate more than 200 days per year, from local sunrise ($\sim17$~UT) until sky conditions preclude coronal observations ($\sim02$~UT).  While a single groundbased facility cannot provide as complete temporal coverage as is achievable by a spacebased platform, the sheer longevity of MLSO observations and the ability to complement these with observations from spacebased instruments make them ideal for this study. 

We combined the MLSO CME observations with spacebased observations from the P78-1 SOLWIND coronagraph  (\cite{sheeley1980}; \url{https://lasco-www.nrl.navy.mil/solwind_transient.list}), the SMM C/P coronagraph (\url{https://www2.hao.ucar.edu/mlso/solar-maximum-mission}) and the LASCO coronagraphs on SOHO (\cite{brueckner1995}; \url{https://lasco-www.nrl.navy.mil}) with the aim of characterizing CME motions from the low into the mid corona as discussed below. When possible, we tracked CMEs out to $\sim10$~R$_s$. 

Although CME parameters (e.g., plane of the sky speed, angular width and direction) obtained using MLSO or spacebased coronagraphs are available from the above sources and linked catalogs, we chose to remeasure the CME height-time profiles for this study using the original images to provide a uniform set of parameters derived in a similar way, for example, to ensure that the same morphological feature was tracked from the low to the mid-corona. (Also, as noted by \cite{richardson2015}, the speeds and widths of SEP-associated CMEs reported in the many available spacebased CME catalogs typically are not in agreement.) To estimate the (plane of the sky) CME speed and acceleration to be used in the analyses discussed below, we required at least four estimates of the CME leading-edge height from MLSO observations.  We also considered the spacing of the measurements in time to ensure that they were relatively uniform across the CME’s transit through the field of view.  This was necessary because, {\bf as described above}, Mk3/Mk4 acquired observations by scanning a linear diode around the inner corona, first rotating clockwise then counter-clockwise, with each rotation requiring three minutes. For the CME height-time measurements, the azimuth of the point measured on the CME front relative to the azimuth of the scan start was used to obtain the offset of the observation time from the start of the scan.  Thus, features near the initial scan azimuth position were observed twice in quick succession (followed by a gap of nearly six minutes), and measurements derived from these observations are too close in time to be considered as independent estimates of the CME height. Unfortunately, for many SEP-associated CMEs, this four image requirement could not be met due to image quality, cloud cover or instrumental issues. However, where at least two or three images were available, we did estimate the speed and if possible, the acceleration, but did not use these in further analysis. 

\begin{figure}

\centerline{\includegraphics[width=0.6\textwidth,clip=]{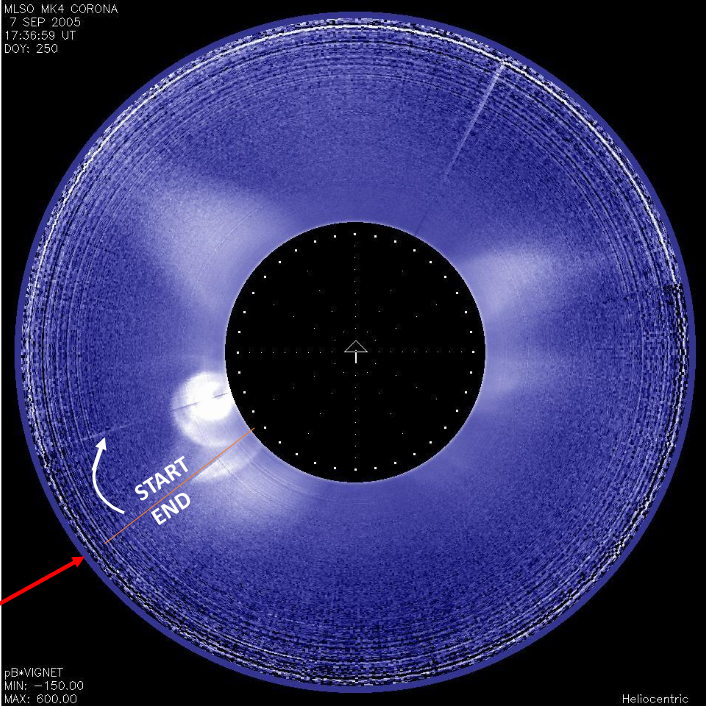}}
    \caption{MLSO Mk4 coronameter image of the exceptionally fast ($\sim2500$~km/s) CME on September~7, 2005 at 17:56:39~UT. The image is formed by scanning a linear diode clockwise starting from the south east (red line/arrow). By the time the scan was completed 3 minutes later, the CME, which was directed close to the scan start/end line, had moved away from the Sun significantly, resulting in the discontinuity in the image on the CME southern flank.  The CME front had moved beyond the field of view in the following counterclockwise scan (not shown). This circumstance of a fast CME close to the scan start/end line is very rare (another case occurred on December 28, 2001) but illustrates how images obtained with the scanning technique used by Mk3/4 differ from those produced by K-Cor and spacebased coronagraphs.  }
    \label{090705}
\end{figure}

A rare issue with observing fast CMEs with Mk3/4 is illustrated by what appears to have been the fastest CME detected by any of the MLSO coronameters in the inner corona, an East limb event on September 7, 2005 that was also associated with an SEP event \citep{cane2010, ling2014}.  Figure~1 shows a Mk4 coronal scan for this CME, which also happened to be directed close to the starting/ending location for this clockwise scan as indicated in the figure.  In the 3 minutes between the start and end of this scan, the CME moved further out in the field of view (FOV), resulting in a line of discontinuity at the southern flank of the CME. Unfortunately, the front of the CME had left the FOV by the time the next (counterclockwise) scan was completed, nearly 6~minutes after the initial imaging of the front in Figure~1. As a result, the speed of the CME front, estimated to be 2500~km/s, can only be determined from the image shown, when the front was at 1.68~R$_s$ and a previous scan completed just a few seconds earlier when the front was at 1.63~R$_s$. Although the CME speed is uncertain because only two closely-spaced measurements are available, a 2500 km/s CME would travel 1~R$_s$ in 4.6~minutes, which is consistent with the CME front leaving the Mk4 FOV (at 2.86~R$_s$) between the scans separated by $\sim6$ minutes.  In addition, three height-time measurements of a trailing prominence associated with this CME indicate that the prominence was traveling at 2600 km/s, confirming the extremely high speed inferred from the two-point measurement.
 
 Considering the maximum average speed of a CME that might be observed by Mk3/4 under ideal (but unlikely) conditions where the CME is observed entering the FOV, extending 1.12-2.44~R$_s$ (1.12-2.8~R$_s$) for Mk3(Mk4), in one scan and exiting in the next scan 3 minutes later, this is $\sim4600$~km/s for Mk3 and 5900~km/s for Mk4. Requiring four images to analyze the CME dynamics would increase the minimum time in the field of view to $\sim9$~minutes for a nominal 3 minute cadence, reducing these respective maximum average speeds to $\sim1500$ or 2000~km/s.          

The individual CME height-time (h-t) measurements from MLSO and spacebased observations, either separately or in combination, were used to provide an estimate of the average CME linear velocity (average acceleration) determined by a first-order (second-order) least squares polynomial fit to all the individual h-t measurements with their associated uncertainties. Such estimates are similar to those typically cited in CME catalogues \citep[e.g.,][]{Gosling1976,Howard1985,Burkepile1993, Yashiro2004}, and frequently used in the heliophysics research community.

There are also numerous ways to utilize these height-time data to extract peak speed and acceleration values, from simple differencing of the individual raw measurements to increasingly complex techniques to smooth the measurements before deriving the higher-order velocity and acceleration values.   Since the results obtained with ``point to point" estimates can be noisy, we smoothed such estimates over three points as in \cite{gopalswamy2013}. 
More complex mathematical techniques include fitting exponential expressions to smoothed measurements \citep{Vrsnak2001,Shanmugaraju2003}.  The most successful method appears to be cubic spline interpolation (CSI) as applied by \cite{Maricic2004} to inner corona height-time measurements  for the 15-May-2001 CME observed by multiple MLSO instruments and SOHO LASCO.  \cite{vrsnak2007} then applied this technique to an expanded sample of more than 20 MLSO and LASCO CMEs, and provided details of the smoothing technique in an appendix. CSI was also applied to more than 90 STEREO CMEs by \cite{Bein2011} and to 59 STEREO CMEs by \cite{Majumdar2020}. The CSI technique is attractive because the smoothed curve remains continuous at the tie points (the “knots”), hence the first and second derivatives can be derived. Future researchers may consider using the CSI technique in near-real-time applications of CME height-time analysis, although they will have to overcome the obstacles to optimal placement of the knots. When applying CSI, we also required measurements from higher altitudes, where the CME velocity has become nearly constant, to optimize the fit.

Other techniques previously used for estimating CME speeds and acceleration in the inner corona rely on measurements made in the middle corona (e.g., by LASCO C2 and C3) combined with various assumptions about the unseen motions of the CME in the inner corona.  \cite{gopalswamy2012} provided several estimates for a set of CMEs associated with SEP ``ground level enhancements" (GLEs) based on flare and CME properties in the middle corona.  One such technique was described by \cite{Zhang2001} as the ``flare-proxy" (F-P) method.   This is based on the CME speed measured in the middle corona and the assumption that the CME is accelerated to this speed during the duration of the rising phase of the associated soft X-ray flare, giving an estimate of the (constant) CME acceleration.  In a subsequent paper, \cite{zhang2006} applied this technique to a sample of 50 LASCO CMEs, and based a correlation between the F-P technique and direct height-time measurements on nine CMEs where the main (peak) acceleration was $<500$~ms$^{-2}$.  A tenth event with direct CME measurements available was much more impulsive and gave disparate accelerations of 4.4~km~s$^{-2}$ for F-P compared with 7.7~km~s$^{-2}$ for the direct observations. In this paper, we compare the directly-measured CME speeds/accelerations with those obtained by the F-P method.

\begin{figure}

\centerline{\includegraphics[width=1.0\textwidth,clip=]{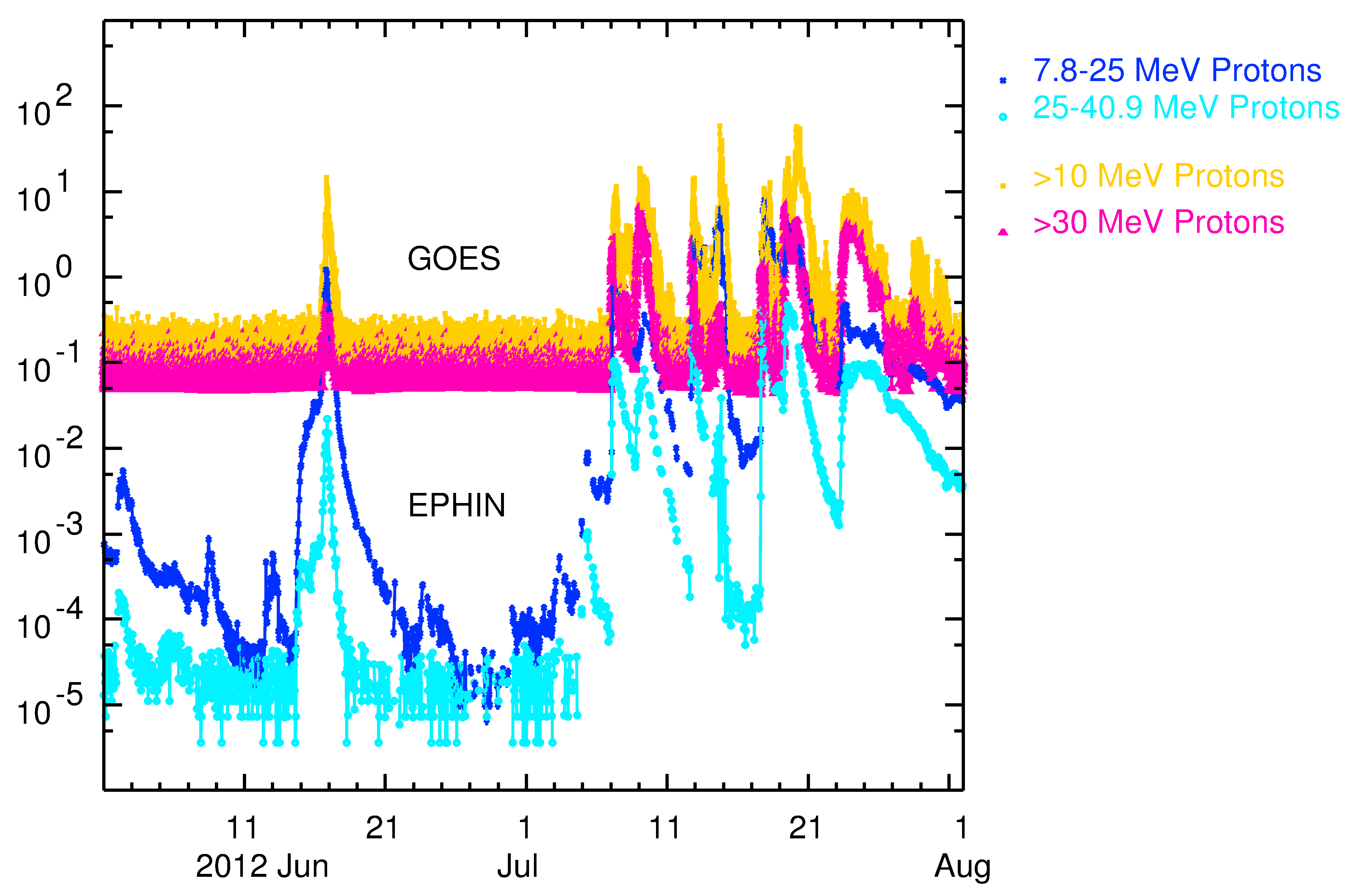}}
    \caption{Comparison of the GOES $>10$~MeV (orange) and $>30$~MeV (magenta) proton fluxes (in pfu) and SOHO/EPHIN 7.8-25~MeV (blue) and 25-40.9~MeV (cyan) proton intensities ((MeV s cm$^2$ sr)$^{-1}$) for a period in 2012. Note that many more SEP events \citep[with identified sources, e.g.,][]{richardson2014} are evident in the EPHIN data and that the widely-used ``SEP event" definition of $>10$ pfu at $>10$~MeV based on GOES data only identifies a small subset of these events.}
    \label{GOESERNE}
\end{figure}

\subsection{Solar Energetic Particle Event Observations}
\label{SS-SEPobs}
To identify SEP events associated with CMEs observed by the MLSO Mk3/4 coronameters, we examined SEP observations over a wide energy range (typically $\sim 1$-100~MeV protons and near-relativistic electrons, as available) made by instruments on several spacecraft near the Earth during the period of operation of these coronameters (1980-2013).  These instruments included: the Goddard Medium Energy (GME) instruments \citep{mcguire1986} on IMPs~7 and 8 in Earth orbit, covering October 1972-December 2005; the VLET and HET \citep{vonrosen1978} on ISEE-3/ICE (launched in August 1978 to L1 before departing for comet Giacobini-Zinner in December 1983 following a $\sim1$ year period predominantly in the geomagnetic tail \citep{tsurutani1984}); the EPHIN \citep{muller1995} and ERNE \citep{torsti1995} instruments on SOHO, the EPACT instrument \citep{vonrosen1995} on Wind \citep{wilson2021}, and the SIS \citep{stone1998} and EPAM \citep{gold1998} instruments on ACE. 

Following past studies \cite[e.g.,][]{VanHollebeke1975, cane1988, cane2010, richardson2014, richardson2017}, we focused on SEP events including at least $\sim20$~MeV protons that were detectable above the instrumental background, typically $\sim10^{-4}$~(MeV cm$^2$ s sr)$^{-1}$ for the above instruments. 
Note in particular that we do not adopt the widely-used definition of an ``SEP event" based on GOES spacecraft observations, i.e., $>10$~pfu (1~pfu=1~(cm$^2$ s sr)$^{-1}$) at $>10$~MeV as followed for example by the NOAA solar proton event list (\url{https://umbra.nascom.nasa.gov/SEP/}, \url{ftp://ftp.swpc.noaa.gov/pub/indices/SPE.txt}). As illustrated in Figure~2, the differential energy channels of the science instruments used, such as SOHO/EPHIN shown here (blue and cyan graphs) for a period in 2012, generally have much lower backgrounds than the integral fluxes from the operations-oriented GOES proton instruments (orange and purple graphs) and hence can detect a larger number of SEP events covering a wider range of intensities.  Even the smallest events in Figure~2 have identified solar sources \citep{richardson2014} and several are associated with MLSO CMEs.  In addition, the standard definition identifies only a small subset of relatively high intensity SEP events, although these are of most concern for space weather. Another reason for choosing a definition based on $\sim20$~MeV protons rather than $>10$~MeV is that SEP event onsets are typically easier to identify than at lower energies where particles associated with interplanetary shocks can obscure the onsets (cf., Figure 1 of \cite{richardson2017}). The GOES integral fluxes also mix low and high energy protons.  For the current study, major advantages are the  greater likelihood that an SEP event will be identified at the time of a MLSO CME than when just GOES proton observations are considered, and that these events will cover a larger range in intensity, allowing a more complete investigation of the relationship between SEP and MLSO CME parameters.  

We then selected the subset of SEP events with onsets in or close to the nominal MLSO observing window (17 UT-02 UT). We checked the online MLSO observations to see if data were available for the day of each SEP event.  If so, all images on that day were scanned as both direct and differenced images to confirm whether there was a CME with timing consistent with the SEP onset. In some cases, initial associations were rejected on closer examination of the timing of the SEP onset and CME, for example if the SEP onset was clearly before the CME eruption. Observations of near-relativistic electrons, which arrive ahead of protons \citep{posner2007} were especially useful for inferring and confirming the time of the associated solar eruption.  If a CME was identified, measurements of its location, size, and motion were made. 
 
 Many of the solar events associated with the SEP-associated MLSO CMEs have been reported in existing SEP catalogs \citep[e.g.,][and the NOAA SEP event list]{cane2010, richardson2014,papaioannou2016, miteva2018} or were established for prior studies \citep[e.g.][]{cane1988}. In other cases, or when rechecking the reported associations, we referred to solar flare observations, e.g., from {\it Solar Geophysical Data} for the earlier events and on-line sources e.g., Solarmonitor.org (\url{https://solarmonitor.org/}), GOES X-ray flare reports (\url{https://swpc-drupal.woc.noaa.gov/products/goes-x-ray-flux}), the Heliophysics Event Knowledgebase (HEK; \cite{Hurlburt2012}; \url{https://lmsal.com/hek/}) and the Solarsoft latest events archive (\url{https://www.lmsal.com/solarsoft/latest_events_archive.html}) for evidence of an associated H-$\alpha$, EUV, and/or soft X-ray flare consistent with the CME time and direction (for example, a CME above the west limb is unlikely to be associated with an eastern hemisphere flare).  In some cases, no flare was observed, possibly indicating a far side source - as \cite{richardson2014} noted using STEREO spacecraft EUV observations of the far side, around 25\% of $\sim20$~MeV proton events originate from eruptions beyond the limbs of the Sun with respect to the observer. When available, we also examined solar radio observations from spacecraft such as Wind and STEREO A/B for evidence of type II and type III radio emissions around the time of the CME that may be indicative of particle acceleration \citep[e.g.][]{cane2002, laurenza2009,richardson2014, winter2015, richardson2018} and can help to confirm the CME-SEP event association.  
 
 In general, the SEP/CME/solar event associations could be made more reliably for well-connected western SEP events with relatively prompt/fast rising onsets.  than for slower rising, possibly poorly-connected events.  In a few cases, SEP observations from a spacecraft away from Earth could be used to verify that an SEP event was associated with a CME that was poorly-connected to Earth.  In particular, a few SEP events associated with MLSO CMEs were identified in observations \citep[e.g.,][]{kallenrode1993} from Helios~1 and 2 (in heliocentric orbits at 0.3-1~AU from the Sun and observing in  December 1974- February 1986), Ulysses (in a heliocentric orbit extending to high latitudes and out to $\sim5$~AU, observing in October 1990-October 2006, e.g., \cite{Lario2008}), or STEREO~A and B in heliocentric orbits near 1~AU (October, 2006-present) \citep{richardson2014}.  It is also likely that some MLSO CMEs associated with SEPs were overlooked because a high SEP background from previous events obscured the new particle event.

 The 84 identified SEP events associated with MLSO Mk3/4 CMEs are listed in Tables~1 to 3. The first column of each table gives the date of the event. Bold type indicates that the SEP event was a ground level enhancement (GLE; \url{https://gle.oulu.fi}) observed by neutron monitors. Column 2 gives the location of the related solar eruption. Estimated or observed farside locations are from studies such as \cite{cane2010} or \cite{richardson2014}. Where no location is listed, this indicates either that the eruption is at some unknown location on the far side of the Sun, the CME onset time is not consistent with any reported flare, or the location is uncertain. Columns 3 and 4 give the peak intensity and onset time of the associated soft X-ray flare, from the GOES spacecraft. Columns 5 to 7 give, for the MLSO CME, the time of first detection, the plane-of-the-sky central position angle and full angular width. The first detection time is corrected to take into account the direction of CME relative to the start/end of the Mk3/4 scan, as discussed in section~2.1. Columns 8 and 9 show the height range over which the CME was tracked below 3~R$_s$ and the number of height-time points available. In a few cases, indicated by `*', the extreme point has been taken from observations made by a spaceborne instrument. In particular, the lower height measurement contains points from SOHO/EIT, SOHO/LASCO C1 or SDO/AIA, while the highest points below 3~R$_s$ contain measurements from SMM or SOHO/LASCO~C2.  Columns 10-12 give the CME average velocity, average acceleration, and the velocity determined using the derived acceleration at the final measurement point (the ``final velocity"). At least four height-time measurements are required except for the average velocity. Column 13 indicates whether a spacebased coronagraph also detected this CME and the time of first detection.  Finally, columns 14 and 15 give the SEP proton spectral index $\gamma$ at $\sim5$ to 60~MeV and intensity at 20~MeV.  The proton intensity is measured near Earth except in a few cases as indicated in the table. Some of these parameters will be explained further in Section~3. 

\newpage

\begin{landscape}
\begin{table}
\label{T-list1}
\renewcommand{\arraystretch}{.7}
\setlength{\tabcolsep}{.02in}
\caption{MLSO Mk3 Coronameter CMEs Associated With $\sim25$~MeV Solar Proton Events. \\
Column~1: Event date (ground level enhancements (GLEs) are indicated in bold type); column 2: Source (typically flare) location. Events without locations may be beyond the limbs (for some far-side events, a reported/estimated location is indicated), the location of the reported soft X-ray flare may be unclear, or none of the reported flares is consistent in time with the MLSO CME; columns 3-4: GOES soft X-ray peak intensity and onset time; columns 5-7: MLSO CME first detection time, central position angle (measured anti-clockwise from north) and full angular width; Column 8: Height range ($<3$~R$_s$) over which the CME is tracked. * indicates that the extreme point is taken from observations made by a spaceborne instrument; Column 9: Number of CME height-time measurements $<3$~R$_s$; Columns 10-12: Average CME speed, average acceleration, and final speed below 3~R$_s$; column 13: Spacebased coronagraph that observed the CME (Sol=P78-1/SOLWIND; SMM=Solar Maximum Mission C/P; LAS=LASCO) and first observation time; Column 14: SEP proton spectral index $\gamma$ for a power-law in energy spectrum ($dJ/dE\sim E^{-\gamma}$) at $\sim5$--60~MeV; column 15: SEP proton intensity at 20~MeV in (MeV~s~cm$^2$~sr)$^{-1}$ near Earth except: Hel=Helios; Uly= Ulysses; STA/B=STEREO A/B. }
\begin{tabular}{llcccccccccclcc}

\hline
1   &   2  &  3       &  4      &  5          & 6  &7& 8     &9 &10         &11          &12        &13     &14      & 15 \\
Date&Source&SXR Peak  &SXR Onset&MLSO 1$^{st}$&CPA&W&Height &N &Av. V      &Av. Acc     &Final V   &Space  &$\gamma$&I(20 MeV)\\
 
[m/d/y]     &Location&[W/m${^2}$]&[UT] & Det. [UT]    &[deg]&[deg]&Range [R$_s$]   &       &[km/s]&[km/s${^2}$]&[km/s]&Coronagraph  &     &  \\

\hline
Mk3\\
05/01/1980   &S22E60& 3.00E-04&18:51&19:20 &108 &41 &1.42--2.23 &7 &$1035\pm167$ &$0.48\pm1.07$ &$1167\pm492$ &... & ...  &2.0~Hel \\
05/03/1980&  ... & ...& ...     &20:50 &289 &20 &1.70--2.07 &4 &$436\pm34$   &$0.24\pm0.28$ &$503\pm111$&Sol 21:14&  ...  &0.001  \\
06/29/1980 &S25W90 &4.20E-05 &18:21 &18:25 &246 &29 &1.22--2.00 &7 &$954\pm53$&$-1.21\pm2.44$ &$675\pm603$&SMM 18:39 &2.88 &0.0028\\
03/25/1981 &N09W89 &2.20E-04 &20:39 &20:43 &293 &31 &1.25--2.10 &9 &$798\pm17$ &$0.08\pm0.13$ &$826\pm62$&Sol 21:42 &2.78 & 0.029 \\
06/27/1982 &N16W66 & ...     & ...  &18:15 &297 &38 &2.35       &1 & ...       & ...          & ...      &Sol 20:02 &1.73 &0.10 \\
11/07/1987 &N31W90 &1.20E-05 &20:28 &19:50 &289 &78 &1.84--2.36 &4 &$603\pm22$ &$0.45\pm0.07$ &$725\pm35$&SMM 20:01 &3.85 &0.67 \\
10/13/1988 &S20W88 &6.40E-05 &20:24 &20:32 &243 &26 &1.70--2.30 &3 &$1004\pm66$ &...         &...       &SMM 20:44 &2.57&0.042 \\
03/17/1989  &N33W60 &6.50E-04 &17:20 &17:37 &307 &24 &1.88--2.47 &3 &$678\pm48$  &... & ... &SMM 17:50 &3.61 &27.5 \\
03/23/1989  &N18W28 &1.50E-04 &19:20 &19:37 &306 &99 &1.35--2.70* &6 &$1015\pm47$ &$1.58\pm0.30$ &$1768\pm131$&SMM 19:45 &2.83 &2.87 \\
07/27/1989 &...     & ...     & ...  &17:49 &285 &56 &1.36--2.90* &6 &$663\pm30$  &$-0.29\pm0.07$ &$443\pm84$ &SMM 18:15 & ...& $>60$~MeV$^1$\\
08/07/1989  &S26W38  &7.60E-05 &20:21 &20:56 &230 &60 &1.59--2.85* &6 &$863\pm115$ &$2.22\pm0.25$&$1875\pm176$&SMM 20:56 &3.07 &0.017\\
09/09/1989  &S15W67  &1.30E-04 &19:28 &19:38 &269 &18 &1.68--2.80* &7 &$299\pm32$  &$-0.15\pm0.03$ &$130\pm6$ &SMM 19:54 &3.21 &0.017 \\
{\bf 10/22/1989} &{\bf S27W31}  &2.90E-04 &17:08& 17:35&222 &100 &1.67--2.22 &6 &$\mathbf{453\pm2}$ & $\mathbf{1.41\pm0.23}$ & $\mathbf {1018\pm95}$ &...&{\bf2.54} &{\bf 129} \\
11/08/1989 &N20W57  &9.80E-05 &18:56 &19:10 &301 &66 &1.73--2.22  &4 &696        &0.31          &777       &SMM 19:48 &4.36 &0.22\\
{\bf 05/26/1990} &{\bf $\sim$W100} & ...& ...    &20:50 &310 &127 &1.31--2.11&3 &$\mathbf{1548\pm12}$&... &...&... &{\bf 1.71} &{\bf 2.44} \\
10/25/1990 &... & ...& ...      &17:24 &316 &95 &1.52--2.13 &2 &$1386\pm68$& ...           & ...       &...&...&0.5~Uly\\
11/15/1991 &S13W19 &1.50E-04 &22:34 &22:50 &193 &95 &1.60--2.20 &4 &$951\pm230$ &$1.98\pm2.39$&$1304\pm664$&...&...&...\\
01/09/1992   &Unclear&4.00E-06 &19:52 &19:56 &278 &48 &1.31--1.97 &7 &$407\pm117$ &$-0.11\pm0.11$ &$353\pm181$&... &3.06 &0.0038\\
02/05/1992 &S15W31  &3.10E-05 &20:15 &20:19  &272 &67 &1.18--2.40 &9 &$562\pm39$ &$-0.01\pm0.31$ &$552\pm288$ &... &2.88 &0.022 \\
07/25/1997 &N16W54 &5.00E-06 &20:17 &20:23  &278 &75 &1.36--1.87 &4 &$690\pm85$ &$-0.40\pm0.34$ &$587\pm176$ &LAS 21:01 &2.6 &0.016 \\
10/21/1997 &N16E07 &3.30E-06 &17:00 &17:44 &95  &40 &2.25--2.26 &2 &$119\pm19$ & ...           & ...        &LAS 18:03 &2.97&0.0033 \\
06/16/1998 &S W115 &1.00E-05  &18:03 &18:00 &268 &64 &1.33--1.90 &5 &$527\pm30$ &$1.56\pm0.60$ &$1058\pm227$&LAS 18:27 &2.67&0.053 \\
08/19/1998 &N32E75 &3.90E-04 &21:35  &21:42 &53  &51 &1.25--1.97 &3 &$1425\pm116$ & ... & ... &... &1.62&0.020 \\
{\bf 08/24/1998} &{\bf N35E09} &1.00E-04 &21:50 &22:07  &50  &40 &1.46--1.71 &3 &$\mathbf{446\pm286}$  &...         &  ...          &... &{\bf 3.38}&{\bf 12.4} \\

\hline

\end{tabular}
$^1$ Possible onset observed at $>60$~MeV; not evident above background from preceding event at 20~MeV.\\
\end{table}

\newpage
\begin{table}
\label{T-list2}
\renewcommand{\arraystretch}{.7}
\setlength{\tabcolsep}{.02in}
\caption{MLSO Mk4 Coronameter CMEs Associated With $\sim25$~MeV Solar Proton Events}
\begin{tabular}{llcccccccccclcc}

\hline
1   &   2  &  3       &  4      &  5          & 6  &7& 8     &9 &10         &11          &12        &13     &14      & 15 \\
Date&Source&SXR Peak  &SXR Onset&MLSO 1$^{st}$&CPA&W&Height &N &Av. V      &Av. Acc     &Final V   &Space  &$\gamma$&I(20 MeV)\\
 
[m/d/y]     &Location&[W/m${^2}$]&[UT] & Det. [UT]    &[deg]&[deg]&Range [R$_s$]   &       &[km/s]&[km/s${^2}$]&[km/s]&Coronagraph  &     &  \\
\hline
Mk4&\\
01/20/1999&NE limb &5.20E-05 &19:06 &18:16  &86 &130 &1.90--2.54 &8 &$112\pm5$ &$0.11\pm0.01$ &$301\pm18$ &... &3.63&0.27 \\
05/09/1999 &N W95 &7.60E-05   &17:53 &18:02  &318 &68 &1.45--2.80 &6 &$859\pm144$ &$-0.33\pm0.17$ &$709\pm229$ &LAS 18:27 &2.52 & 0.055 \\
06/01/1999 &N W120 & ...      & ...  &18:43 &335 &70  &1.60--2.20 &3 &$908\pm149$ & ... & ... &LAS 19:37 &2.41&1.09 \\
08/28/1999 &S26W14 &1.10E-04 &17:52 &18:05 &193 &85 &1.42--2.52 &7  &$498\pm121$ &$-0.21\pm0.06$ &$353\pm171$ &LAS 18:26 &2.55&0.0024 \\
02/17/2000 &S29E07 &1.30E-05 &20:17 &20:31 &173 &145 &1.31--2.50 &8 &$606\pm67$ &$0.13\pm0.08$ &$690\pm52$   &LAS 21:30 &2.04&0.028 \\
03/22/2000 &N14W57& 1.10E-04 &18:34 &18:42 &307 &51  &1.21--2.53&10 &$565\pm55$ &$-0.30\pm0.15$ &$336\pm172$ &LAS 19:31 &3.59 &0.022\\
05/10/2000 &N14E20 &8.70E-06 &19:26 &19:51 &41  &70  &1.41--2.90* &3 &$510\pm4$ &... &...     &LAS 20:06 &2.8&0.0029  \\
05/17/2000 &...    & ...& ...&18:43 &260 &50 &1.35--1.98 &4 &$589\pm94$ &$-0.02\pm0.46$ &$581\pm248$ &LAS 19:26 &4.82&0.012 \\
05/23/2000 &N23W41 &9.50E-06 &20:48 &20:56 &312 &26 &1.50--2.40 &6 &$698\pm100$ &$-0.20\pm0.08$ &$609\pm125$ &LAS 21:30 &3.8 &0.0022 \\
06/02/2000 &N16E60 &7.60E-05 &18:48 &19:57  &97 &84  &1.15*--2.28 &14 &$354\pm57$ &$0.18\pm0.05$&$563\pm100$ &LAS 20:30 &3.93 &0.0011 \\
06/10/2000 &N22W38 &5.20E-05 &16:40 &16:58 &295 &100 &1.39--2.40 &5 &$986\pm134$&$0.97\pm0.79$ &$1285\pm397$ &LAS 17:08 &2.2&1.3 \\
06/28/2000 &N~~W95? &3.70E-06 &18:48 &18:42 &294 &32 &1.22*--2.40 &9 &$602\pm73$ &$0.68\pm0.13$ &$1065\pm143$ &LAS 19:31 &3.51 &0.0029\\
09/07/2000 &N06W47 &7.20E-06 &20:32 &19:40   &274 &59 &1.41*--1.50 &2 &$248\pm331$ & ... & ...                 &LAS 21:30 &2.97&0.0082 \\
12/18/2000 &N17W71 &2.90E-06 &18:17 &18:07 &303 &12 &1.08*--2.30 &3 &$1080\pm247$ & ... &... &LAS 18:30 &2.21&0.0007 \\
01/20/2001 &S07E40 &1.20E-05 &18:30 &21:13 &96 &56 &1.25*--2.18 &4 &$1262\pm107$ &$5.45\pm1.11$ &$2528\pm356$ &LAS 21:30 &3.09&0.055\\
09/12/2001 &S16W62? &9.60E-06 &21:05 &21:23 &243 &77 &1.33--2.69* &13 &$441\pm58$ &$0.36\pm0.02$ &$784\pm32$ &LAS 22:06 &3.11&0.0032 \\
11/22/2001 &S25W67 &3.80E-05 &20:24 &20:22 &240 &61 &1.27*--2.58 &6  &$1226\pm30$ &$-0.73\pm0.33$ &$971\pm142$&LAS 20:30 &... &0.3 \\
12/28/2001 &S~~E100 &3.40E-04 &19:46 &19:53$^1$ &114 &$>67$ &2.14--2.60 &7 &$310\pm7$ &$0.68\pm0.71$ &$641\pm340$&LAS 20:06 &3.06 &0.94\\
01/08/2002   &N~~E120 & ...     & ...   &17:47 &86  &52 &1.20*--2.56* &6 &$488\pm43$ &$0.06\pm0.13$ &$539\pm164$ &LAS 17:54 &3.15 &2.08\\
07/15/2002 &N14E01   &3.00E-04 &19:56&19:55 &40 &75  &2.06--2.38 &3 &$570\pm83$ &... &... &LAS 20:30 &4.14 &2.36\\
08/03/2002& S16W76    &1.00E-04 &18:59 &19:08 &293 &62 &1.41--2.14 & 5 &$698\pm76$ &$0.26\pm0.36$ &$773\pm146$ &LAS 19:31 &3.54 &0.0048\\
10/13/2002 &S07W54 &4.70E-06  &17:46 &17:55 &260 &85 &1.46--2.47 &9  &$464\pm122$ &$-0.11\pm0.09$ &$384\pm110$ &LAS 19:35 &3.72 &0.0010\\
12/02/2002 &$>$W90 & ...      & ... &17:43 &316 &82 &1.51*--2.40 &7 &$261\pm103$ &$0.21\pm0.08$ &$488\pm165$ &LAS 17:54 &4.01 &0.0004\\
12/19/2002 &N15W09 &2.70E-05 &21:54 &21:56 &308 &115 &1.66--2.33 &6 &$524\pm123$ &$-0.20\pm0.35$ &$446\pm273$   &LAS 22:06 &2.6 &0.10\\
01/27/2003 &S17W23 &2.40E-06 &21:50 &$\sim21$:43 &195 &122 & 1.40--1.93 &4 &$520\pm74$ &$0.28\pm0.09$ &$615\pm44$&LAS 22:23 &2.8 &0.0007\\
03/17/2003 &S14W39  &1.50E-04 &18:50 &19:24   &275 &70 &1.68--2.15 &2 & $928\pm197$     & ...          & ...      &LAS 19:54 &3.89 &0.017\\
06/17/2003 &S07E55 &6.8E-05  &22:27 &22:42   &115 &110 &1.18--1.89 &4 &$1210\pm173$ &$0.81\pm2.96$ &$1358\pm712$ &LAS 23:18 &4.7  &0.39\\
10/04/2003 &...    & ...      & ...  &19:12   &273 &16  &1.30*--2.33 &3 &$1818\pm248$ &... & ... &LAS 19:31 &3.72 &0.0011\\
10/26/2003 &N02W38 &1.20E-04 &17:21 &17:23   &273 &129 &1.27*--2.50 &9 &$629\pm54$&$0.77\pm0.22$ &$1186\pm202$ &LAS 17:54     &2.98 &12.5\\
{\bf 10/29/2003} &{\bf S15W02} &1.00E-03 &20:37 &20:48   &201 &123 &1.42--2.24 &4 &$\mathbf{1335\pm166}$ &$\mathbf{0.33\pm2.17}$ &$\mathbf{1412\pm347}$ &LAS 20:54 &{\bf 2.34}&{\bf 48.1}\\
{\bf 11/02/2003} &{\bf S14W56} &8.30E-04  &17:03 &17:17   &249 &133 &1.30--2.90* &6 &$\mathbf{1507\pm220}$ &$\mathbf{1.01\pm2.43}$ &$\mathbf{1825\pm978}$ &LAS 17:30 &{\bf 2.2} &{\bf 42.4}\\
\hline
\end{tabular}
$^1$CME slowly rises for $\sim120$ minutes prior to this time, followed by the rapid expansion of a loop that forms high in corona. The CME parameters relate to this rapid expansion.

\end{table}

\newpage
\begin{table}
\label{T-list3}
\renewcommand{\arraystretch}{.7}
\setlength{\tabcolsep}{.02in}
\caption{MLSO Mk4 Coronameter CMEs Associated With $\sim25$~MeV Solar Proton Events (Cont.)}
\begin{tabular}{llcccccccccclcc}

\hline
1   &   2  &  3       &  4      &  5          & 6  &7& 8     &9 &10         &11          &12        &13     &14      & 15 \\
Date&Source&SXR Peak  &SXR Onset&MLSO 1$^{st}$&CPA&W&Height &N &Av. V      &Av. Acc     &Final V   &Space  &$\gamma$&I(20 MeV)\\
 
 [m/d/y]    &Location&[W/m${^2}$]&[UT] & Det. [UT]    &[deg]&[deg]&Range [R$_s$]   &       &[km/s]&[km/s${^2}$]&[km/s]&Coronagraph  &     &  \\
\hline
08/18/2004 &S14W90 &1.80E-04  &17:29 &17:45   &247 &72  &1.38*--2.80 &9 &$570\pm119$ &$-0.17\pm0.13$ &$458\pm223$  &LAS 17:54 &2.77 &0.004\\
09/12/2004 &N03E49 &4.80E-05  &00:04 &00:13   &99  &107 &1.25--2.91* &12 &$537\pm35$ &$0.38\pm0.12$ &$938\pm166$   &LAS 00:36 &3.84 &4.95\\
04/19/2005 &S12E57 &8.00E-07  &21:35 &21:48   &95 &96 &1.26--2.45* &12 &$604\pm99$ &$0.02\pm0.24$ &$632\pm231$ &LAS 22:06 &1.74 &0.0004\\
05/11/2005 &S10W47 &1.10E-05  &19:22 &19:33  &229 &62 &1.39--2.40 &4  &$720\pm39$ &$-0.59\pm0.38$ &$476\pm197$ &LAS 20:13&2.86 &0.013\\
05/13/2005 &N12E11 &8.00E-05  &16:13 &17:19  &149 &258 &1.75--2.26 &7 &$333\pm55$ &$0.07\pm0.01$ &$373\pm48$ &LAS 17:12&4.39& 0.26\\
06/12/2005 &N06W21 &3.50E-06  &01:50 &02:21  &294 &33 &1.62--2.78* &4 &$523\pm11$ &$0.23\pm0.12$ &$696\pm80$ &LAS 02:36 &2.91 &0.0016\\
06/16/2005 &N08W90 &4.00E-05  &20:01 &19:53  &285 &77 &1.19--2.31 &7 &$554\pm35$ &$1.00\pm0.20$ &$1247\pm165$ & ...    &1.83 &1.12\\
07/09/2005  &N12W28 &2.80E-05  &21:45 &22:03  &298 &75 &1.38*--1.99 &5 &$537\pm117$ &$0.85\pm0.50$ &$860\pm329$ &LAS 22:30 &2.78 &0.061\\
07/12/2005 &N09W67 &1.50E-05  &15:47 &16:54  &278 &75 &2.00--2.60 &6 &$489\pm154$ &$0.14\pm0.21$ &$537\pm69$ &LAS 16:54 &3.23 &0.0055\\
07/13/2005 &N08W90 &1.20E-05  &21:49 &21:58  &263 &30 &1.60--2.15 &5 &$495\pm129$ &$-0.47\pm0.04$ &$330\pm 143$ &LAS 22:30 &3.23 & 0.20\\
08/22/2005 &S11W54 &2.60E-05  &00:44  &00:50  &278 &76 &1.17*--2.86* &9 &$448\pm41$ &$0.13\pm0.02$ &$609\pm16$ &LAS 01:31 &3.69 &0.26\\
08/31/2005 &S,W120 & ...      & ...  &22:14  &182 &105 &1.50--2.25 &3 &$1503\pm236$ & ... & ... &LAS 22:30 &1.56 &0.035\\
09/07/2005  &S11E77 &1.70E-03  &17:17 &17:34  &105 &104 &1.36--1.62 &2 &2483        & ...           & ...        &...       &2.54 &30.4\\
11/05/2006 & ...   & ...      & ...  &17:32  &75  &48  &2.09--2.19 &2 &$1140\pm570$ & ...           & ...        &LAS 17:54&2.03 &0.0002\\
11/06/2006 &N00E89 &8.20E-06  &17:43 &17:27  &75  &54  &1.22--1.82 &6 &$599\pm8$ &$0.06\pm0.15$ &$619\pm44$      &LAS 17:54 &1.62 &0.0021\\
06/12/2010  &N23W43 &2.00E-05 &00:30 &$\sim00$:59 &303 &66 &1.36--2.86 &8 &$549\pm23$ &$-0.05\pm0.03$ &$502\pm4$ &LAS 01:31 &2.79 &0.015\\
08/31/2010  & W145  & ...     &...   &21:00  &242 &33 &1.34--1.64 &3 &$579\pm104$ & ... & ... &LAS 21:17 &2.81 &0.009\\
09/08/2010  &N21W87 &3.30E-06  &23:05 &$\sim23$:14 &300 &55 &1.14--2.95* &18 &$515\pm47$ &$0.34\pm0.05$ &$940\pm107$ &LAS 23:26 &3.02 &0.0077\\
01/28/2011 &N16W88 &1.30E-05  &0:44 &$\sim00$:59 &305 &50 &1.36--2.68 &9 &$588\pm62$ &$-0.15\pm0.32$ &$464\pm209$ &LAS 01:25 &2.55 &0.12\\
03/16/2011 &N17W72 &3.70E-06  &17:52 &18:34  &304 &68 &1.37--2.67 &11 &$293\pm41$ &$0.15\pm0.02$ &$535\pm25$ &LAS 19:12 &3.99 &0.013\\
09/06/2011  &N14W07 &2.10E-04  &22:12 &01:51  &130 &50 &1.78--2.19 &3 &$820\pm308$ & ...          & ...       &LAS 02:24 &1.72 &0.047\\
09/08/2011 &W143    & ...      & ...  &21:54  &314 &60 &1.31--2.41 &7 &$774\pm72$ &$0.64\pm0.06$ &$1072\pm102$ &LAS 22:12 &...&0.003 STA\\
11/12/2011 &E103  & ...      & ...  &18:26  &69  &32 &2.04--2.46* &3 &$489\pm124$ &... &... &LAS 18:36 &...&0.0002 STB\\
05/26/2012 &N05W110 & ...     & ...  &20:43  &290 &130 &1.42--2.06 &4 &$1014\pm294$ &$-2.29\pm1.57$ &$550\pm626$ &LAS 20:57 &4.36 &0.21\\
06/01/2012 &N12W89  &3.30E-06 &22:30 &22:19  &279 &57 &1.21*--2.23 &13 &$275\pm80$ &$0.18\pm0.05$ &$515\pm124$ &LAS 23:12 &3.67  &0.0021\\
06/03/2012 &N16E38  &3.30E-05  &17:48 &18:00  &50  &71 &1.42--2.78* &4 &$662\pm28$ &$0.21\pm0.02$ &$806\pm19$ &LAS 18:12 &2.3 &0.030\\
07/06/2012 &S13W59 &1.10E-04&23:01 &23:07  &239 &86 &1.12*--1.95 &6 &$1387\pm706$ &$-2.43\pm3.31$ &$994\pm1308$ &LAS 23:24 &2.31 &0.40\\
07/13/2012 &W130  & ...       & ...  &19:47  &218 &70 &1.34--2.71* &6 &$419\pm16$ &$-0.02\pm0.06$ &$402\pm47$ &LAS 20:00 &2.61 &0.015 \\
05/10/2013 &W140 & ...        & ...  &18:09 &260 &80 &1.45--2.15 &5 &$644\pm234$ &$1.24\pm0.67$ &$1034\pm454$ &LAS 19:12 &2.55 &0.011\\
\hline

\end{tabular}

\end{table}

\end{landscape}

\begin{figure}
   \centerline{\includegraphics[width=1.0\textwidth,clip=]{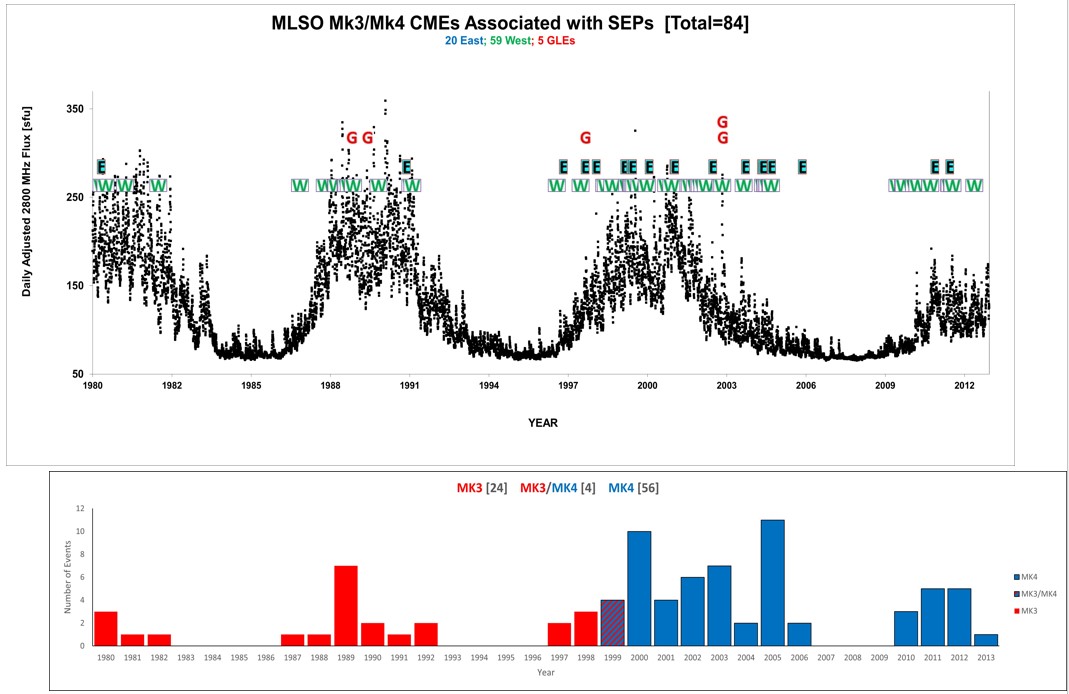}
              }
              \caption{The times of 84 MLSO CMEs associated with SEP events shown relative to the daily adjusted Penticton 10.7~cm 
solar radio flux. W (E) indicates 59 (20) cases where the solar event was on the western (eastern) hemisphere with respect to the observing spacecraft, and G that the
SEP event was a ``Ground Level Enhancement" (GLE) observed by neutron monitors. The bottom panel shows the annual number of SEP-associated CMEs observed by Mk3 (red) or Mk4 (blue); in 1999 (purple), both instruments observed the CMEs. The performance of Mk4 was {\bf greatly} reduced after a lightning strike in 2009. }
   \label{MLSO}
   \end{figure}

 Figure~3 summarizes the distribution in time of these 84 SEP events associated with MLSO Mk3/4 CMEs over four solar cycles (21-24) as indicated by the 10.7~cm solar radio flux (\url{https://www.spaceweather.gc.ca/forecast-prevision/solar-solaire/solarflux/sx-en.php}) in the top panel.  As expected, these events are predominantly associated with higher solar activity levels.  The number (59) of western hemisphere events with respect to the observing spacecraft, indicated by `W' exceeds that of eastern events (`E') (20).  As discussed above, this largely reflects the greater ability to associate CMEs and SEP events unambiguously for western events (events originating behind the respective limb are also included here) in addition to the western bias introduced by the connection of the spiral interplanetary magnetic field to the western hemisphere.  `G' indicates five SEP events that are GLEs and occurred at the peak of cycle~22 or during the ascending and declining phases of cycle~23.    
 
 The annual number of SEP-associated CMEs in the bottom panel of Figure~3 is also influenced by other factors. Considering MLSO operations, Figure~4 shows the MLSO observing periods on the days when SEPs-associated CMEs were detected. In 2003, an additional observer was added to the Observatory staff, resulting in an extension of the daily observing time from $\sim5$~hours to $\sim9$~hours as is evident in the top part of the figure.   Another factor is that, in the 1980s until 1991, large format tapes (recording one hour of observations/20 images) were used to transport Mk3 observations to HAO for processing; the tapes were then returned to MLSO for reuse. Due to the high costs of tape transport, after the failure of SMM in September, 1980, only two images/day were retained if the on-duty observer did not notice a CME in the data for that day. However, without the benefit of subtraction images or movies, it is likely that some CMEs were overlooked. \cite{stcyr2015} give details of the influences on the Mk3 duty cycle and how discarding data has severely impacted the Mk3 CME archive. This policy was changed in 1989, when all images were gathered. The use of more compact tapes from 1991 also increased the data coverage.    Another factor is the improved capabilities of Mk4 (the blue histogram in Figure~3) relative to Mk3 (red histogram).  However, the performance of Mk4 was degraded following a lightning strike in 2009 that ``fried" the detector and damaged ancillary electronic equipment. The optics were moved in 2010 to try to improve the signal but this caused the outer field-of-view to shift down to 2.5~R$_s$. Unfortunately, the signal continued to seriously degrade from 2010 through 2013 with increased electronic noise level in the detector and impact on the ability to provide an absolute brightness calibration. 
 
 There were $\sim900$ $\sim25$~MeV proton events during the period in Figure~3 \citep[cf.,][and references therein]{richardson2017}, suggesting that around 9\% of these events could be associated with an MLSO CME. This is reasonably consistent with the $\sim20$\% expected taking into account both the $\sim9$~hours (or less) MLSO daily observing time and typically $\sim200$~days of observations/year, with further reductions, for example, for instrumental and operational issues, poor viewing conditions and the greater difficulty of making reliable SEP associations for eastern events. 

\begin{figure}
   \centerline{\includegraphics[width=0.8\textwidth,clip=]{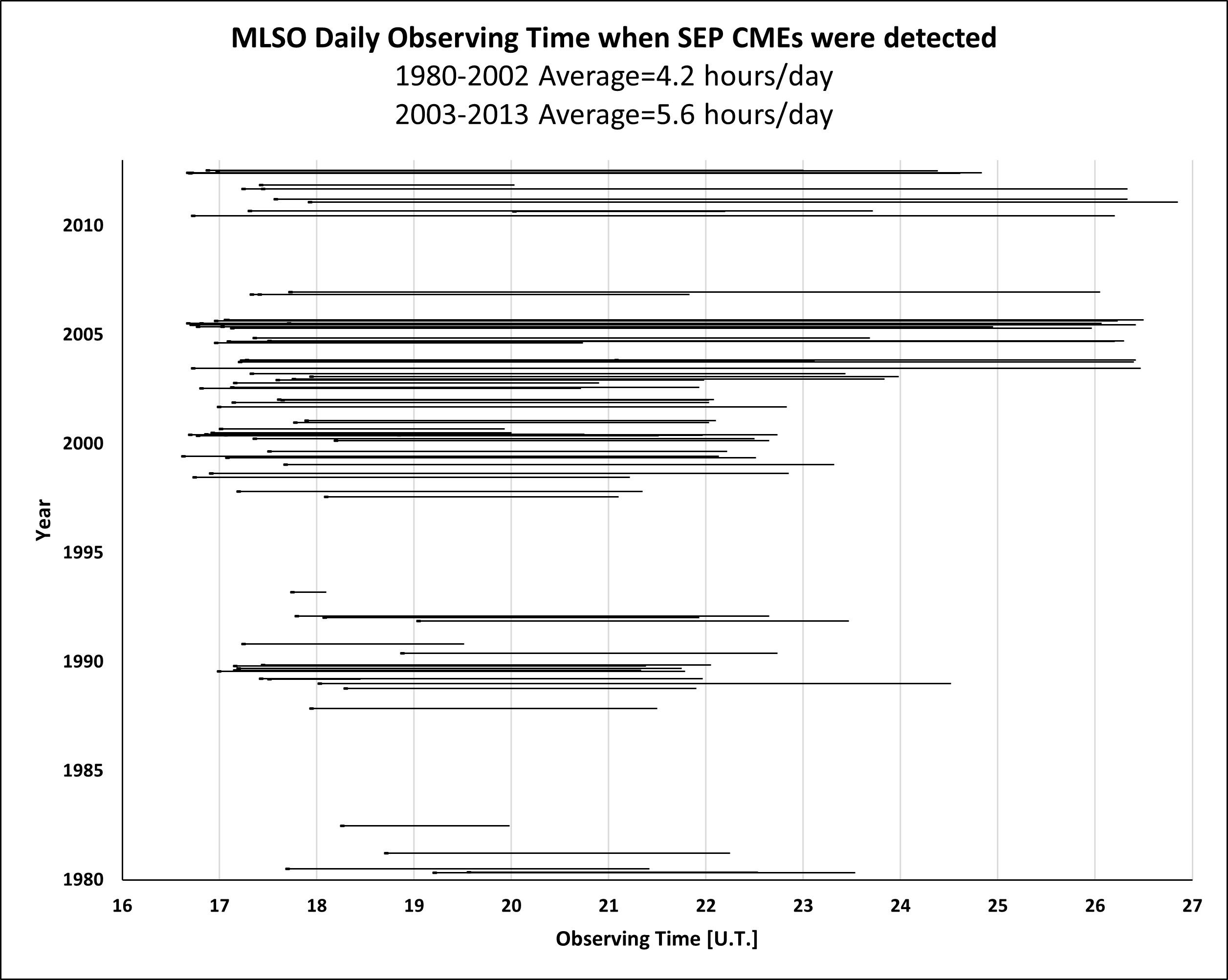}
              }
              \caption{Daily observation periods at MLSO on days when SEP-associated CMEs were detected, expressed as Universal Time versus Year. Note the increase after an observer was added in 2003.}
   \label{ops}
   \end{figure}

 \cite{Webb1994} determined instrument-specific ``visibility functions" to compare CME rates deduced from different coronagraphs and combined those results with their respective duty cycles \citep[e.g.,][]{macqueen1991}.  This concept was recently updated for modern coronagraphs by \cite{Vourlidas2020}.  Because of the longevity and high duty cycle of the SOHO LASCO coronagraphs, we can determine the Mk3/Mk4 visibility function for SEP-associated CMEs by comparing the 1997-2013 observations from both sets of instruments.  There were around 499 $\sim25$~MeV proton events in May 1997-May 2013 \citep[cf.,][]{cane2010, richardson2014} although no LASCO observations are available for 45 of these events. Most of these SEP events have an associated SOHO and/or STEREO CME as well as an identification of the associated solar source region.  We then selected a subset of 187 SEP events where the associated CME appeared between 17-02 UT, the nominal observing window for MLSO. (Since 9/24 hours of observations per day is a duty cycle of $\sim38$\%, we might expect $\sim190$ CMEs, which agrees well with the number actually observed.) Of these 187 SEP events, 111 events occurred during MLSO data gaps.  For another 64 events, MLSO made at least one observation of the CME, and in a further 6 cases, MLSO observed the CME, but no measurements of the CME were possible.  Finally, in 6 cases, MLSO was observing but the CME was not detected. Thus, {\it when making observations}, Mk3/4 detected at least 70/76 (92\%) of the SEP-associated  CMEs observed by SOHO/LASCO in 1996-2013. Also, four of the six missed events occurred after the lightning strike in 2009 which degraded the performance of Mk4 and so these could also be reasonably removed from this comparison.  This analysis indicates that if a $\sim25$~MeV proton event occurs and MLSO is making observations, then the associated CME is highly likely to be observed in the low corona.

\begin{figure}
   \centerline{\includegraphics[width=1.0\textwidth,clip=]{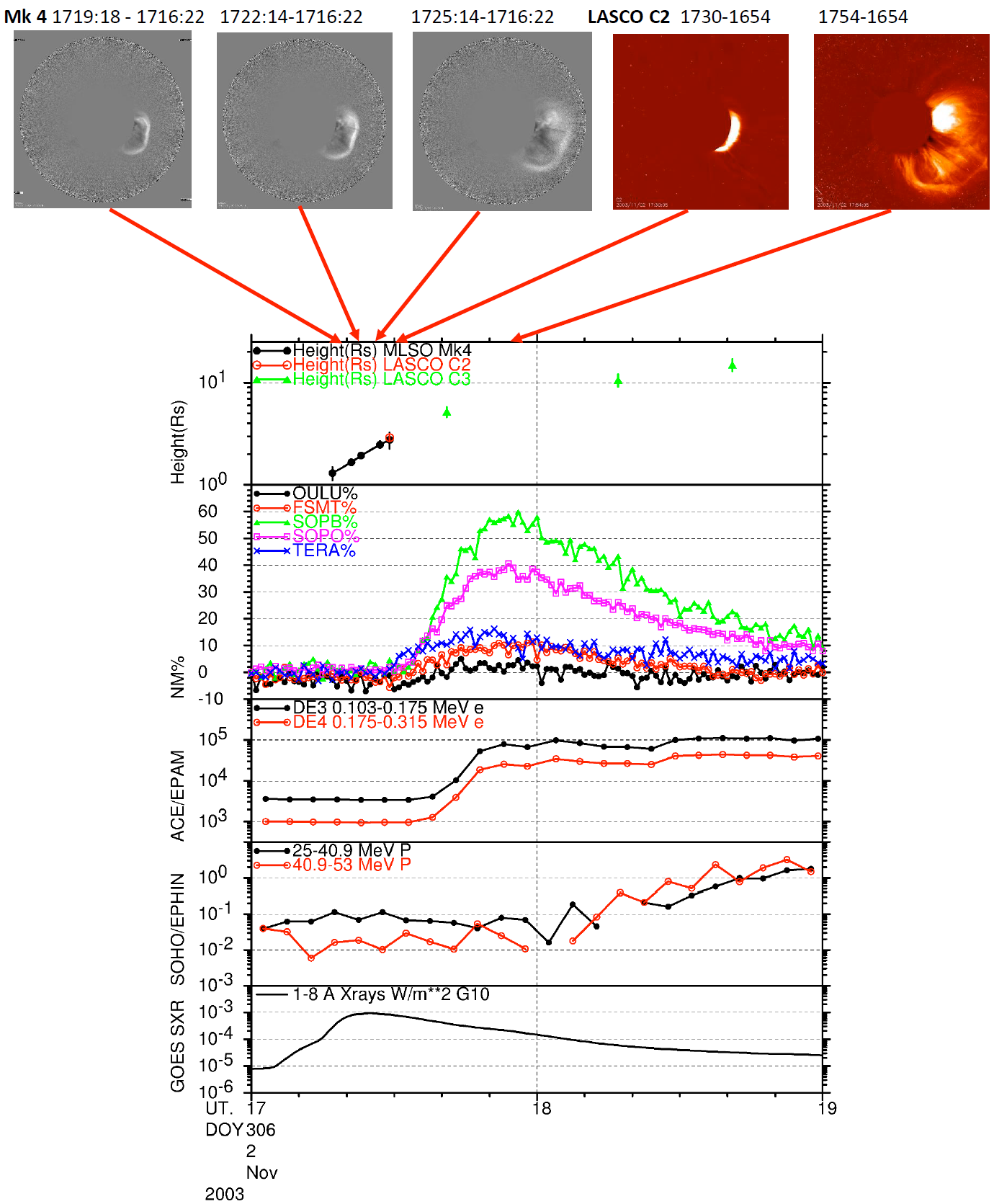}
              }
              \caption{MLSO Mk4 coronameter (gray) and SOHO/LASCO C2 coronagraph  (orange) observations of the CME associated with GLE~67 on November 2 2003. The coronal images are base-difference images obtained by subtracting a pre-event image at the times indicated. The main figure shows: the plane of the sky height of the leading edge of the CME in Mk4 (black) or LASCO C2 (red) and C3(green); The percentage counting rate increase (1 minute averages, normalized at 17:00UT) in several neutron monitors (Oulu, Fort Smith, South Pole, South Poles Bares and Terre Adelie) from \url{https://www.nmdb.eu/nest/}; Near relativistic electron intensities (5~minute averages) from ACE/EPAM; 25-53 MeV proton intensities (5~minute averages)  from SOHO/EPHIN; and the GOES 10 soft X-ray intensity. Note that the CME was observed in several MLSO images before the onset of the particle event was detected at Earth. All times shown are as observed at Earth with no corrections for propagation time from the Sun.}
   \label{GLE67}
   \end{figure}

\subsection{Examples of SEP events associated with MLSO CMEs}
\label{SS-examples}
Figure~5 shows an example of how MLSO can observe CME motion well before the onset of an SEP event, even if this event is a GLE.  In this case, we show GLE~67 \citep[e.g.,][and references therein]{mishev2021} on November~2, 2003 associated with an X8.3 flare at S14$^\circ$W56$^\circ$. At the top of the figure are examples of difference images from MLSO Mk4 (gray) and SOHO/LASCO C2 (orange) related in time by the arrows to the SEP observations during the two-hour interval shown below. The SEP observations are from several neutron monitors (see the figure caption for more details), near-relativistic electrons from ACE/EPAM and 25-53 MeV protons from SOHO/EPHIN.  The GOES soft X-ray intensity is also shown. The CME leading edge height is shown in the top panel of the main figure using a logarithmic scale that emphasizes observations close to the Sun. Note that MLSO provides several measurements of the CME height and speed (black), commencing before the GLE particles and near relativistic electrons arrive at Earth. The first CME height estimate is available at 17:17~UT (corrected for CME direction relative to the scan stop/start location).  The height-time profile is then extended using the LASCO C2 and C3 observations. The energetic proton intensity observed by SOHO/EPHIN finally rises above the existing elevated background nearly an hour after the CME was first observed at MLSO. An interesting aside is that \cite{mishev2021} conclude that the GLE particles had a bidirectional (sunward/anti-sunward) flow during onset.  Such flows often occur in interplanetary coronal mass ejections \citep[e.g.][]{richardson1996, richardson2000}, and we note that  the onset of this SEP did indeed occur during passage of an ICME \citep{richardson2010}.  

A first-order fit to the CME leading edge height vs. time from the combined MLSO and LASCO coronagraph observations gives an average CME speed in the low-mid corona of $1922\pm274$~km/s, compared with a mid-corona speed of 2598~km/s (with no error quoted) in the CDAW CME
catalog (\url{https://cdaw.gsfc.nasa.gov/CME_list/}) which is based on a manual fit to LASCO data alone. The second-order MLSO-LASCO fit gives a final speed of $2100 \pm271$~km/s and an average acceleration of $81\pm 28$~m/s$^2$.  This contrasts with the 
CDAW catalog deceleration of -32.4 m/s$^2$ in the LASCO field of view, illustrating the effect of including the low corona observations on the inferred CME motions. 
Using a cubic spline interpolation (CSI), we estimated a maximum CME speed of 2487~km/s, and a maximum acceleration of $6.03 \pm 0.67$~
km/s$^2$ (considerably larger than the average acceleration) at a height of $2.29\pm 0.65$~R$_S$. The combined MLSO-LASCO height-time observations for GLE~67 have also been discussed by \cite{gopalswamy2012} (see their Figure 15). Assuming a second-order fit to the MLSO and initial
LASCO observations, they obtained a smaller acceleration of 2.4~km/s$^2$,
or 2.79~km/s$^2$ when including a correction for projection due to the flare longitude. Considering just the six observations below 3R$_s$, we obtained (Table~2) an average velocity of $1507\pm220$~km~s$^{-1}$ and average acceleration of $1.01\pm2.43$~km~s$^{-2}$. These disparate results illustrate the considerable variations in the CME motions inferred by applying different fitting techniques to the observations, as will be discussed further in Section~3.

\begin{figure}  
   \centerline{
               \includegraphics[width=0.48\textwidth,clip=]{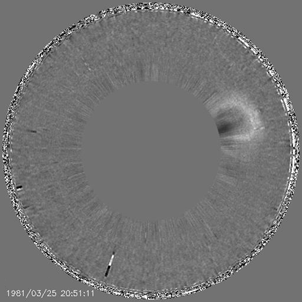}
               \hspace*{0.01\textwidth}
               \includegraphics[width=0.48\textwidth,clip=]{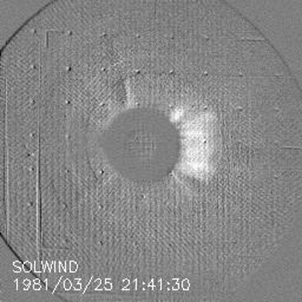}
              }
    \centerline{\includegraphics[width=0.8\textwidth,clip=]{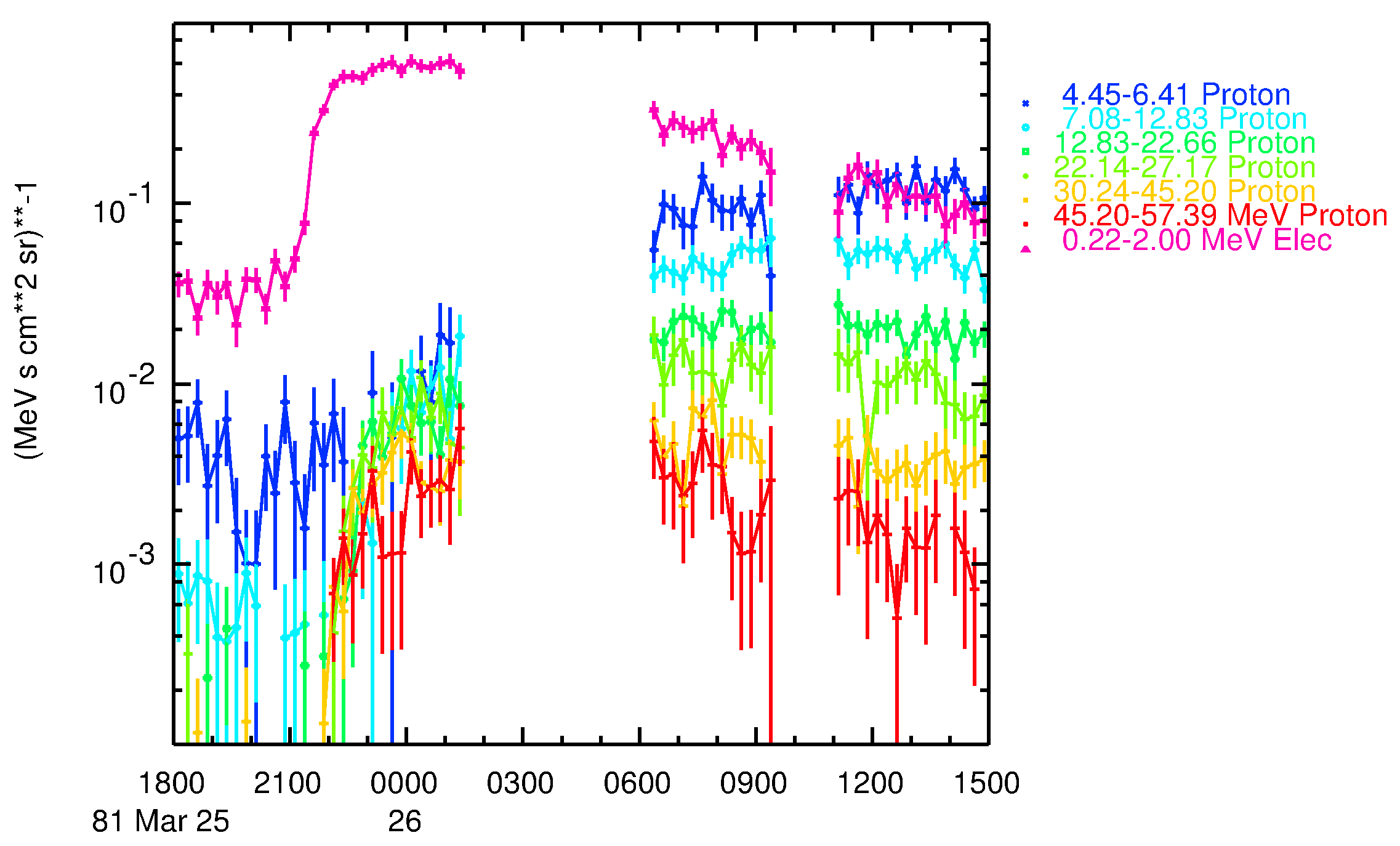}
    }          
              
\caption{MLSO Mk3 coronameter (top left) and SOLWIND coronagraph (top right) observations (at 20:51:11 and 21:41:30~UT, respectively) of a CME associated with an SEP event on March~25, 1981. The ISEE--3 observations (bottom panel) show that 0.22-2~MeV electrons were first detected by ISEE-3 at 21:15-21:30~UT, around $39\pm7$~minutes after the CME was first observed by Mk3 at 20:43 UT, followed by protons extending to at least $\sim50$~MeV.     
        }
\label{19810325}
\end{figure}

Figure~6 shows an example of a CME associated with an SEP event early in our study period, on March~25, 1981, observed by the MLSO Mk3 coronameter and SOLWIND coronagraph.  This event was associated with an X2.2 flare at W87$^\circ$ peaking at 20:48~UT.  The CME was first observed by Mk3 at 20:43 UT and tracked over 1.25-2.1~R$_S$ through nine images giving an average speed of $798\pm17$~km/s and average acceleration of $80\pm 130$~m/s$^2$.  ISEE--3/ICE detected the onset of 0.22-2~MeV electrons in the 15~minute average at 21:15-21:30~UT, around $39\pm7$~minutes after the CME was first observed by MLSO. The CME was first observed by SOLWIND at 21:41 UT and tracked over 4.2-7~R$_S$ with an average speed of 712~km/s, and acceleration of 138~m/s$^2$. This event illustrates that, notwithstanding the poorer performance of these earlier instruments, CME motions can still be inferred in the low-mid corona from these observations.
 
\begin{figure}
    \centerline{\includegraphics[width=1.0\textwidth,clip=]{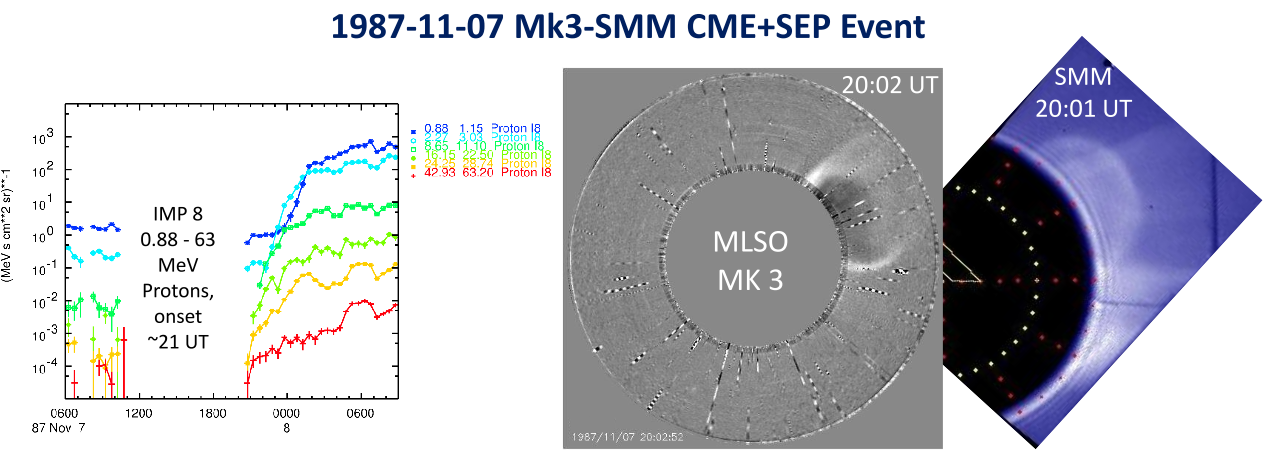}
    }
    \caption{MLSO Mk3 coronameter (center) and SMM coronagraph (right) observations of a CME on November 7, 1987 associated with an SEP event detected by IMP~8 (left).}
    \label{fig:19871107}
\end{figure}

 The CME shown in Figure~7 was observed by Mk3 and the SMM coronagraph on November~7, 1987. In this case, the images from each instrument shown are nearly coincident in time, illustrating the complementary observations from the ground and space. The CME and SEP event were associated with an M1.2 flare at W90$^\circ$ with peak intensity at 20:30~UT. MLSO first observed the CME at 19:51~UT and tracked it over 1.84-2.36~R$_s$. From the four MLSO frames available, we estimate a speed of $603\pm22$ km/s and an acceleration of $0.45\pm0.07$~km/s$^2$.  Unfortunately, we could not estimate the CME speed from the SMM observations. The IMP~8 GME proton data (30 minute averages) show an energy-dispersive onset that extended to at least 50~MeV, commencing at around 21~UT following a data gap.

\begin{figure}
    \centerline{
    \includegraphics[width=1.0\textwidth]{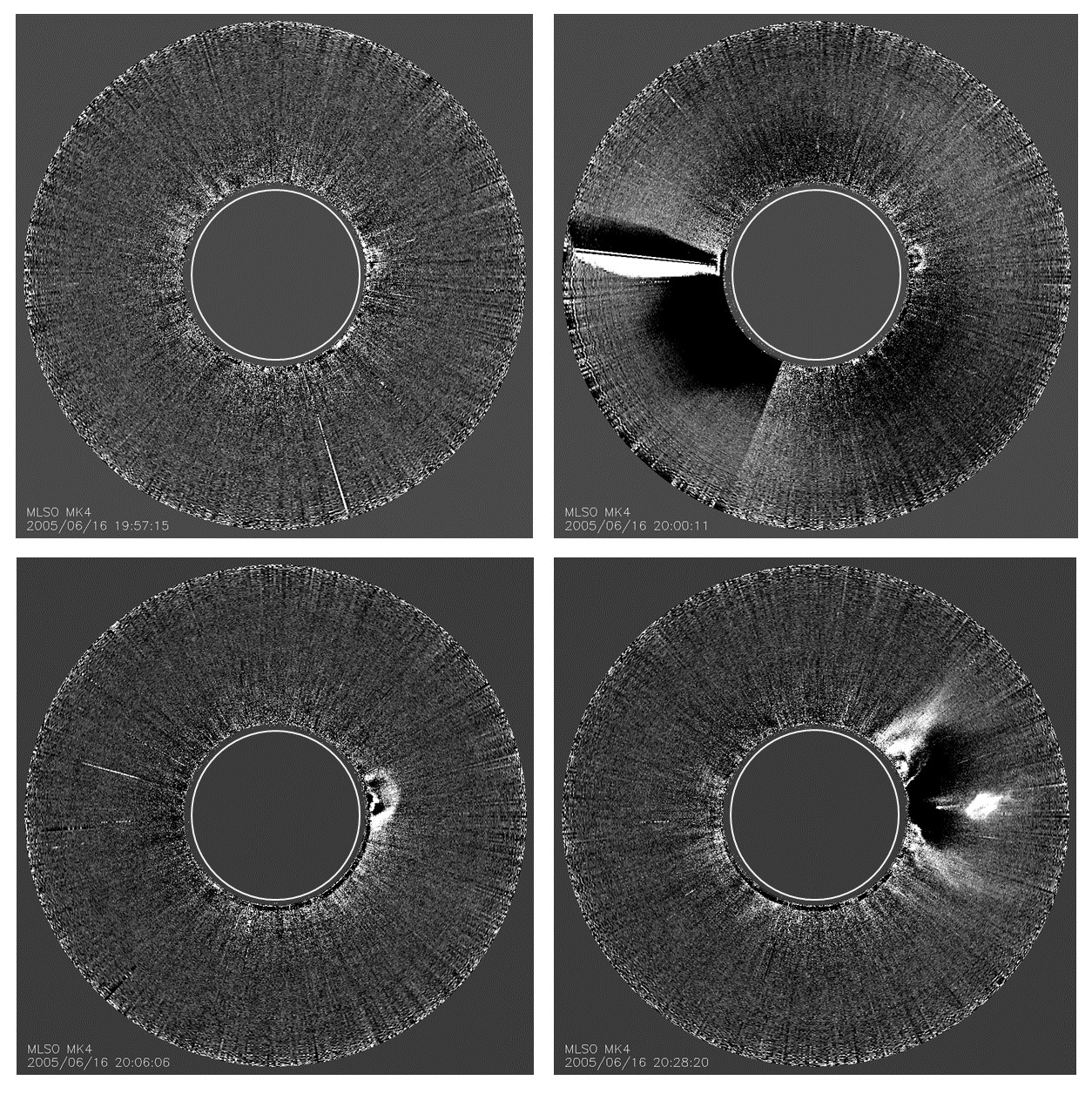}
    }
          
\caption{MLSO Mk4 observations of the SEP-associated CME on June 16, 2005. In the first (top left) image at 19:57~UT (with the image at 19:39~UT subtracted), the CME is barely visible in the field of view above the west limb at a position angle of $\sim270^\circ$. In the next image (20:00~UT, with the 19:42~UT image subtracted), the CME is again just visible at a similar height. In the image (bottom left) at 20:06~UT (with the 19:42~UT image subtracted), the CME has started to expand rapidly. By 20:28~UT (bottom right, with the 19:42~UT image subtracted), the CME leading edge has left the Mk4 field of view.  The skirt of the CME and a bright prominence knot at about 1.7~R$_s$, with a current sheet trailing behind it, are evident. A prompt SEP electron onset (not shown) was detected at Earth at 20:26~UT.
        }
\label{20050616}
\end{figure}

\begin{figure}  
   \centerline{
               \includegraphics[width=0.48\textwidth,clip=]{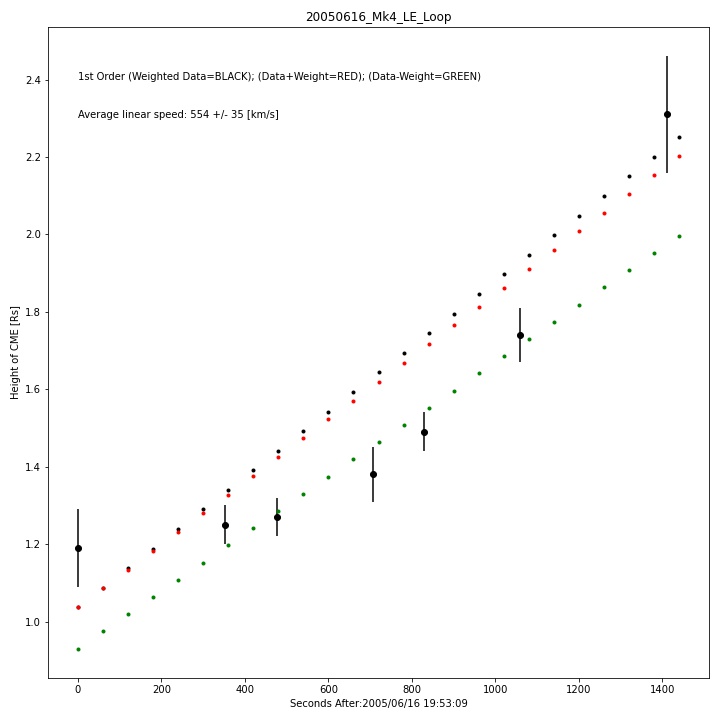}
               \hspace*{0.01\textwidth}
               \includegraphics[width=0.48\textwidth,clip=]{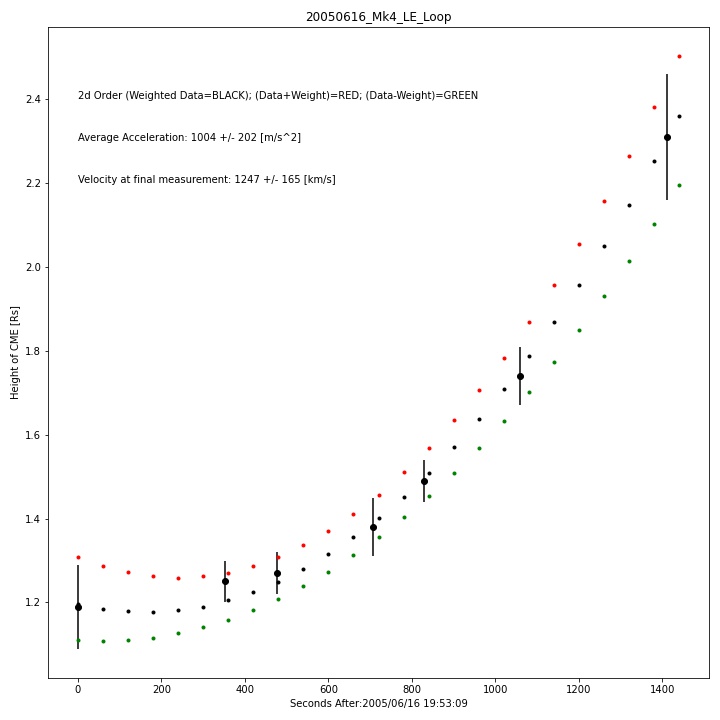}
              }        
              
\caption{First order (left) and second-order (right) fits to the Mk4 height-time profile for the June 16, 2005 SEP-associated CME in Figure~8. The strong acceleration in the low corona following an initial interval of slow expansion is evident.}
\label{20050616fit}
\end{figure}

 A final example of an SEP-associated CME, observed by MLSO Mk4 on June 16, 2005 and related to a flare at the west limb, is shown in Figure~8. The associated prompt SEP electron onset at Earth was at 20:26~UT.  The CME is just visible above the west limb at a similar height in the differenced images at 19:57~UT and 20:00~UT in the top row. In the image at 20:06~UT (bottom left), the CME has started to expand rapidly (following the onset of soft X-ray emission at 20:01~UT; Table~3), and by the time of the bottom right image at 20:28~UT, the leading edge had left the Mk4 field of view; there is evidence of a prominence knot and a trailing current sheet \citep{webb2016} in this image.  Overall, seven estimates of the CME height are available from the Mk4 observations.   No LASCO mid-corona observations are available for this event due to a data gap, so this event gives an example of how MLSO observations alone can provide information on the dynamics of an SEP-associated CME. In particular, this is a clear example of a CME that initially expands slowly and then accelerates substantially in the low corona.  Figure~9 shows first-order (left) and second-order (right) fits to the CME leading edge height-time profile from Mk4. The first-order (linear) fit gives an average speed of $554\pm35$~km/s but this is clearly a poor fit to the height-time profile for this strongly-accelerating CME.  The second-order fit gives an acceleration of $1.004\pm0.202$~km/s$^2$ and a speed at the final point of $1247\pm165$~km/s that is substantially higher than the average speed.

\begin{figure}
    \centering
    \includegraphics[width=0.7\textwidth]{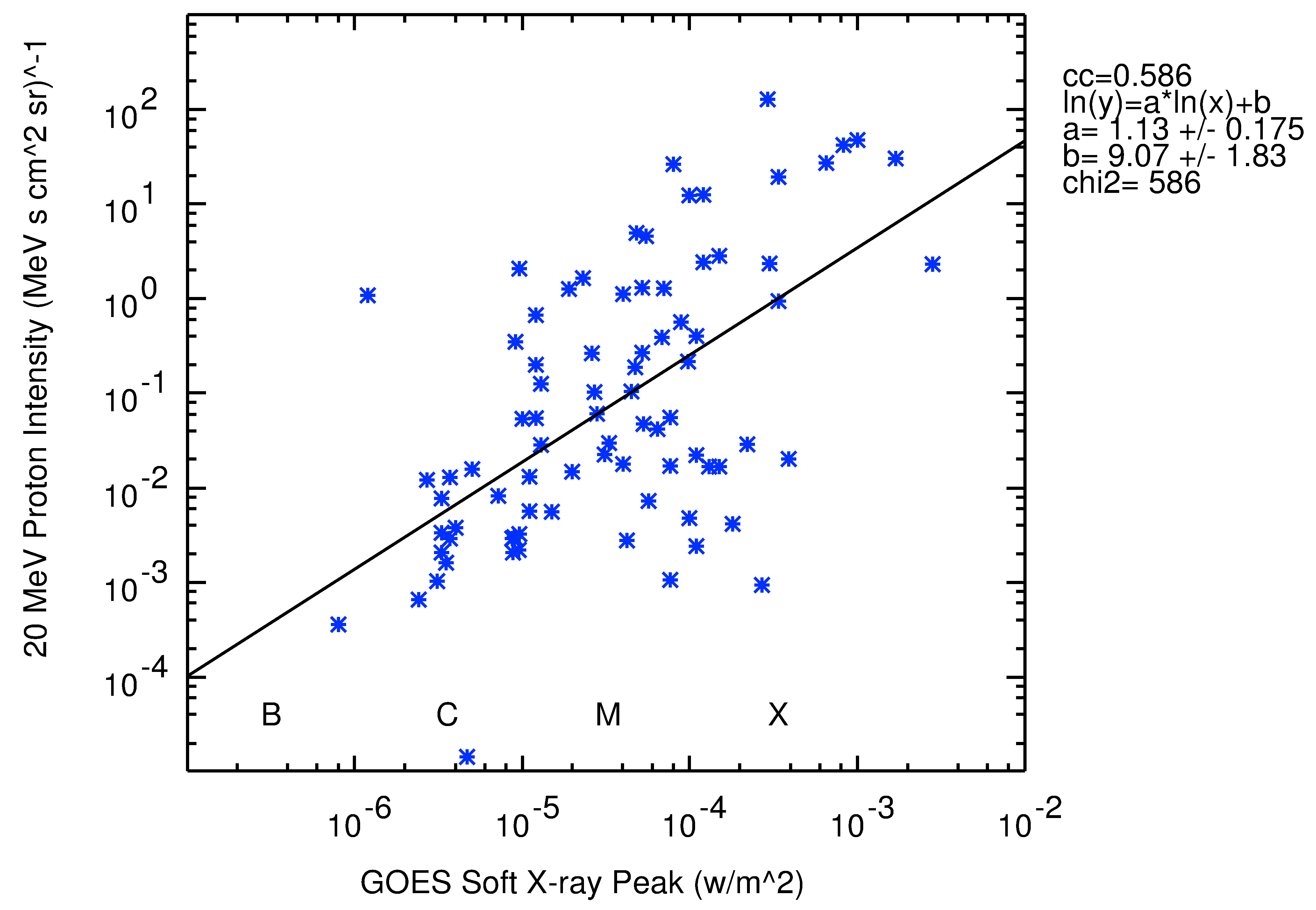}
    \caption{Peak 20~MeV proton intensity vs. peak soft X-ray intensity (W/m$^2$) of the associated flare from GOES for the SEP events associated with MLSO Mk3/4 CMEs. The SEP events cover a wide range of intensities and are associated with B to X10 class flares. Events at all longitudes are shown. }
    \label{fig:intxray}
\end{figure}
   
\section{Results}
\label{S-results}
\subsection{General properties of SEP events associated with MLSO Mk3/4 CMEs}
\label{SS-genprop}
We first summarize the characteristics of the SEP events associated with MLSO Mk3/4 CMEs.  Figure~10 demonstrates that these SEP events cover a wide range of peak proton intensities at 20 MeV and are associated with GOES soft X-ray flares ranging from B ($10^{-7}-10^{-6}$~W/m$^2$) to X ($>10^{-4}$~W/m$^2$) class. In Figure~10 (and also column~15 of Tables 1--3), the 20 MeV proton intensity is derived from an inverse power
law in energy fit ($dJ/dE\sim E^{-\gamma}$) to peak intensity observations in selected instrument channels, for example covering
4.2-63~MeV for the IMP 8 GME (with poorly calibrated channels removed), 4.3-53~MeV for SOHO/EPHIN and 8-67~MeV for SOHO/EPHIN. For most events, the power-law fit is a good representation of the peak intensity spectra in these energy ranges. In occasional cases where there is a spectral break in this energy range, the fit is restricted to include the energy channels that best represent the spectrum at around 20 MeV.  Where spectra are available for the same event from different spacecraft, we generally use the spectral fit with the smallest error. Values of the power law index $\gamma$ are shown in column 14 of Tables~1--3. The fit is then used to calculate the intensity at 20~MeV. 

For 18 SEP events, likely to be from the far side, no associated flare was recorded so they do not appear in Figure~10. In addition, in 8 cases, the peak proton intensity spectrum could not be obtained. Reasons include data gaps, a subsequent unrelated, more intense event obscured the peak of the event in question, or, in a few cases, the SEP event was detected by a spacecraft away from 1~AU, so the peak intensity will be influenced by the spacecraft location and is not plotted.  Of the 84 SEP events associated with MLSO CMEs, 22 (26\%) were associated with X-class flares, 29 (35\%) with M-class flares, 17 (20\%) with C-class flares and 1(1\%) with a B-class flare while, as already noted, 18 (21\%) had no associated flare and probably originated on the farside of the Sun.  Thus, the SEP events associated with CMEs detected by MLSO Mk3/4 are related to a wide range of solar eruptive events and are not, for example, strongly biased towards major flares, though such events are certainly well represented in our sample.  Note also that the 20 MeV proton intensity in Figure~10, even though events at all longitudes are included, shows the typical positive correlation with flare soft X-ray intensity \citep[e.g.,][]{cane2010, richardson2017}).

\begin{figure}
   \centerline{\includegraphics[width=1.0\textwidth]{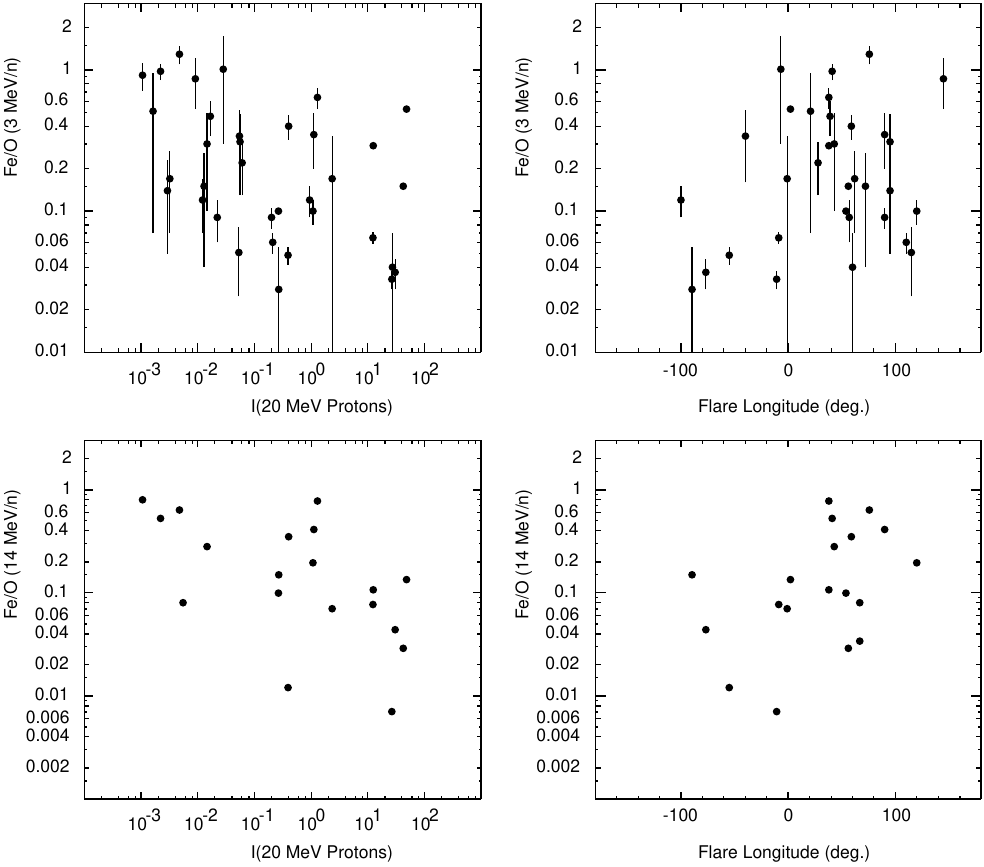}
              }
              \caption{Fe/O ratio vs. (left) peak 20 MeV proton intensity ((MeV s cm$^2$ sr)$^{-1}$) and (right) flare longitude at $\sim3$~MeV/n (top) and $\sim 14$~MeV/n (bottom).  The SEP events associated with the MLSO CMEs cover a wide range in Fe/O and show, in both energy ranges, the typical decrease in Fe/O with increasing peak proton intensity and tendency for Fe-rich events to originate on the western hemisphere.}
   \label{feo}
   \end{figure}

Figure~11 shows the Fe/O ratios for the SEP events associated with MLSO CMEs, when available, and their variation with the proton intensity at 20~MeV and the longitude of the associated flare.  The ratios at $\sim14$~MeV/n are taken from ACE/SIS as in \cite{cane2010} with a few later additions. Those at $\sim3$~MeV/n are generally from Wind/EPACT at 2.4-3.1 MeV/n with a few cases from STEREO/LET at 4-6 MeV/n where the event is better observed at STEREO.  The ratios are calculated using peak intensities, or as close as possible if the peak is missed, also avoiding abundance variations due to transport effects early in the event and local intensity peaks associated with interplanetary shocks.  The main point here is to again illustrate that the SEP events associated with the MLSO Mk3/4 CMEs have a wide range of properties.  In particular, they show (left panels) the typical general decrease in Fe/O with increasing proton intensity, including events that are Fe-rich compared to the typical ratio of $\sim0.1$ for large ``gradual" SEP events \citep[e.g.,][]{reames1999} and (right panels) the tendency for Fe-rich events to originate on the western hemisphere \citep[e.g.,][]{cane2006}. Though not shown here (but discussed later in Section~3.4), the electron to proton intensity ratios also cover a wide range from above $10^6$ to below $10^4$ where, following \cite{cane2010}, the e/p ratio is calculated using $\sim0.5$~MeV electron and $\sim25$~MeV proton intensities.

\subsection{Properties of SEP-associated MLSO Mk3/4 CMEs}
\label{SS-CMEprop}
Figure~12 summarizes some of the basic properties of the SEP-associated CMEs observed by Mk3/4.  The top left panel shows the distribution of the apparent (in the plane of the sky)  central position angle (PA) of the CMEs. The position angle is measured anti-clockwise with $0^\circ$ directed to the north.   The western (PA$\sim270^\circ$) bias, due to preferred IMF connection to the western hemisphere and selection bias, is clearly evident. In contrast,  a general population of more than 500 Mk3 CMEs reported in \cite{stcyr1999} and \cite{stcyr2015} shows an equal division between eastern and western events.  The middle left panel shows the CME central latitude. For our selected events, there is a bias towards the northern hemisphere.  This apparently reflects when the selected events occurred in relation to the time-varying northern-southern hemisphere asymmetries in SEP event sources during solar cycles \citep[e.g.,][]{richardson2016,richardson2017} and is not a general property of SEP-associated MLSO CMEs. The bottom left panel shows the distribution of CME full angular widths as measured in Mk3/Mk4.  The mean width is 73$^\circ$, compared to 37$^\circ$ for the general population of Mk3 CMEs, suggesting that MLSO CMEs associated with SEPs are on average wider than is typical. The widest SEP-associated CME had a width of 258$^\circ$, and no cases of full halo (360$^\circ$) CMEs were found, which is not unexpected for an instrument measuring polarized Brightness (pB) in the inner corona.

\begin{figure}
    \centerline{
\includegraphics[width=1.0\textwidth]{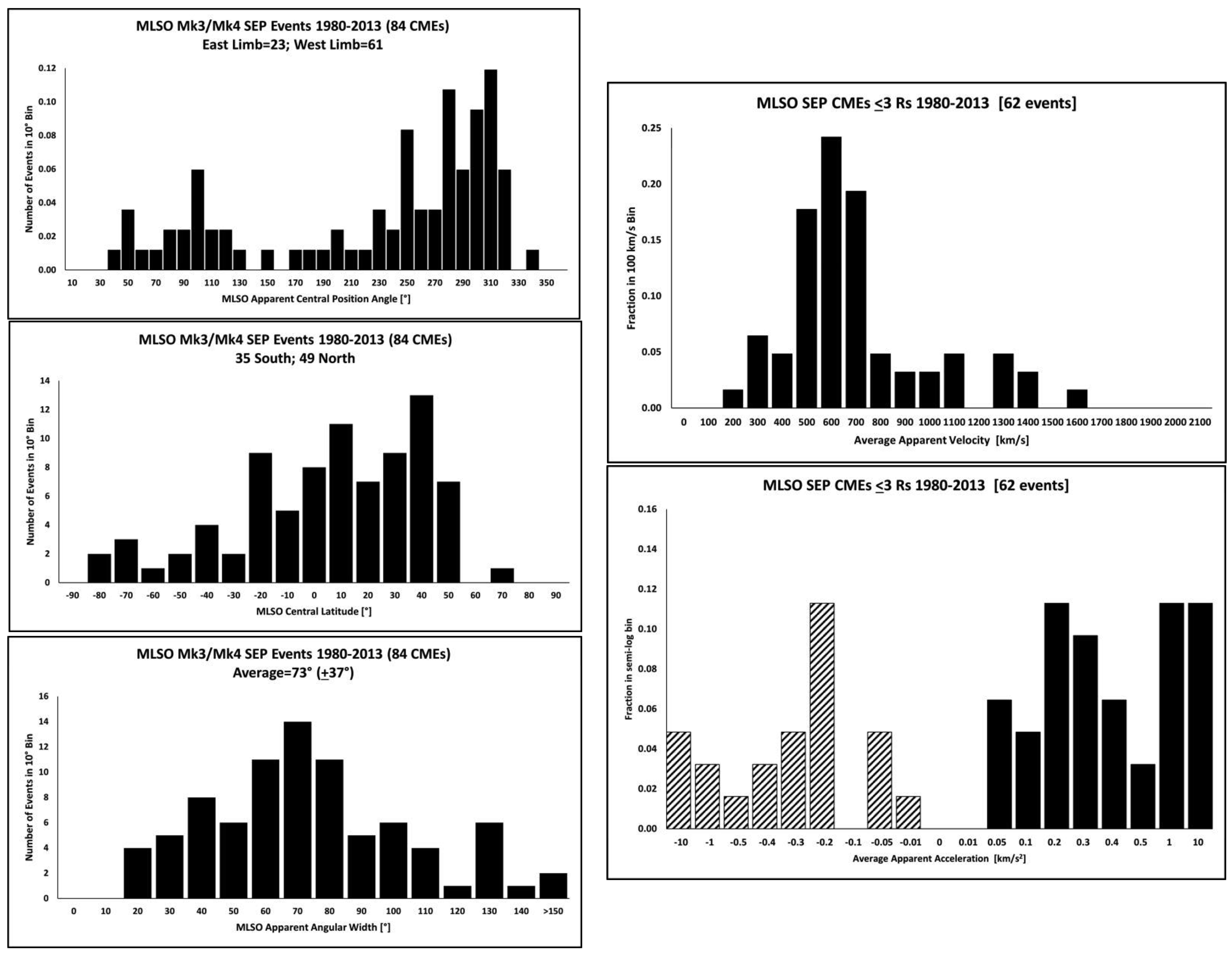}
}
    \caption{Summary of the properties of the SEP-associated CMEs observed by the MLSO Mk3/4 coronameters. Left panel: Top: The apparent central position angle with 0$^\circ$ = north, and measured anti-clockwise. Note the excess of western events (PA$\sim270^\circ$); Middle: The CME central latitude showing, for these events, a bias towards the northern hemisphere; Bottom: The apparent angular width (mean=73$^\circ$, maximum=258$^\circ$). Right panel: Top: The average apparent velocity.  The mean value is 650 km/s; Bottom: The average apparent acceleration. Note the semi-logarithmic scale and that the CMEs may be accelerating or decelerating in the MLSO field of view, with accelerations of the order of km~s$^{-2}$. }
    \label{props}
\end{figure}

The top right panel of Figure~12 shows the distribution of average apparent velocities in the MLSO field of view for the subset of 62 SEP-associated CMEs with at least four height-time measurements available.  Where possible, the MLSO white-light measurements were augmented by those from other telescopes when it was clear that the same CME feature could be tracked.  The instruments providing measurements near the inner boundary of the field-of-view included: SOHO LASCO C1 (1 event); SOHO EIT (20 events); and SDO AIA (5 events).  Also, some spacebased coronagraph measurements were included if they were inside the $\sim 3$~R$_S$ outer boundary of the MLSO field of view, including  SMM C/P (2 events) and SOHO LASCO C2 (22 events).  Although many of the post-2010 SEP-associated CMEs were also observed by the SECCHI instruments on STEREO A and B, these observations were not included in this study since the angular separation between the STEREO spacecraft and Earth was already $>70^\circ$.  Combining those unique observations with MLSO and other measurements into 3-D views will be addressed in a future study. The mean average velocity is 650 km/s for the SEP-associated CMEs, compared with 390 km/s for the general Mk3 CME population \citep{stcyr1999,stcyr2015}, indicating that SEP-associated CMEs on average have higher speeds in the MLSO field of view. The distribution has a tail reaching to velocities of $\sim1600$~km/s but, on the other hand, there are cases of SEP-associated CMEs with average velocities of only $\sim200$-300~km/s. The bottom right panel shows the average CME accelerations (black) or decelerations (hatched) in the MLSO field of view. Note that the scale is semi-logarithmic to accommodate the wide range of values.  Both the accelerations and decelerations in the low corona tend to be of the order of km~s$^{-2}$, indicating that large changes (increases or decreases) in the CME velocity may occur, as was also found for the general population of CMEs observed by Mk3 \citep{stcyr1999} where around a third of the CMEs were found to be decelerating.  

\begin{figure}
    \centering
\includegraphics[width=0.7\textwidth]{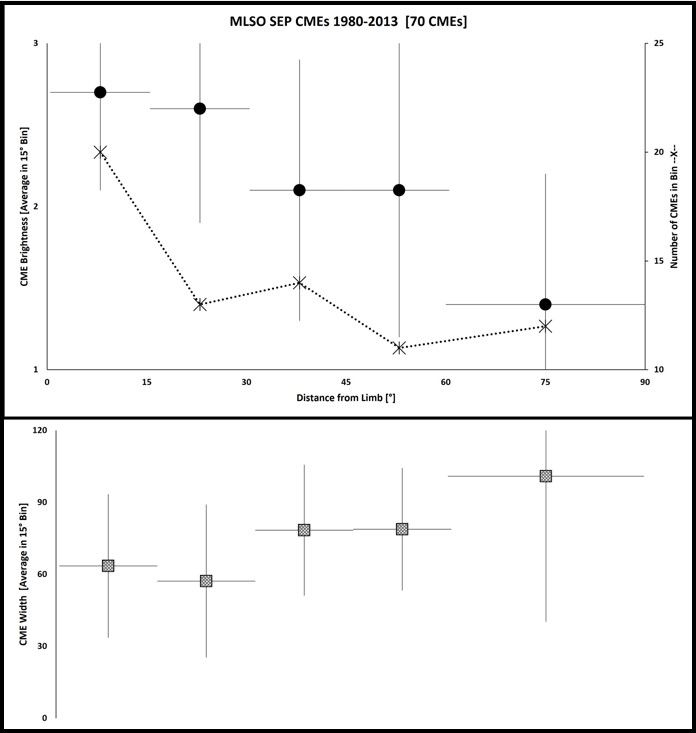}
    \caption{Average MLSO SEP-associated CME brightness (top panel) and average width (bottom panel) as a function of distance from the sky-plane.  The number of events in each $15^\circ$ bin is shown in the top panel connected by the dashed line.  Horizontal error bars indicate the range of longitudes over which the average was derived; Vertical error bars are the standard deviations of the averages. The average CME brightness falls off with distance from the limb while the average projected CME width increases.}
    \label{bright}
\end{figure}

We have also quantified the brightness of each of the MLSO SEP-associated CMEs, as described in \cite{stcyr2015}.  The categories are ``Very Faint" (only detected in difference images and assigned “1” numerically); ``Faint" (initially detected in difference images, but then identified in direct images, “2”); and ``Bright" (easily seen in direct images, “3”).  There are several factors that influence CME brightness including distance from the sky-plane and solar cycle variations in CME mass \citep[e.g.,][]{macqueen2001,Vourlidas2010}.  It has also been reported that SEP-related CMEs are brighter than the general population \citep[e.g.,][]{Kahler2005}. In addition, the MLSO polarized-Brightness observations are more sensitive to distance from the limb than those of most spacebased coronagraphs that observe total-Brightness (B) \citep[see Appendix~A in][]{Hundhausen1993}.  The black points in the top panel of Figure~13 show how this rudimentary brightness measure decreases as distance from the limb, based on the longitude of the associated solar event, increases.  The more limited visibility of CMEs in groundbased pB observations is also reflected in the smaller number of events in each bin as distance from the limb increases, as indicated by the crosses and dashed line in the top panel of Figure~13.  The bottom panel shows that the average plane-of-the-sky size of these events increases with distance from the limb.  Although there were no full-360$^\circ$ halo CMEs detected by Mk3/Mk4, the decreasing average brightness and increasing average size when the associated activity is located closer to the Sun’s central meridian are both obvious in Figure~13.

\begin{figure}
    \centering
\includegraphics[width=0.7\textwidth]{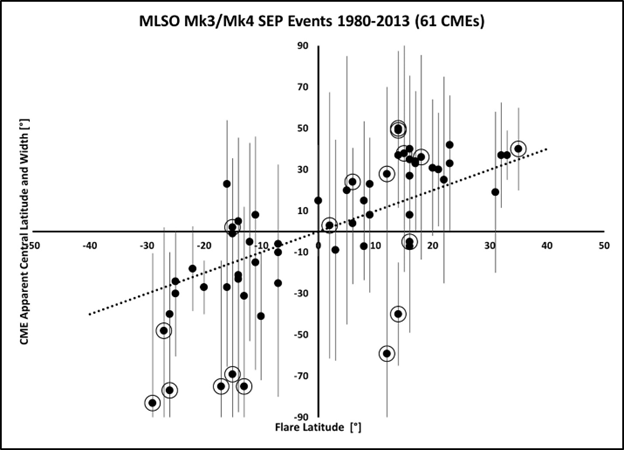}
    \caption{The CME apparent central latitude (vertical axis) versus the latitude of the flare (or other associated activity; horizontal axis).  The dashed equality line passes through almost all of the CME widths (shown as the vertical bars). 
 Events where the associated activity was $>50^\circ$ from the sky-plane are circled and tend to lie away from this line.}
    \label{lat}
\end{figure}

Another aspect to consider is where the CME appeared in latitude with respect to the location of the photospheric activity, as discussed by \cite{Harrison1990} and \cite{yashiro2008}.  Figure~14 shows the CME apparent central latitude (dot) and width (vertical bars) plotted against the latitude of the related photospheric activity.  The line indicates where CME central latitude equals the flare latitude, and it is clear that some part of the CME passes through this line for almost all events.  The exceptions can be explained by noting that the circled points are events where the associated activity was $>50^\circ$ from the sky-plane, so the CMEs appear extremely foreshortened and displaced.

For 42 of the MLSO SEP-associated CMEs, there are estimates of the mass from LASCO observations reported in the CDAW CME catalog.  About a quarter of these estimates are considered to be ``good", while the remainder are ``uncertain".  The average mass density for these 42 events is $3.5\times10^{13}$~ gm/degree, which is well above the average for all CMEs in the catalogue, and intermediate between those of CMEs classified as ``SEP-poor" and ``SEP-rich" in Table 4 of \cite{Kahler2005}. (We also note that in that study, the proton intensities (also at 20~MeV) of the SEP rich and poor events, defined relative to a CME speed-intensity fit, span a similar range to those of the MLSO CME-associated SEP events.) 

\begin{figure}
    \centering
\includegraphics[width=0.7\textwidth]{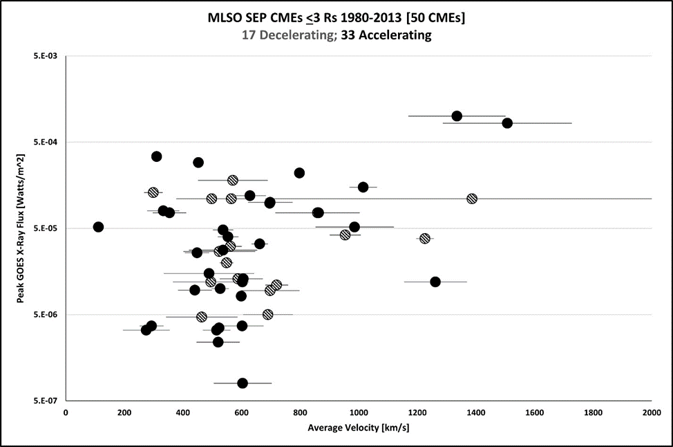}
    \caption{GOES soft X-ray flare peak intensity vs. average CME speed in the low corona. Solid (cross-hatched) points indicate accelerating (decelerating) CMEs. }
    \label{vxray}
\end{figure}

We have also examined (Figure~15) whether there was any relationship between the peak soft X-ray flux measured by GOES for the flare associated with a MLSO CME and the CME average velocity in the inner corona.  The CMEs indicated by solid data points were accelerating while the cross-hatched were decelerating. There does not appear to be a clear distinction between those two populations, and overall, there is only a weak correlation between the CME average velocity in the inner corona and X-ray flare intensity. In particular, while the fastest CMEs in this sample do appear to be associated with predominantly more intense X-ray flares and not with weak flares, relatively slow CMEs are associated with a wide range of flare sizes.  Thus, the occurrence of a strong flare is not necessarily associated with a SEP-associated CME with a high average speed in the low corona. 

\begin{figure}
    \centerline{
\includegraphics[width=0.5\textwidth]{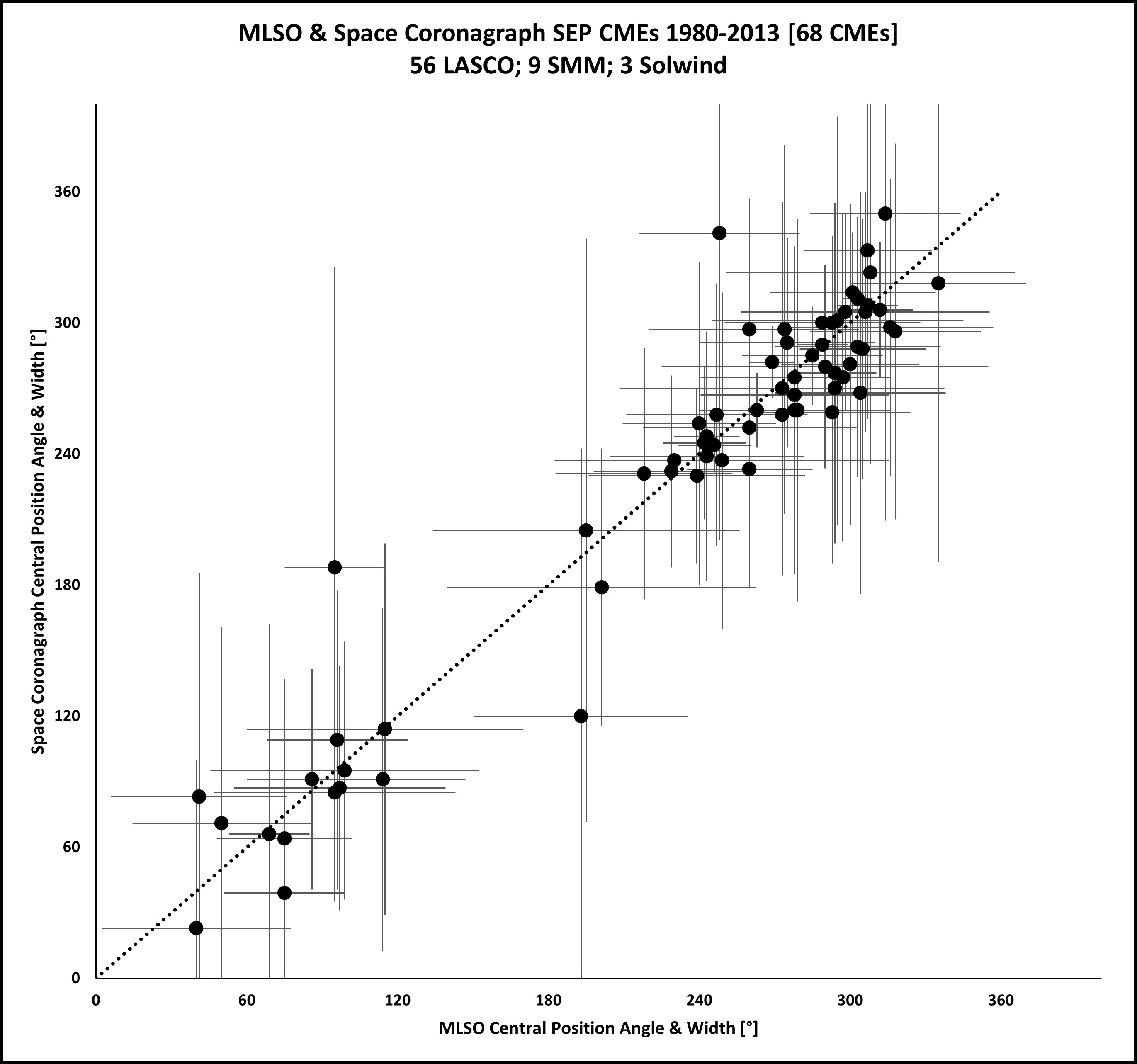}
\includegraphics[width=0.45\textwidth]{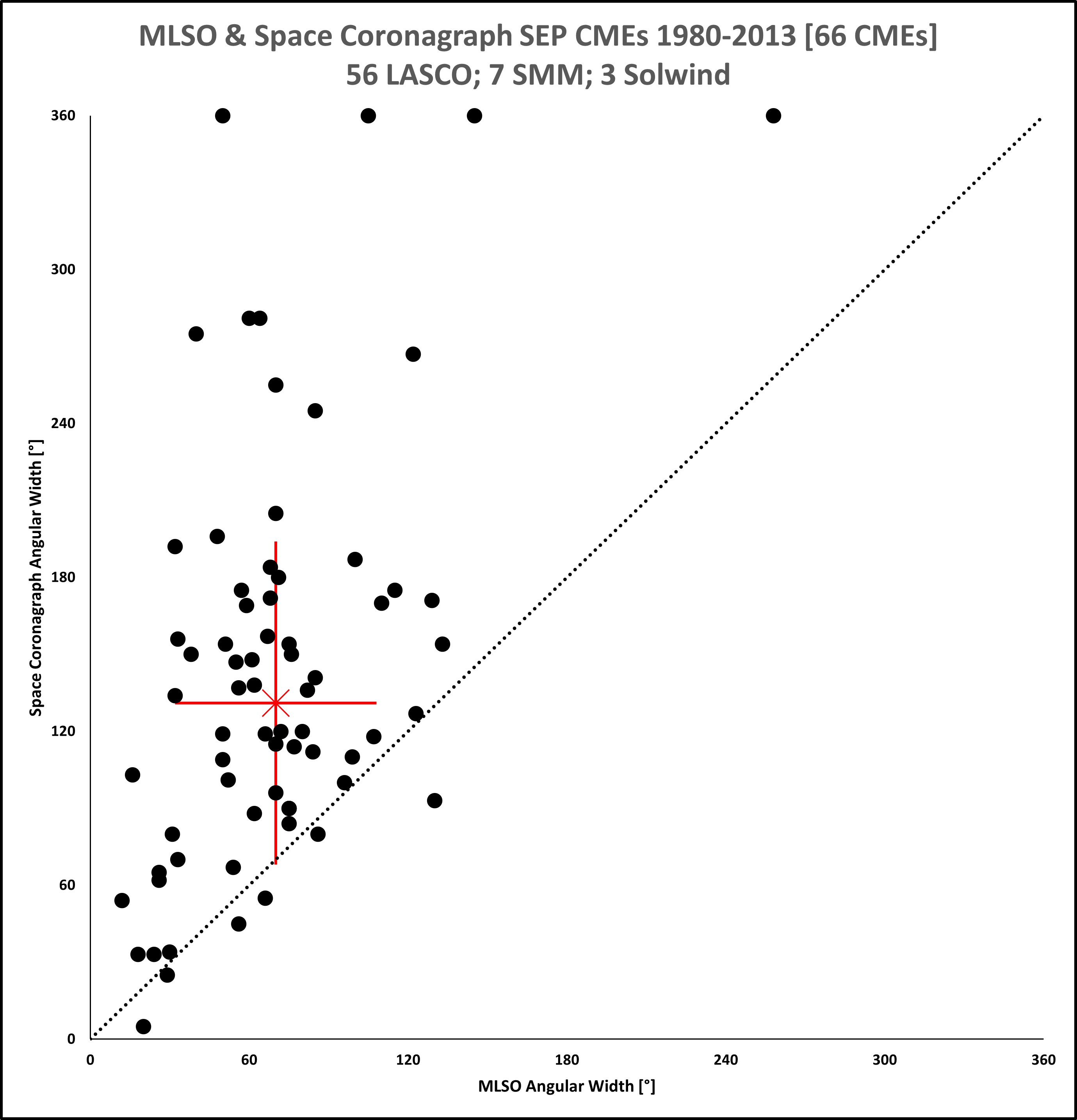}
}
    \caption{Left: Comparison of the central position angle of the MLSO SEP-associated CMEs in the inner vs. middle corona.  The error bars represent the width of the CME. Shock wings (see text) have been removed in the spacebased observations. Right: Comparison of the apparent angular width of the MLSO SEP-associated CMEs in the inner vs. middle corona.  Angular expansion is evident for almost all events.}
    \label{MLSOspace}
\end{figure}

We have also combined MLSO and spacebased CME observations to compare the CME parameters in the low-mid corona with those in the low corona.  Spacebased coronagraph observations were available for 72/84 (86\%) of the MLSO SEP CMEs, and all of the MLSO SEP-CMEs were detected in the middle corona when spacebased measurements were available. For 68 of the events, we could measure the position angle in both MLSO and spacebased observations.  For 66 events, we were also able to measure a width, and for 55 events, we were able to combine the height-time measurements from MLSO and the spacebased instrument.  When the time gap was too large or when the feature being tracked could not be reliably identified in both instruments, no attempt was made to combine the height-time information.  Rather than rely on catalogued values for the characteristic quantities for the spacebased CME detections, we measured each event in the middle corona to ensure that the same morphological feature was matched in observations from each instrument.  

The left panel of Figure~16 compares, for 68 CMEs, the plane-of-the-sky central position angles for CMEs observed by MLSO and for the same CMEs observed by spacebased coronagraphs.  The spacebased observations are from SOHO/LASCO in 56 cases, with nine (three) other cases using observations from SMM (SOLWIND).   The CME directions in the low- and mid-corona are generally similar despite, for example, the different dynamic ranges of MLSO and the spacebased instruments, illustrating that the MLSO and spacebased CME observations can be confidently associated for the majority of our sample of events. The error bars indicate the width of the CME.  Because of the tendency for fast CMEs to generate shock ``wings", compressive waves extending beyond the boundaries of the CME driver, in the middle corona, CME catalogues often show a significant number of ``halo" events in the middle corona, particularly in the CDAW LASCO catalog \citep[e.g.,][]{gopalswamy2010}.  For this study we have excluded obvious deflections/extensions in the angular width measurements in LASCO, measuring only the extent of the primary flux rope, thus reducing the number of CDAW catalogued entries of ``halo CMEs" from 27 down to four. 

The right panel of Figure~16 shows that, with a few exceptions, the plane of the sky CME width is equal or larger in the mid-corona than in the low corona, suggesting that the CMEs may still be expanding as they exit the low corona and enter the middle corona; the difference between instruments measuring B and pB may also contribute. The red cross indicates the average of the widths in the low- and mid-corona.  Additionally, Figure~17 shows that the increase in angular size for these SEP-associated CMEs as a function of limb distance found in the inner corona (black points in Figure~17, from the bottom panel of Figure~13) is even more magnified in the middle corona (red points). This is also likely to be due to a difference between B and pB measuring instruments - the full width extent of the CME may be too faint to be seen by MLSO in pB, and therefore the width may be underestimated, for events near disk center where LASCO is observing severely-projected CMEs.

\begin{figure}
    \centering
\includegraphics[width=0.7\textwidth]{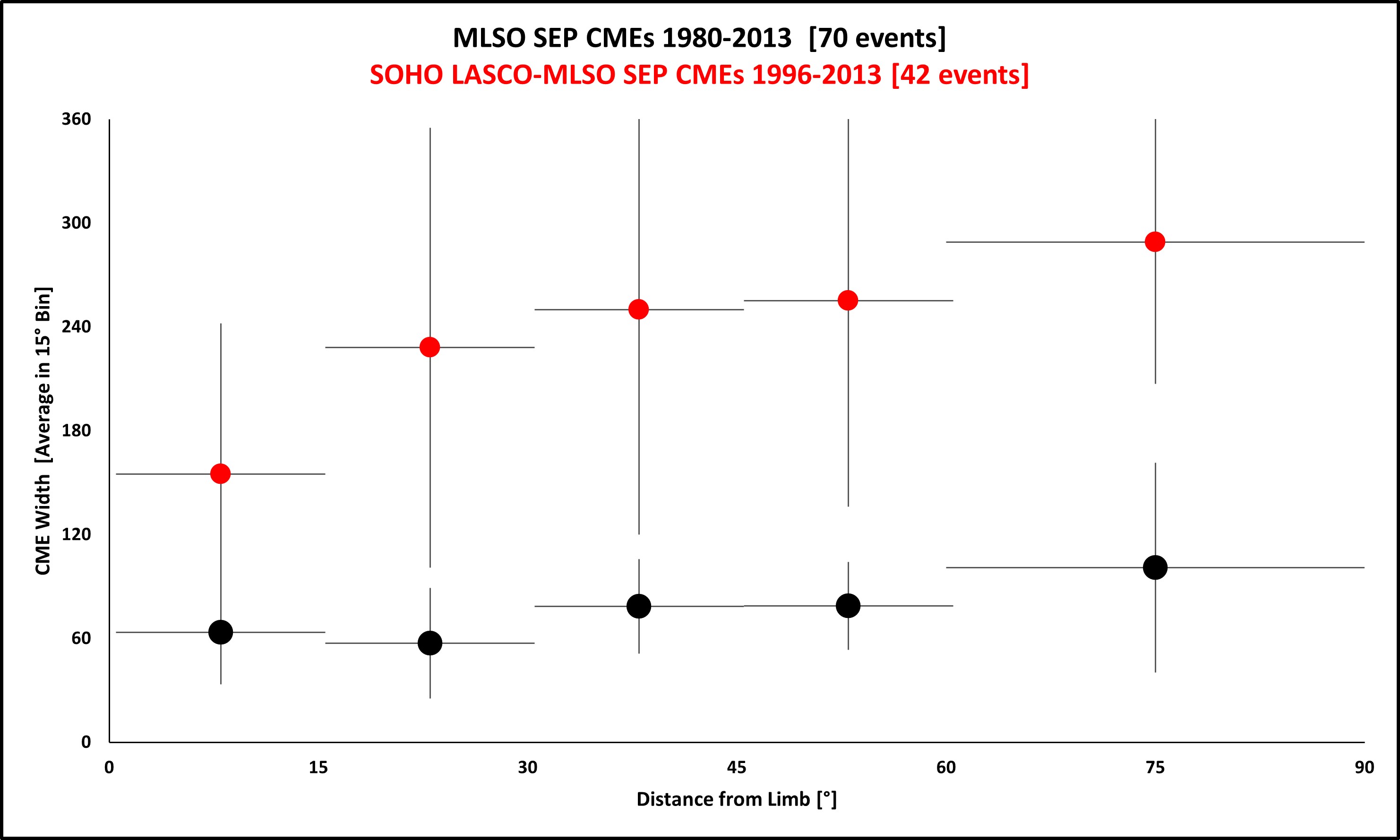}
    \caption{Comparison of the apparent angular width of the MLSO SEP CMEs in the inner (black circles) versus middle corona (red circles) as a function of distance from the limb.  Here the MLSO events are compared with the subset of SOHO LASCO measurements. }
    \label{width}
\end{figure}

Figure~18 compares, where possible, the apparent linear speed of the SEP-associated CMEs in the low to mid-corona, from a fit to MLSO and spacebased observations, with that in the inner corona from MLSO observations. A line of equality is shown for reference, with 33 (22) CMEs lying above (below) the line having accelerated (decelerated) between the inner and mid-corona; the average change in speed from the low to mid-corona for the accelerating (decelerating) CMEs is 385 km/s (-126 km/s).  There is not a clear correlation between the average CME speeds in the low and low-mid-corona. While the fastest CMEs in the inner corona tend also to be fast in the mid-corona, the converse is not always the case.  In particular, CMEs with moderate speeds ($\sim500$~km/s) in the inner corona can have a wide range of speeds in the middle corona.     

\begin{figure}
    \centering
\includegraphics[width=0.7\textwidth]{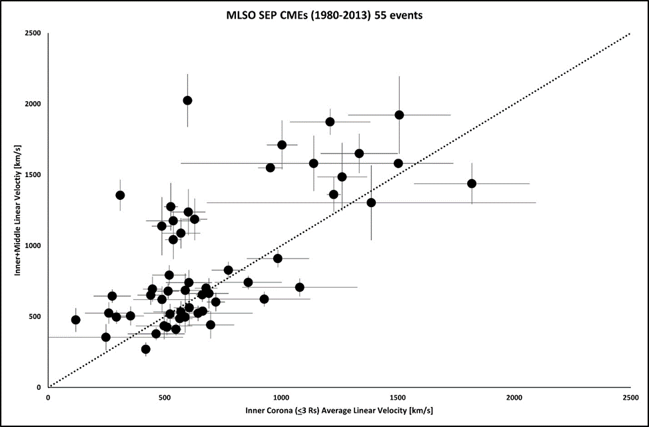}
    \caption{Comparison of the apparent linear speed of the MLSO SEP-associated CMEs in the low versus low plus middle corona.  A line of equality is shown for reference.  }
    \label{vels}
\end{figure}

Figure~19 compares histograms of the apparent average CME accelerations (green) or decelerations (red) in the inner corona from MLSO (top, from Figure~12) and in the low-middle corona from combined MLSO and spacebased observations (bottom), using a semi-logarithmic acceleration scale.  These distributions clearly illustrate that the CMEs typically have larger accelerations/decelerations ($\sim0.1-10$~km~s$^{-2}$) in the inner corona than in the low-mid corona (typically less than $\sim0.2$~km~s$^{-2}$) where, as noted previously, CMEs tend to reach their terminal speeds. The average accelerations in the low and low-middle corona for individual CMEs are compared in Figure~20, again illustrating the larger accelerations/decelerations in the low corona. In particular, 86\% (23\%) of the inner (middle) corona accelerations/decelerations exceed $\pm0.1$~km~s$^{-2}$. In our sample, the SEP-associated CMEs that accelerated in the inner corona outnumber those that were decelerating by approximately two to one (40 to 22).

\begin{figure}
    \centering
\includegraphics[width=0.7\textwidth]{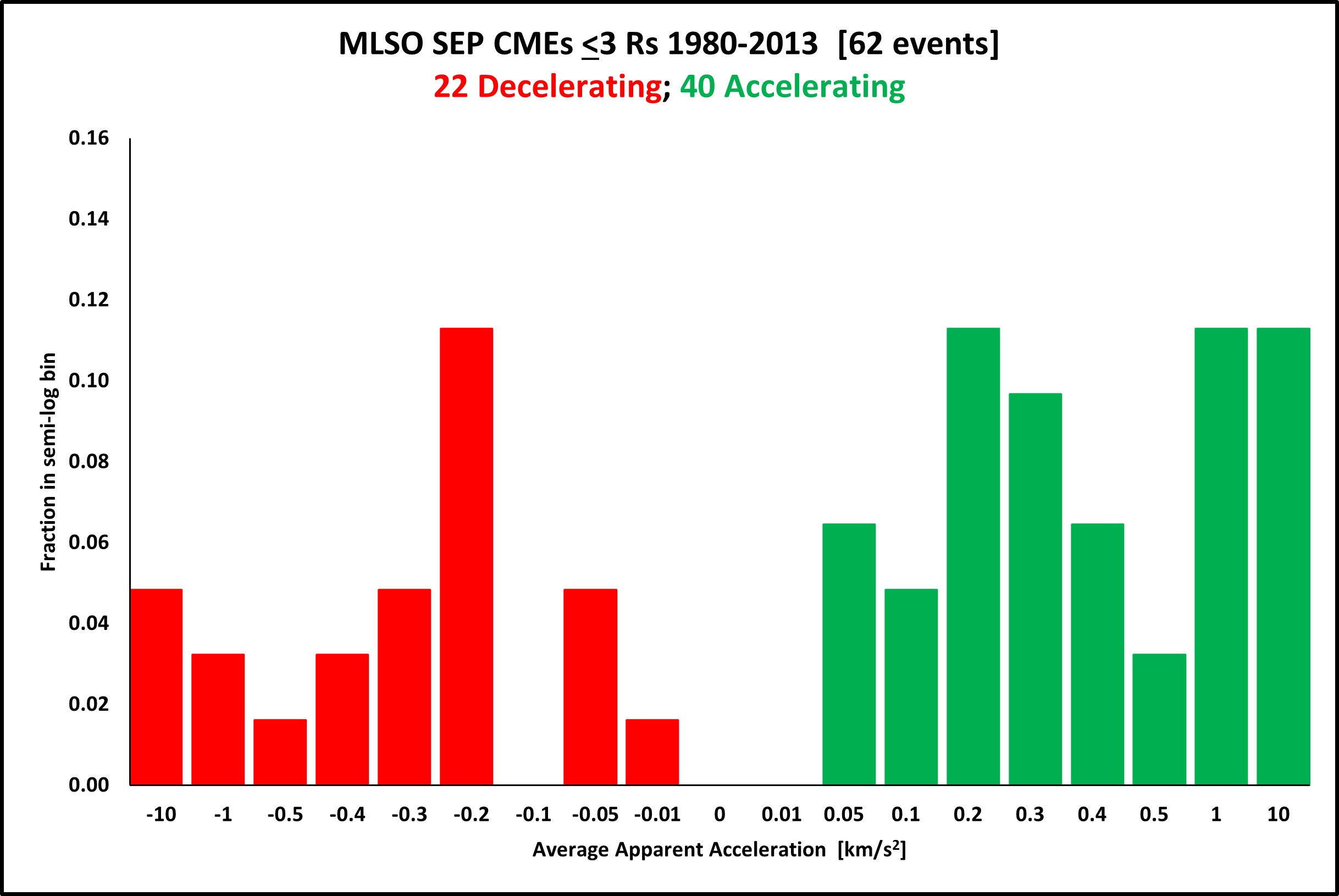}
\includegraphics[width=0.7\textwidth]{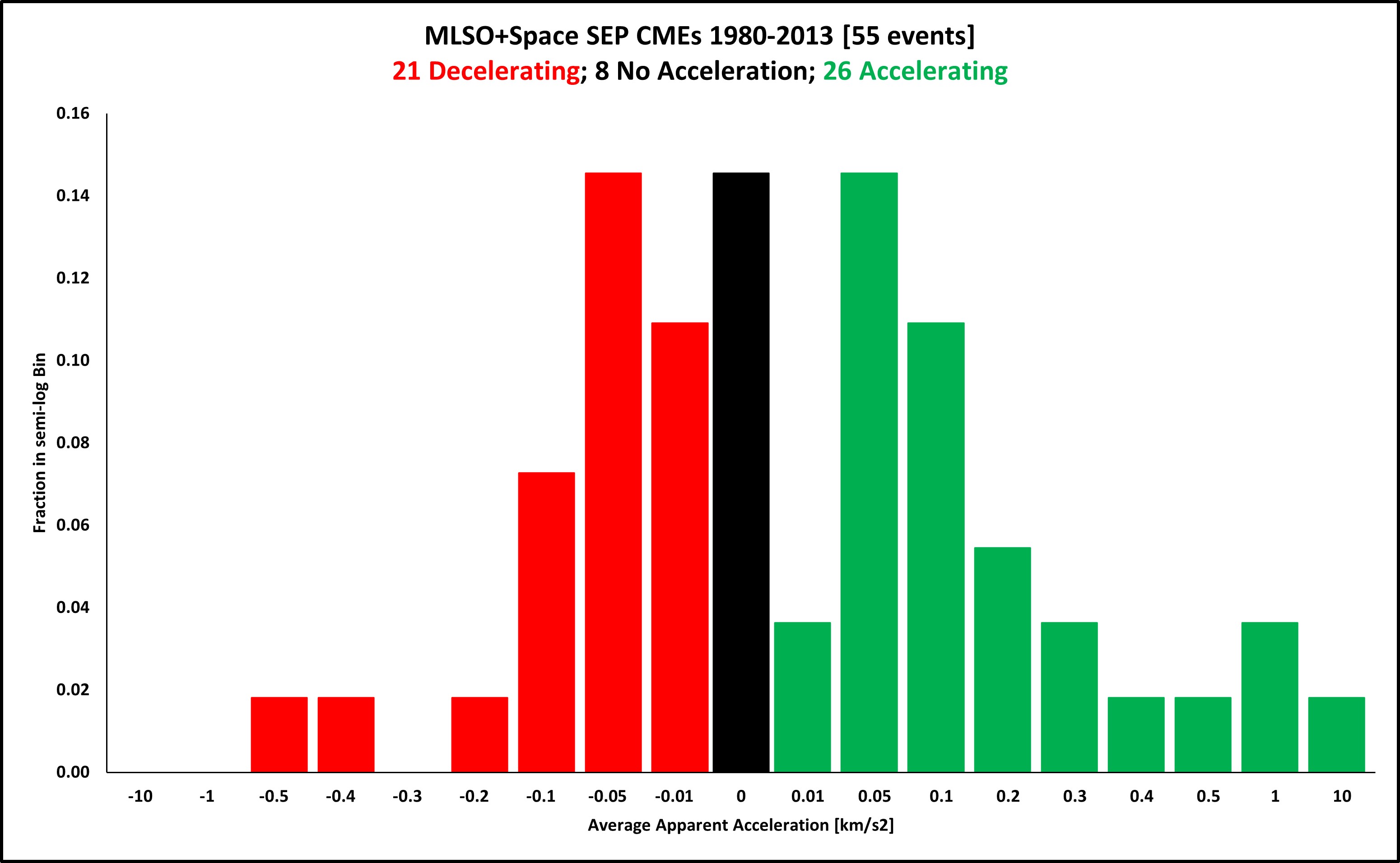}
    \caption{Histograms of the MLSO SEP-associated CME apparent accelerations (green) or decelerations (red) from MLSO observations (top, from Figure~12) and in the low-middle corona from combined MLSO and spacebased observations (bottom). The horizontal scale is semi-logarithmic. Note that CME accelerations/decelerations are typically larger in the low corona.   }
    \label{acel}
\end{figure}

\begin{figure}
    \centering
\includegraphics[width=0.7\textwidth]{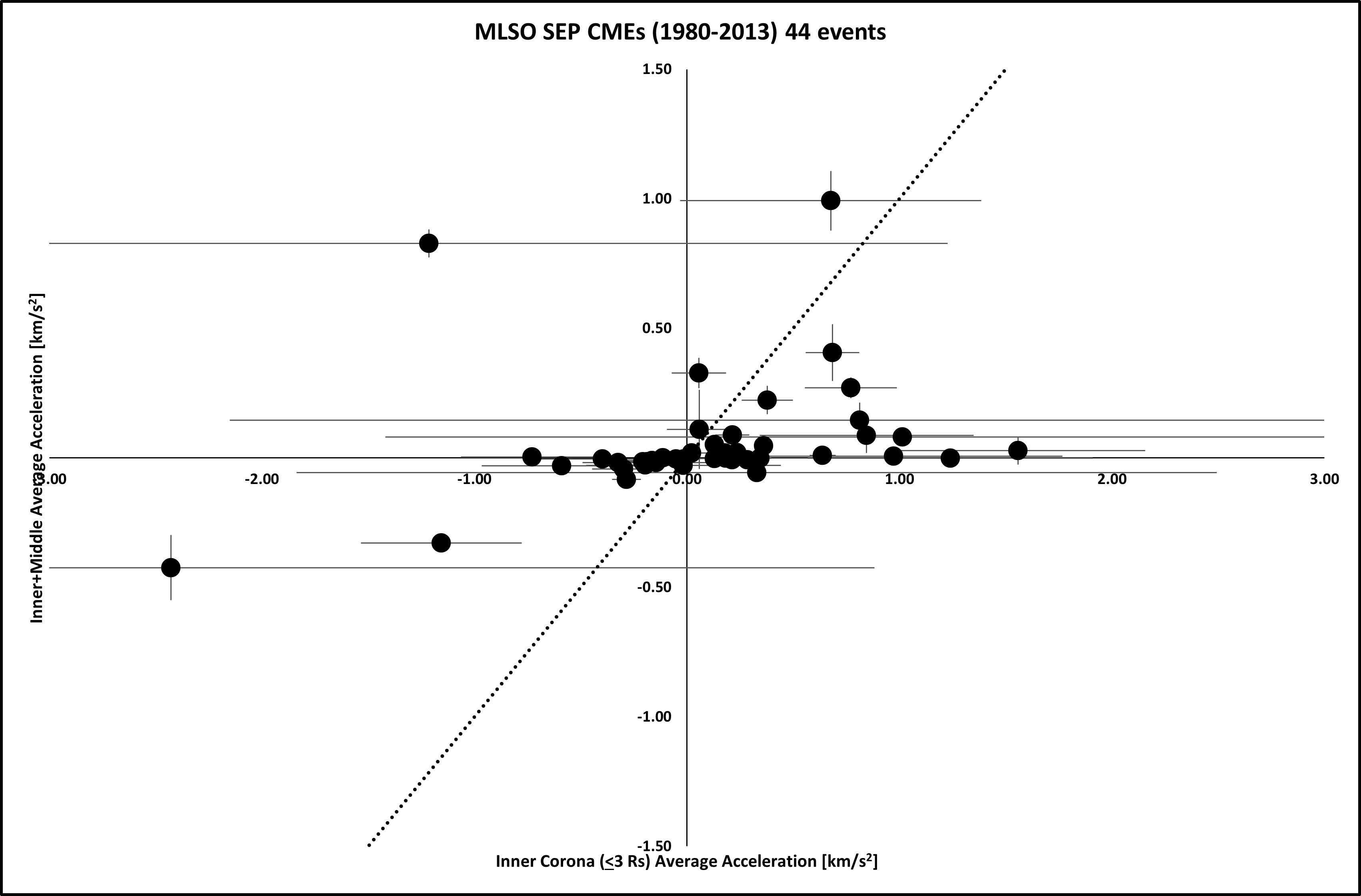}
    \caption{Comparison of the average acceleration of the SEP-associated MLSO CMEs in the inner corona (horizontal axis) versus the low-middle corona (vertical axis).  A line of equality is shown for reference. 86\% (23\%) of the inner (middle) corona accelerations are larger than $\pm0.1$~km~s$^{-2}$. }
    \label{inmidacc}
\end{figure}

\begin{figure}
    \centering
\includegraphics[width=0.7\textwidth]{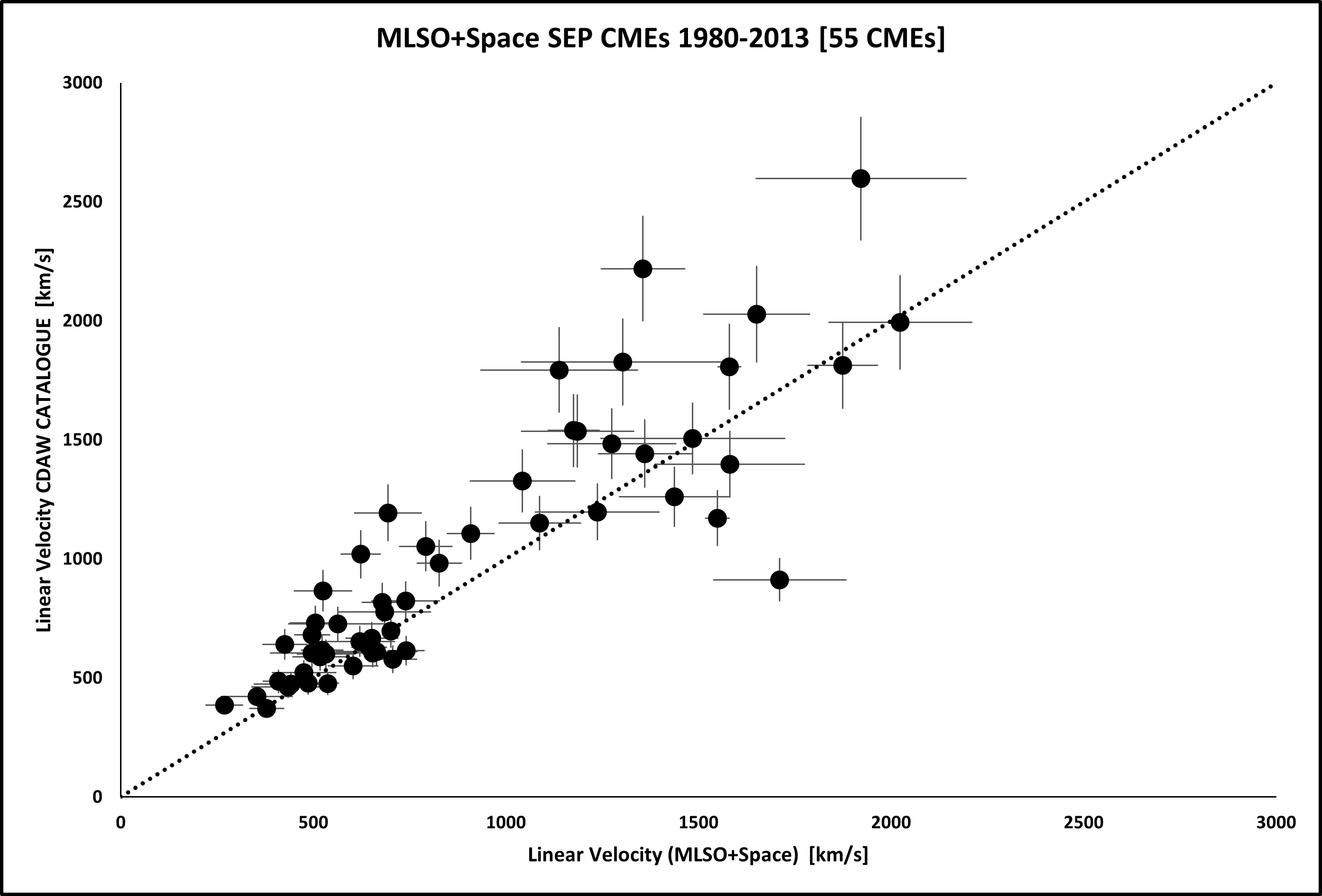}
    \caption{Comparison of the apparent linear speed of the SEP-associated MLSO CMEs in the inner+middle corona with CDAW catalogue values in the middle corona for the same CMEs. A 10\% uncertainty is assumed for the CDAW CME speeds.  A line of equality is shown for reference. }
    \label{cat}
\end{figure}

Finally, Figure~21 compares the average speeds inferred from combined MLSO and spacebased CME observations with the CME speeds in the middle corona reported in the CDAW CME catalog for the same CMEs.  While these speeds tend to be correlated, as expected, the CDAW speeds do trend higher for the faster CMEs, often by several hundred km/s.

\subsection{Estimating CME dynamics}
As discussed in Section~2.1, we have explored several methods of analyzing CME dynamics in the low-mid corona.  The results above use the average apparent (plane of the sky) speed or acceleration in the coronameter/coronagraph fields of view obtained using a least-squares-weighted polynomial fit. Other methods outlined in Section~2.1 include  point-to-point determination (P2P) and cubic spline interpolation (CSI), as well as the flare proxy method using only observations of the CME in the middle corona and the rise time of the associated soft X-ray flare.

\begin{figure}
    \centering
\includegraphics[width=0.7\textwidth]{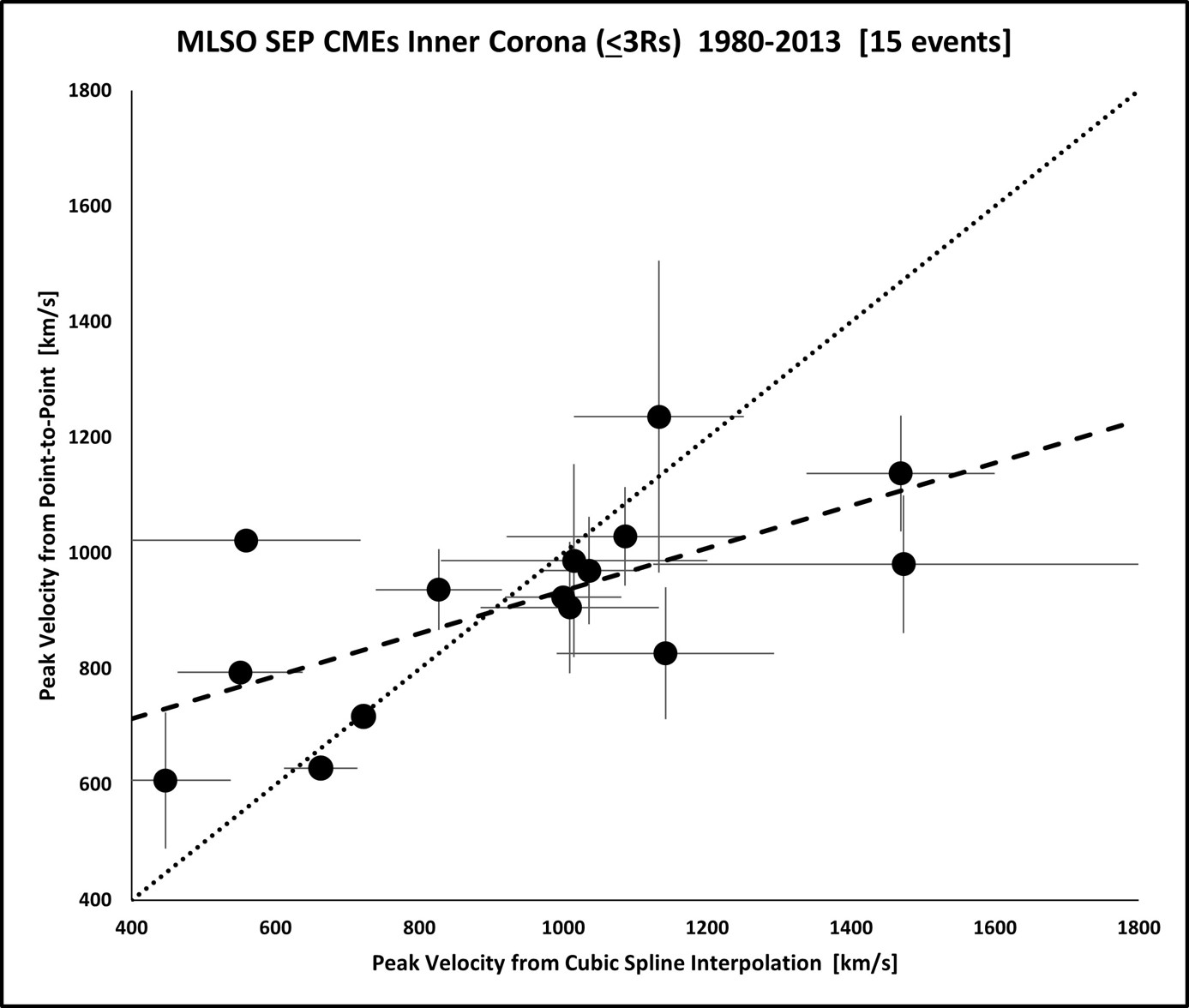}
    \caption{Comparison of the peak CME velocities in the inner corona obtained from cubic spline and point-to-point fitting. The dashed and dotted lines indicate the fit to the points and the line of equality, respectively.}
    \label{pvcsip2p}
\end{figure}
\begin{figure}
    \centering
\includegraphics[width=0.7\textwidth]{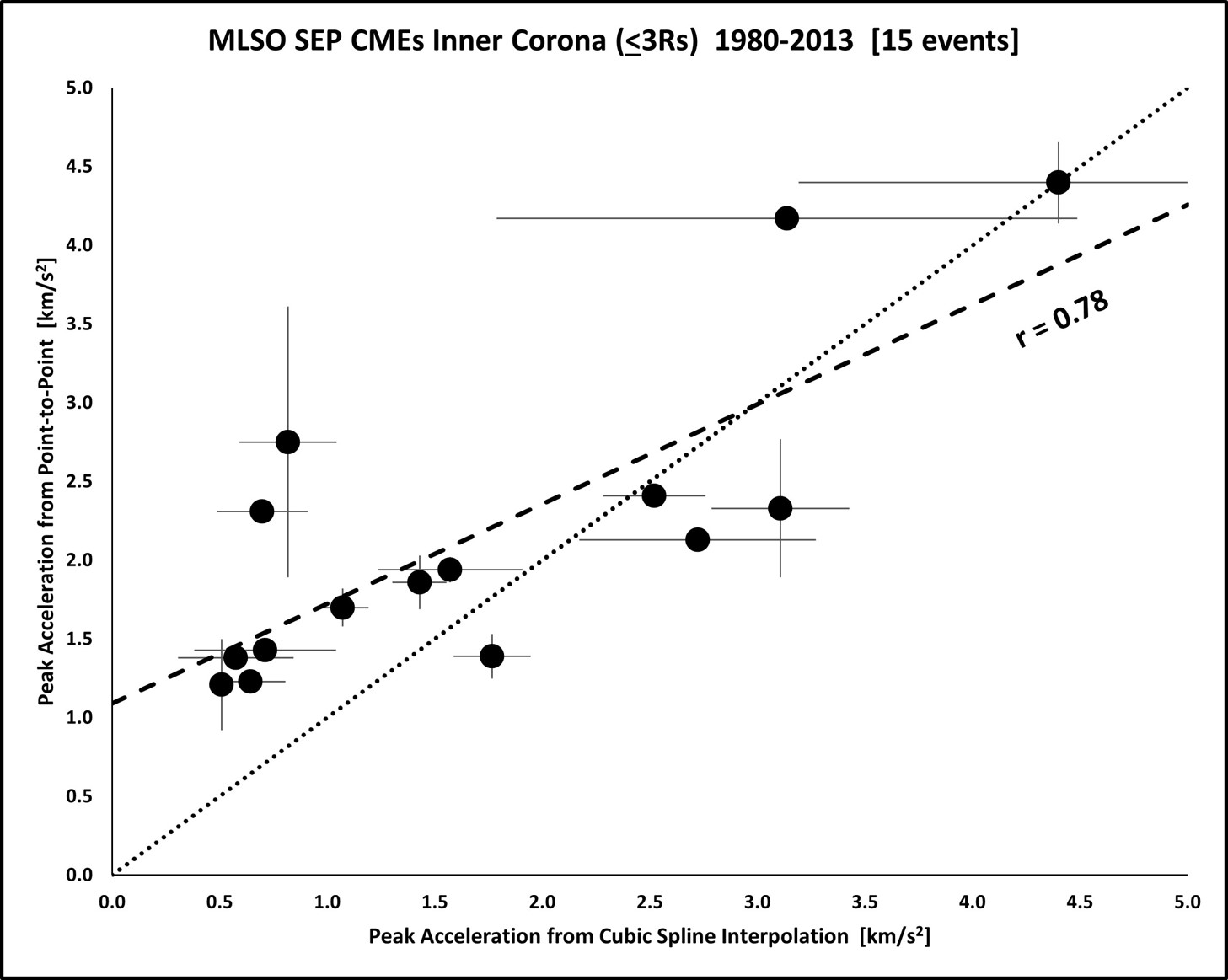}
    \caption{Comparison of the peak CME accelerations obtained from cubic spline and point to point fitting. The dashed and dotted lines indicate the fit to the points and the line of equality, respectively.}
    \label{pacsip2p}
\end{figure}

Unfortunately, as already noted in Section~2.3 in relation to the CME associated with GLE~67, we find that the different methods can give quite different results for CME speeds/accelerations. For example, Figure~22 compares the peak velocities in the inner corona determined by CSI and P2P for subsets of events where these methods can be applied.  Although there is some correlation between the points, as indicated by the fitted dashed line, in general they do not lie along (and can differ significantly from) the line of equality (dotted line). Figure~23 compares the peak acceleration in the inner corona determined as by CSI and P2P for the subset of events where both methods can be applied.  Again, while there is some correlation between the points, in general they do not lie along the line of equality. There are also large errors, in particular in the largest CSI accelerations.

\begin{figure}
    \centerline{
\includegraphics[width=0.5\textwidth]{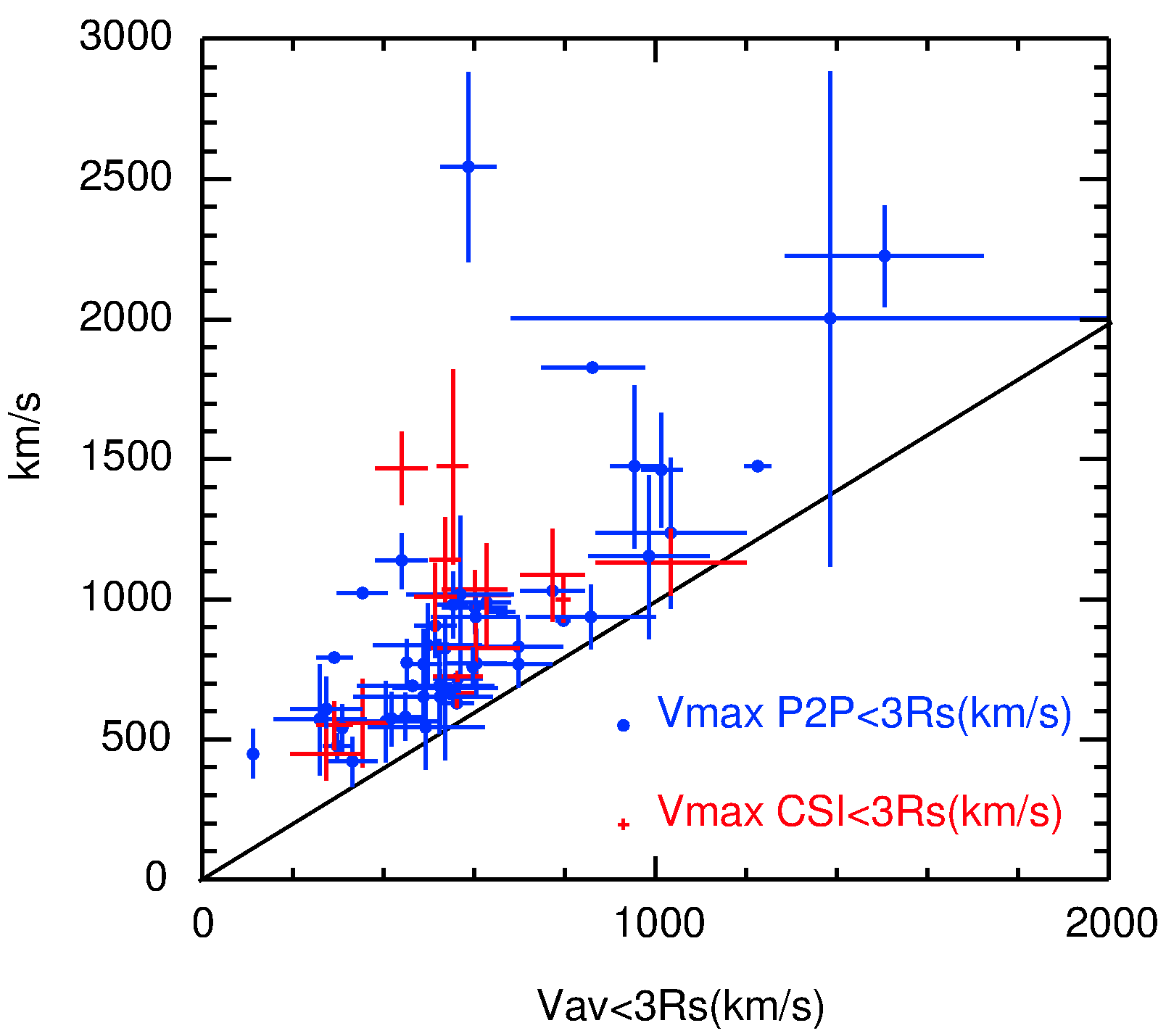}
\includegraphics[width=0.5\textwidth]{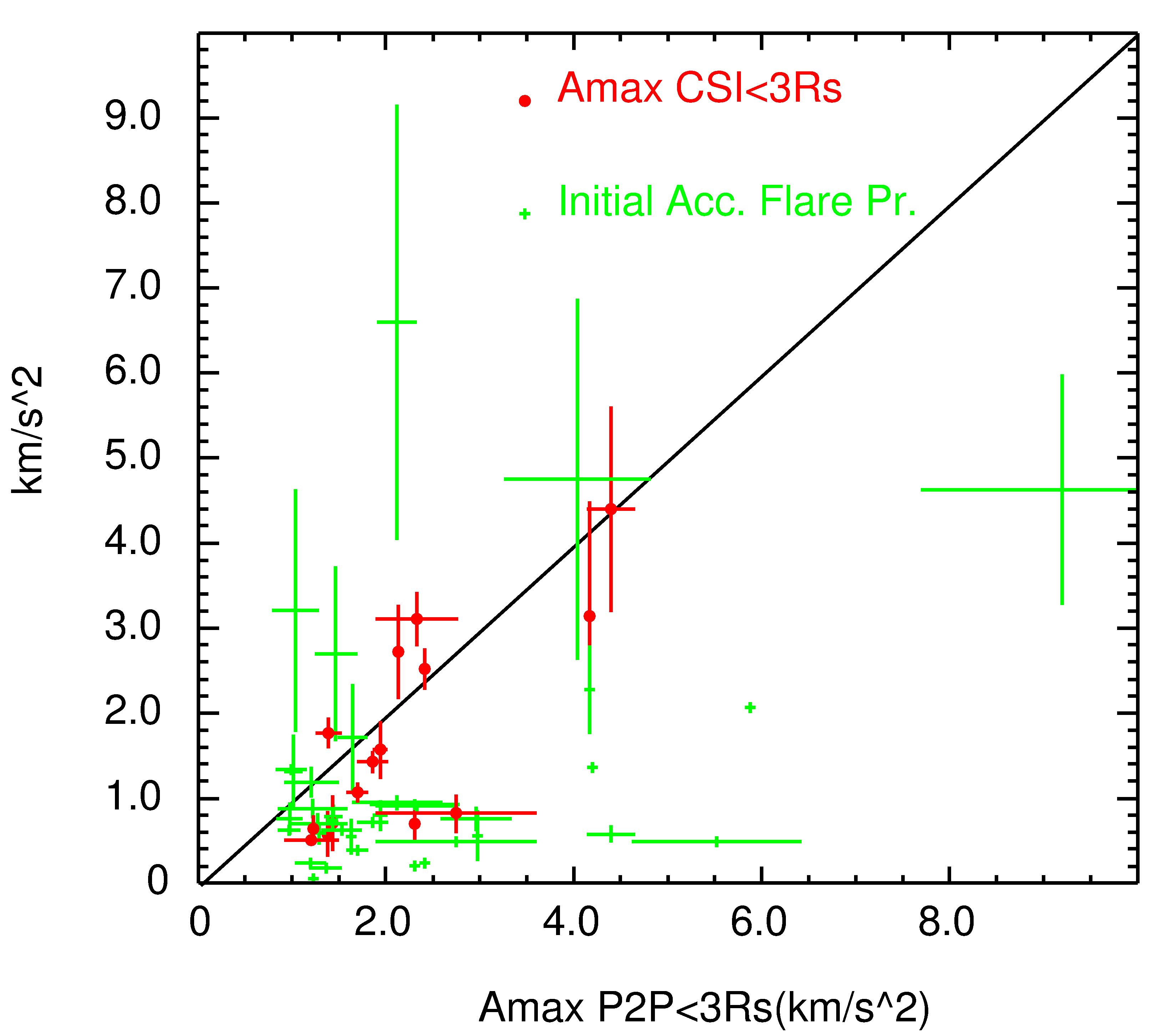}
}
    \caption{Left: CSI (red points) and P2P peak CME speeds (blue points) vs. the average speed at $<3$~R$_s$. Right: The peak acceleration from CSI (red points) and the initial acceleration from the flare proxy method (green points) vs. maximum acceleration from P2P, showing the generally poor agreement between different estimates of the CME acceleration in the low corona. Black lines indicate equal values of the parameters.}
    \label{cmespeedsacc}
\end{figure}

The left panel of Figure~24 compares the maximum CME speeds below 3~R$_s$ inferred from P2P (blue points) or CSI (red points) with the average speed.  Again, there is some correlation between these speeds, and as expected, the peak speeds exceed the average speeds (the black line indicates equality). However, there also are large differences and uncertainties for individual events, in particular for faster CMEs. The right panel of Figure~24 compares the peak acceleration from CSI (red points) or the initial CME acceleration from the flare proxy method (green points) with the peak acceleration from P2P, showing the generally poor agreement between these different estimates of the CME acceleration.  In particular, initial accelerations derived from the flare proxy method are typically lower than the other estimates.

Our overall conclusion from such comparisons and considerable experimentation in fitting the observations is that we were unable to determine which, if any, of these methods consistently gives the most reliable estimates of CME parameters such as the peak speed or acceleration, and associated heights, in the MLSO field of view.  We will therefore continue to focus in this paper on average speeds or accelerations, as well as the final speed of the CME, inferred from the MLSO Mk3/4 observations. In addition, using average values has the advantage that they can be derived for a larger subset of our events than when other fitting methods are used.   

\begin{figure}
    \centering
\includegraphics[width=1.0\textwidth]{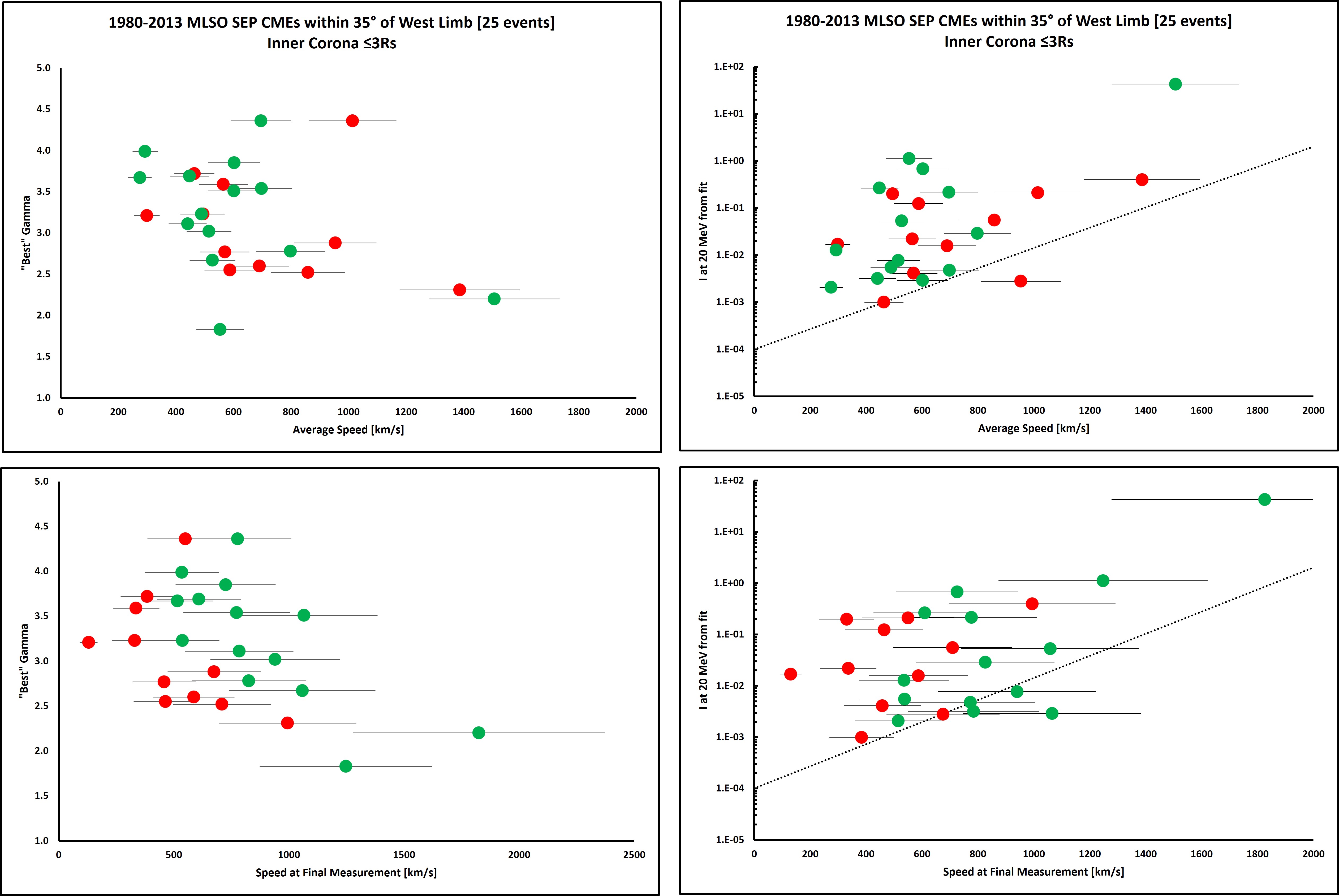}
    \caption{The SEP proton spectral index ($\gamma$; left panels) or the peak 20 MeV proton intensity (right panels) vs. the average (top plots) or final speed (bottom plots) in the low corona for events originating within $35^\circ$ of the west limb. Green (red) points indicate accelerating (decelerating) CMEs.  The dotted lines in the right panels indicates the intensity-mid-coronal CME speed relationship from \cite{richardson2014}. There is an indication that faster CMEs in the low corona are associated with more intense and harder SEP proton events, but the sparse data, in particular for fast CMEs, precludes a clear conclusion.  } 
    \label{ipv}
\end{figure}

\subsection{CME dynamics and SEP event properties}
\label{SS-CMESEP}
We now examine the dynamics of a subset of 25 ``best case" SEP-associated MLSO CMEs, originating near (within $35^\circ$ of) the west limb, where CMEs tend to be brighter (Figure~13) and projection effects in the plane-of-the-sky CME motion are reduced, and with a sufficient number of measurements available.  We will then consider whether there is any relationship between the CME dynamics and the properties of the associated SEP events. 

The right panels of Figure~25 compare the peak proton intensity at 20 MeV with the average CME speed in the low corona (top panel) or the final CME speed in the MLSO field of view (bottom panel).  Green (red) points indicate accelerating (decelerating) CMEs. There is an indication of a trend towards higher SEP proton intensity with increasing CME speed, in particular for the final speed estimates.  However, the limited number of particularly fast CMEs (e.g., $>1000$~km/s) in this sample of near-west limb events precludes a clear conclusion as to whether a fast CME in the low corona may be predictive of a large SEP event, as is typically found in studies using mid-corona CME speeds. In addition, the relatively slow CMEs in the low corona are associated with a range of SEP event sizes including some that are nearly comparable to those associated with relatively fast CMEs. Whether the CME was accelerating or decelerating in the low corona also does not appear to be a significant factor relating to SEP intensity, though we note that CMEs that were deccelerating would have, at some point in the low corona, reached speeds that were higher than suggested by the average or final speeds. Thus overall, this limited sample of events suggests that the average speed in the low corona of an SEP-associated CME may not be a clear predictor of the peak intensity of the associated SEP proton event at $\sim20$~MeV. The dotted lines in these panels indicate, for comparison, the proton intensity-CME speed relationship obtained by \cite{richardson2014} using {\it mid-coronal} CME speeds.   

We next consider the hardness of the SEP proton spectrum for the near west limb events.  The left panels of Figure~25 show the SEP proton spectral index ($\gamma$) vs. the average (top panel) or final (bottom panel) CME speeds in the low corona. (As described in Section~3.1, $\gamma$ is obtained from a power law spectral fit ($dJ/dE\sim E^{-\gamma}$) to proton observations at energies of $\sim5-60$~MeV.) There is an indication of a trend towards lower values of $\gamma$ (i.e., harder proton spectra) with higher CME speeds, but again the limited number of fast CMEs makes it difficult to assess whether the CME speed in the low corona may be used to predict the SEP spectral index.  We do note that the two CMEs with the highest average speeds in the low corona and among the hardest spectra are associated with GLEs. (Although five MLSO CMEs associated with GLEs were identified in Section~2.2, only two meet the criteria to be included in this sample of near west limb events.) Also, there is an indication, especially in the ``final speed" plot, that, for a given CME speed, the decelerating CMEs tend to be associated with harder proton spectra. Again, presumably, the speeds of these CMEs in the low corona would have been higher than indicated by the final speed, which might help to account for the harder spectra if the highest energy SEPs are accelerated low in the corona as suggested for example by \cite{kahler1994} and \cite{reames2009}. 

\begin{figure}
    \centering
\includegraphics[width=0.7\textwidth]{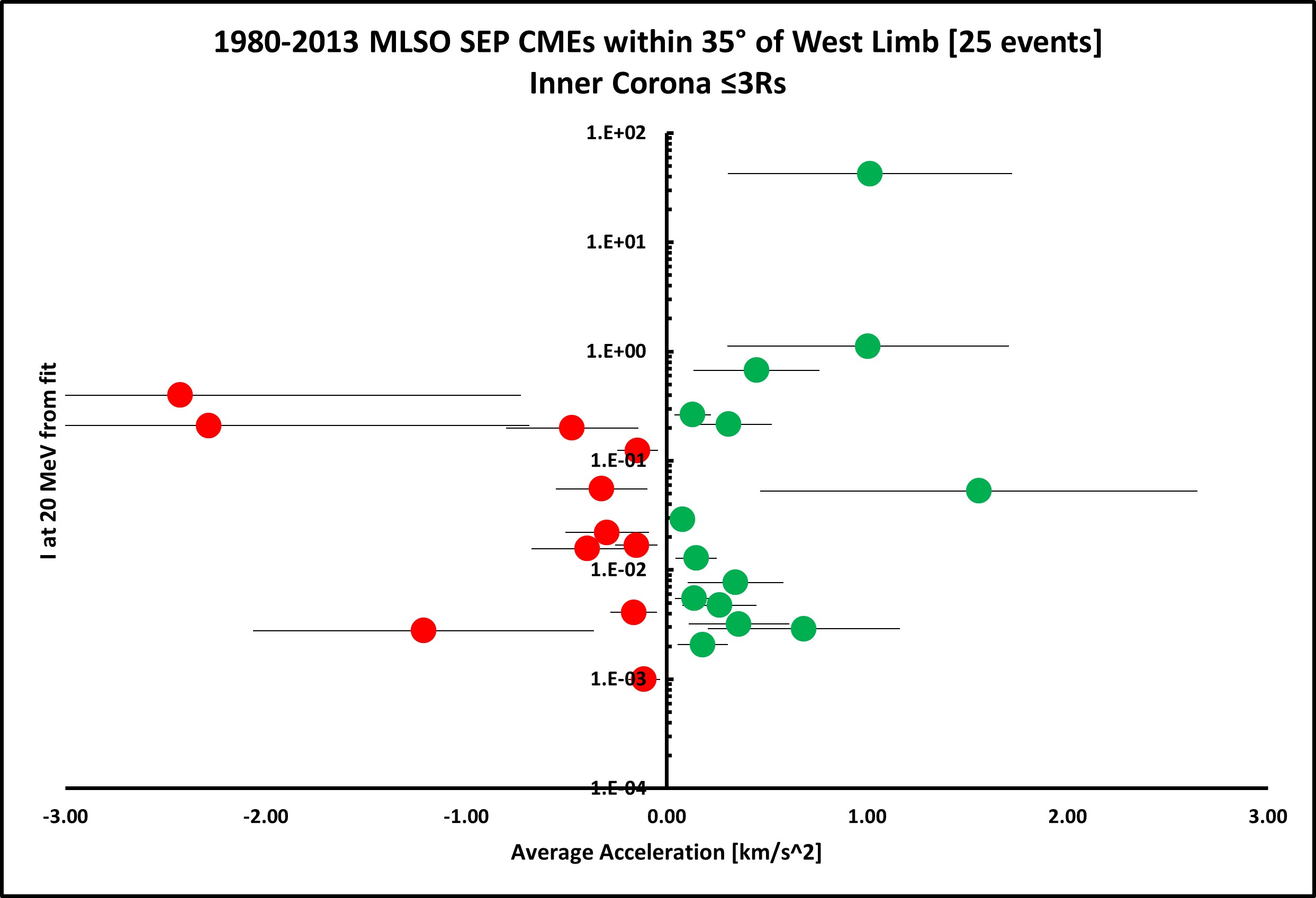}
    \caption{ SEP peak proton intensity at 20 MeV vs. CME acceleration (green symbols) or deceleration (red symbols) in the inner corona 
for near west limb events.} 
    \label{iaccwest}
\end{figure}

Considering the CME acceleration in the inner corona, Figure~26 suggests that the average CME acceleration (green symbols) might show a correlation with the peak SEP proton intensity at 20~MeV, but the data are too sparse to reach a definite conclusion.  The CME deceleration (red symbols) appears to show no relation with the proton intensity. While two of these events with the highest proton intensities also have the largest decelerations, the uncertainties in these decelerations are also large.  Figure~27 suggests, in the right panel, that the proton spectral index $\gamma$ might decrease with increasing CME acceleration (as has also been reported by \cite{gopalswamy2016} using the flare proxy method), but does not show any relationship with the CME deceleration (left panel).  However, again in our view, the data are too sparse to come to a clear conclusion.

\begin{figure}
    \centering
\includegraphics[width=1.0\textwidth]{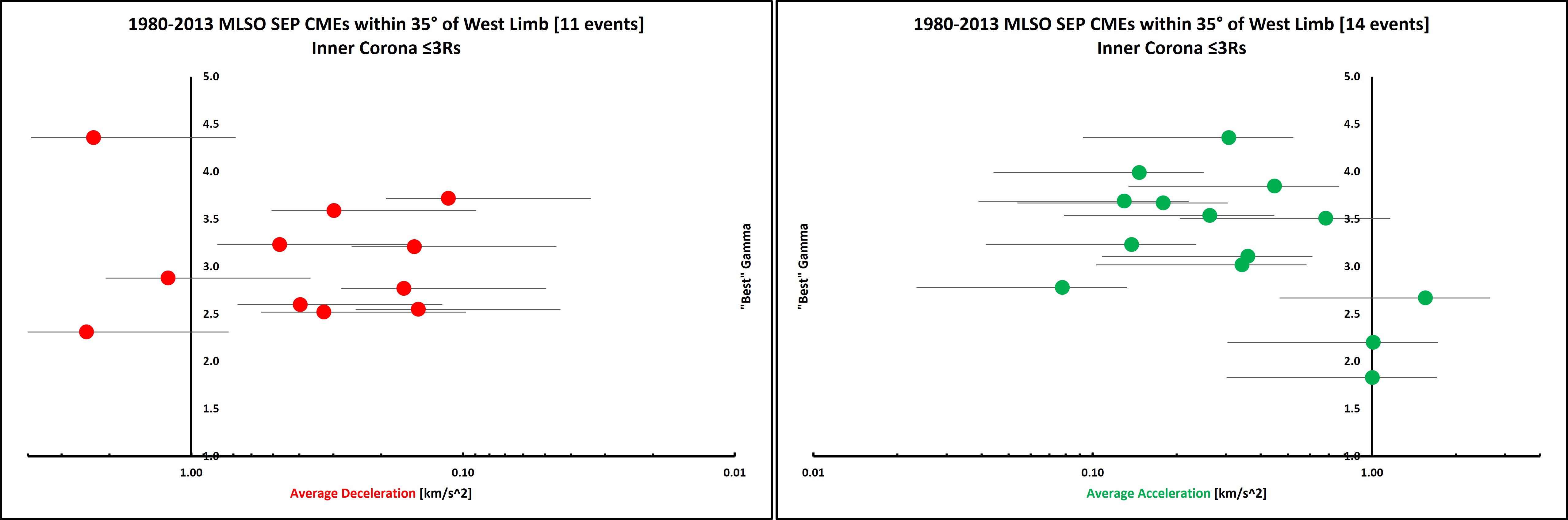}
    \caption{ SEP proton spectral index vs. CME acceleration (right) or deceleration (left) in the inner corona for near west limb events. Note the logarithmic acceleration scale.} 
    \label{gamaccwest}
\end{figure}

\begin{figure}
    \centering
\includegraphics[width=1.0\textwidth]{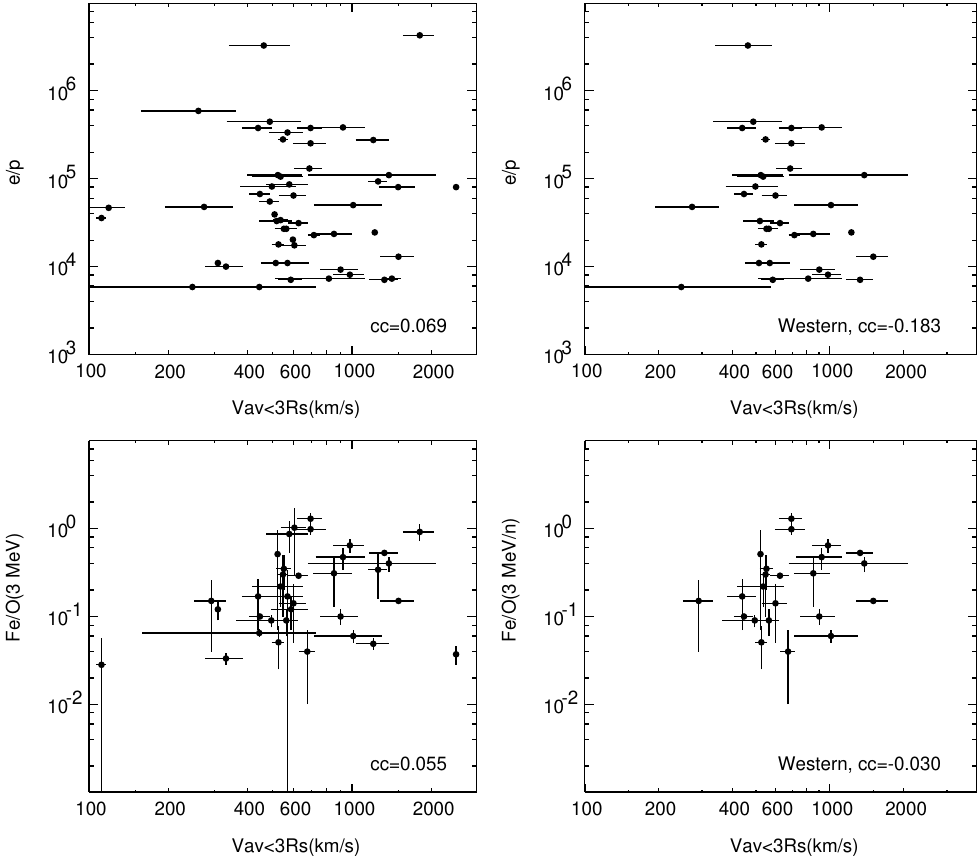}
    \caption{The electron/proton ratio (top) and Fe/O at $\sim3$~MeV/n (bottom) vs. average CME speed at $<3$~R$_s$ for all flare longitudes (left) and western longitudes (right).  No significant correlations with the average CME speed are evident. } 
    \label{vavepfeo}
\end{figure}

Figure~28 compares the SEP electron/proton ratio (top panels) and Fe/O at $\sim3$~MeV/n (from Figure~11; bottom panels), which may provide insight into the SEP acceleration process, with the average CME speed in the low corona. Results for events at all longitudes are shown in the left panels, and for western hemisphere events in the right panels. In addition, here we use CME speeds from Tables 1-3 even if less than four height-time measurements are available.  No significant trends are evident, as indicated by the small correlation coefficients.  As noted in relation to Figure~11, there is a source longitude-dependence in Fe/O, but just considering events with western hemisphere sources (bottom right panel) does not reveal any clearer relationship between the CME speed and Fe/O than considering events at all longitudes (bottom left panel). Similarly for the e/p ratio, where the top right panel shows results for western events, while events at all longitudes are included in the top left panel. We have also examined whether there is any relationship between the CME acceleration/deceleration and these SEP parameters (and also the Fe/O ratio at $\sim14$~MeV/n).  Again do not find any significant correlations, so the related figures are not shown here.

\begin{figure}
    \centerline{
\includegraphics[width=0.5\textwidth]{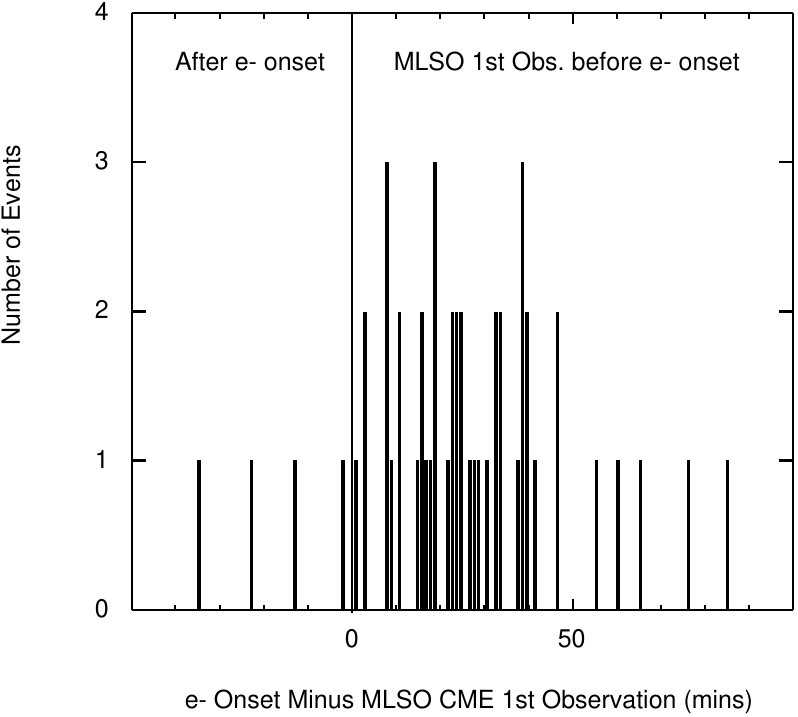}
\includegraphics[width=0.5\textwidth]{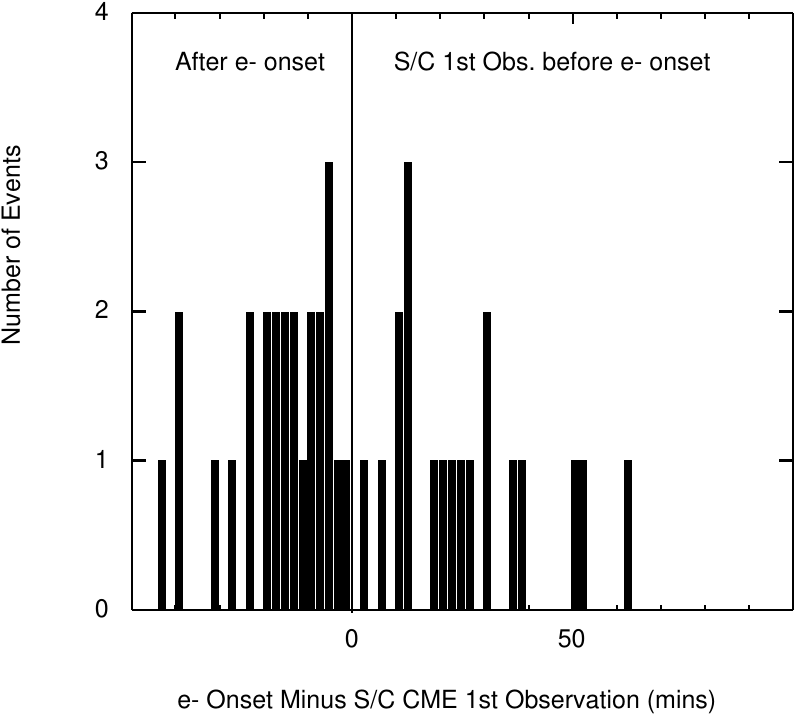}
}
    \caption{Left: Delay in the arrival at the observing spacecraft of near-relativistic electrons relative to the first CME observation by MLSO Mk3/4 for 50 events with delays less than 100 minutes. In all but four of these cases (92\%), the CME is observed before the electron onset. Right: The corresponding delays for the first CME detection by spacebased coronagraphs for 44 events. For 25 of these events (57\%), the electrons arrived before the first CME detection. Note that data latency is not included here.}
    \label{onset}
\end{figure}

\subsection{Comparison of MLSO CME first observation and SEP electron onset times}

 As discussed in Section~1, a motivation for using MLSO observations is that MLSO should detect a CME and provide estimates of its properties earlier than spacebased coronagraphs, and potentially before the onset of the associated SEP event \citep{stcyr2017}. To demonstrate this, Figure~29 compares (left) the distribution of the interval between the
first MLSO observation of a CME and the onset of the associated SEP near-relativistic electron increase at the observing spacecraft, with similar results (right) for spacebased coronagraphs. Here, a positive (negative) interval indicates that the first CME observation is before (after) the electron onset. We use the electron onset time because, as is well-established, SEP electrons arrive before protons \citep[e.g.,][]{richardson2014}, and may be used as a forewarning of a proton event \citep{posner2007}, and so provide the earliest detection of an SEP event except in rare GLE events (cf., Figure~5).  In Figure~29, only events where the delay can be calculated (i.e., with electron observations available without a data gap at SEP event onset) are included, and a few events
with large SEP delays beyond the range shown are excluded. 
 
Figure~29 clearly demonstrates the advantage of MLSO over spacebased coronagraphs as proposed by \cite{stcyr2017} since, for 92\% of the events, the CME is first observed by MLSO before the SEP electron onset. At least another $\sim9$~minutes would then be required to acquire the four MLSO Mk3/4 images we require to estimate the CME dynamics (neglecting issues related to scanning discussed in relation to Figure~1), but this would still occur before the SEP onset for 39 (78\%) of these events.  In contrast, SEP electrons arrived before the first CME observation by spacebased coronagraphs in 25 (57\%) of the events in the right panel.  Other events have delayed SEP onsets that are later than the first spacebased observation of the CME.  Note that data latency, the interval between an observation being made and becoming available for analysis, is not considered here.  For example, for MLSO, as noted in Section~2.2, this might include the time taken to ship tapes to HAO for processing.

In four cases in Figure~29, the
SEP electron onset apparently preceded the first MLSO CME observation. Briefly summarizing the circumstances of these events: The May 3 1980 SEP onset at Earth ($35\pm15$ minutes before the first CME observation at 20:50~UT) was not associated with an identified flare, so it may have been from a solar event behind the west limb.  It might also have been a delayed onset associated with an earlier, larger SEP event observed at Helios 1 (onset at 14:00~UT $\pm7$~minutes) that is unrelated to the MLSO CME but may have been associated with an M2.1 flare at E43$^\circ$ relative to Earth with a peak at 13:04~UT. So the SEP-CME association may be uncertain for this event.  For the September 9 1989 event, the CME was already at 1.68~R$_s$ when first observed by MLSO, and the $\pm15$~minute uncertainty due to SEP averaging is comparable to the 23~minute delay between the estimated SEP onset and first CME observation. For  March 17, 2003, the first MLSO observation was also made when the CME had reached 1.68~R$_s$, $4\pm1$~minutes after the SEP onset, and only two images are available (the second at 2.15~R$_s$). Therefore, the MLSO coverage was poor.  For the November 2, 2003 event (GLE~67), the EPHIN electron channels show a counting rate increase at 17:15~UT, while the MLSO observation window opened at 17:16~UT and the first estimate of the CME leading edge height may be made at 17:17~UT (taking into account the CME direction relative to the scan start/end location) when the CME was at 1.30~R$_s$. However, examining the EPHIN intensity-time profile, it appears that the initial increase in the electron channels is due to contamination from the associated strong (X8.3) flare (see \cite{posner2007} for a discussion of issues with determining SEP onsets with EPHIN). As Figure~5 shows, the ACE/EPAM electron onset was well after MLSO first observed the CME.  It is possible that a few other electron onset times in Figure~29 are affected by similar issues, but nevertheless, this does not change the main conclusion that MLSO typically observes CMEs before the onset of the associated SEP event.

\begin{figure}
    \centering
\includegraphics[width=0.7\textwidth]{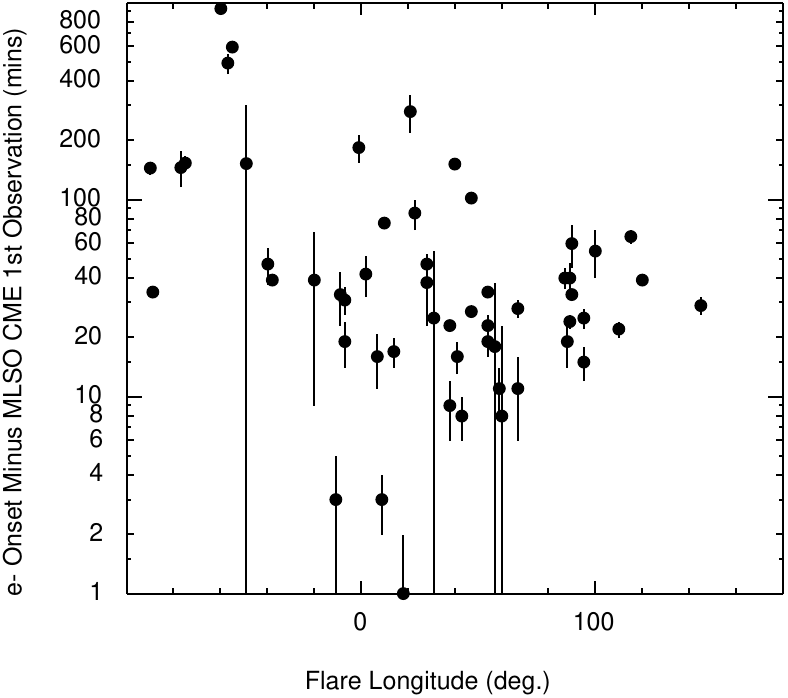}
    \caption{SEP near-relativistic electron arrival delay relative to the first CME observation by MLSO Mk3/4 vs. the longitude of the associated flare relative to the observing spacecraft (West (East) = positive (negative) longitudes). The smallest delays tend to be for well-connected western hemisphere flares.  } 
    \label{long}
\end{figure}

Figure~30 shows the dependence of the interval (if positive) between the first MLSO CME observation and
the near-relativistic electron onset on the flare longitude with respect to the observing spacecraft. The shortest intervals between CME observation
and electron onset occur for well-connected western hemisphere (positive longitude) events. This is consistent with previous observations showing a general increase in SEP electron onset delays with increasing difference in longitude between the solar event/flare and the footpoint of the magnetic field line passing the spacecraft \citep[e.g.,][]{richardson2014}.
In most cases, the minimum intervals are around 9~minutes which, as noted above, would be just sufficient to acquire the four Mk3/4 images that we require for analyzing the CME dynamics.  The error bars generally reflect uncertainty in the SEP onset time due to the time-averaging of the SEP data used or in estimating the onset time for a slowly rising event.

\section{Summary}
\label{S-summary}
\begin{itemize}
\item The work reported here is the first comprehensive study of the relationship between CMEs observed by the MLSO Mk3/4 coronameters in 1980--2013 and SEP events, and covers two full (22, 23) and two partial (21, 24) solar cycles. 
\item The SEP observations used were predominantly from near-Earth spacecraft, with a few events observed by more remote spacecraft including Helios 1/2, STEREO A/B and Ulysses. We consider $\sim25$~MeV proton events observed by instruments on scientific spacecraft rather than just the operations-oriented GOES proton detectors, which have much higher backgrounds, resulting in a more comprehensive list of MLSO CMEs associated with SEP events, with a larger dynamic range in SEP intensity, than would be possible using GOES data alone.
    \item Eighty-four CMEs associated with SEP events that extend to at least $\sim25$~MeV have been identified in the MLSO Mk3/4 coronameter observations during their period of operation in 1980-2013. 
    \item These SEP events have a wide range of properties such as peak intensity, spectral index, Fe/O and electron/proton ratios, and peak soft X-ray intensity of the associated solar flare. In particular, MLSO does not just observe the CMEs associated with, for example, the largest SEP events and most energetic solar eruptions. 
    \item The requirement for MLSO to be operating and viewing the Sun (nominal viewing window 17-02~UT) considerably limits the number of SEP events that can be associated with MLSO CMEs. We estimate that only $\sim9$\% of the $\sim25$~MeV proton events in our study period could be associated with a MLSO CME. However, {\it when operating}, MLSO detected at least 92\% of the SEP-associated CMEs observed by the SOHO/LASCO coronagraphs, and most of the missed events occurred after lightning degraded the performance of Mk4.  The detectability of CMEs by MLSO also falls with distance of the source from the limbs of the Sun.  
    \item SEP-associated CMEs are on average faster (650 vs. 390 km/s) and wider (73$^\circ$ vs. 37$^\circ$) compared to a survey of around 500 MLSO Mk3 CMEs.
\item CME average accelerations/decelerations are larger in the inner corona ($\sim$ km~s$^{-2}$) than in the middle corona observed by space-based coronagraphs, where the CMEs are nearly at their terminal velocity.
\item Methods to refine the CME dynamics in the low corona, such as point-to-point estimation or cubic spline interpolation, do not generally produce consistent results with the three-minute cadence Mk3/4 images.  This complicates efforts to identify parameters characterizing the CME dynamics in the inner corona that may be related to the properties of the associated SEP events. 
\item Estimates of CME dynamics in the low corona inferred from the flare proxy method appear to be inconsistent with those estimated directly from the MLSO CME observations. We also note that the flare proxy method may lead to accelerations that are beyond the range of the calibration of this method ($\sim0.5$~km/s$^2$), as was noted by \cite{zhang2006}.   
\item Considering events originating within $35^\circ$ longitude of the west limb (to reduce projection effects), there is an indication of a correlation between the CME average or final velocity in the inner corona and SEP peak intensity at $\sim20$~MeV, but the data are too sparse (including a limited number of fast ($>1000$~km/s) CMEs) to make a definitive conclusion and to quantify this possible correlation. 
\item The SEP proton spectral index (at $\sim5$--60~MeV) shows evidence of an anti-correlation with the average CME speed or acceleration in the low corona, but again, there are insufficient data to come to a definitive conclusion.  
\item The two GLEs in the near-west-limb sample have hard spectra and are associated with the highest average CME speeds in the low corona.  This may be consistent with the proposal that the highest energy SEPs are accelerated by CME-driven shocks low in the corona.
\item The SEP Fe/O and electron/proton ratios do not appear to be correlated with the average CME speed or acceleration in the low corona.
\item For $\sim92$\% of the selected events, MLSO Mk3/4 first observed the CME before the arrival of non-relativistic SEP electrons at 1~AU.    
\end{itemize}

\section{Discussion}
\label{S-discussion}
As outlined in Section~1, the main motivations for this first comprehensive study of the association of SEP events with CMEs observed in the low corona by the MLSO Mk3/4 coronameters  were (i) to assess whether MLSO CME observations can be used to provide insight into the acceleration of SEPs in the  low corona, (ii) to determine whether there are characteristics of CMEs in the low corona that may be predictive of an SEP event, and hence (iii) to assess whether a network of ground-based coronagraphs might be able to provide a useful early warning of SEP events before the associated CMEs have entered the field of view of spacebased coronagraphs. 

Although we have identified a significant sample of 84 CMEs observed by MLSO Mk3/4 that were associated with SEP events with a wide range in peak intensities, we have been able to address these motivations only to a limited extent. We encountered several significant problems: 

(i) It is difficult to obtain consistent estimates of the CME dynamics in the low corona from the 3~minute cadence Mk3/4 observations.  Estimates of the CME speed and acceleration can differ significantly using different techniques (e.g., P2P, CSI) to fit the observations, and these methods may only be applied to limited subsets of the events. Therefore, we have focused in this paper largely on average speeds and accelerations in the low corona;

(ii) CMEs often have substantial accelerations or decelerations ($\sim$~km/s$^2$) in the low corona, in contrast to the near-terminal speeds found in the mid corona.  Thus, the average and final speeds in the MLSO field of view can differ significantly, and for a strongly decelerating CME, neither speed may reflect a much higher but undetermined maximum speed that might be relevant for understanding particle acceleration in the low corona.  

(iii) The majority of our SEP-associated CMEs have average speeds well below 1000 km/s in the low corona.  The sparseness of exceptionally fast CMEs in our sample (even more so when selecting subsets of events, such as those close to the west limb) makes it more difficult to assess whether there is any relationship between CME speed in the low corona and SEP peak intensity similar to that found for mid-corona CME speeds that might be the basis for an SEP intensity prediction scheme similar to, for example, SEPSTER \citep{richardson2018}.  In addition, if protons at moderate energies (e.g., a few tens of MeV) are predominantly accelerated by shocks in the mid-corona above several $R_s$ \citep[e.g.,][]{kahler1994}, then we might not expect a simple relationship between CME speeds in the low corona and the peak proton intensities at $\sim20$~MeV considered in this study. Unfortunately, none of the SEP events in 2006-2014 with high energy ($\sim80$~MeV-few GeV) proton spectra from PAMELA reported by \cite{bruno2018} occurred at a time when MLSO Mk4 was making observations.  Therefore, we could not compare the CME dynamics in the low corona, where the higher energy SEPs are believed to be accelerated, with the parameters of the PAMELA proton spectra.  

There is a suggestion in our results of a relationship between SEP proton event hardness (measured by the spectral index $\gamma$) and CME speed or acceleration in the low corona, possibly similar to that reported by \cite{gopalswamy2016}, but there is not a sufficiently large range in these CME parameters to make a conclusive case. Since we have made a comprehensive search for CMEs that are associated with SEP events, it is unlikely that additional events remain to be identified that might help to improve the statistics.  Comparing with a large general sample of MLSO CMEs, the SEP-associated CMEs are, on average, wider and faster than typical CMEs.  Hence, the speed and width of an MLSO CME may provide some indication as to whether or not it is likely to be associated with an SEP event. However assessing whether CME speed/width in the low corona provide a useful discriminator of SEP/non-SEP-associated CMEs would require a further study including a ``control" set of MLSO CMEs without SEP events.   

Considering whether a network of ground-based coronagraphs might be used for SEP monitoring, we first note that the availability of MLSO Mk3/4 observations from a single site significantly restricted the number of MLSO CMEs associated with SEP events identified in this study.  We estimate that only $\sim9$\% of the $\sim25$~MeV proton events in our study period could be associated with CMEs observed by Mk3/4. Thus, if Mk3/4 had been used as monitoring instruments, the CMEs associated with a large majority of our SEP events would not have been detected. The major limitations resulted from operational constraints, in particular only $\sim200$~days of observations/year due to lower staffing prior to 2003, instrumental problems with Mk4 following a lightning strike in 2009 and poor sky conditions. Since that time, MLSO has employed a full staff to maximize operations and has made significant upgrades to the electrical and grounding systems and observatory infrastructure. Nevertheless, from 2014 through 2022, K-Cor only acquired, on average, 264 days of observations a year, though it has been providing unique near-real-time CME observations to the CCMC SEP Scoreboard (\url{https://ccmc.gsfc.nasa.gov/scoreboards/sep/}). Thus, we conclude that, in addition to  requiring a worldwide network of observatories to provide 24 hour coverage due to the limited daily solar observation time of a single observatory, multiple overlapping observatories will be required to provide cover for weather and operational interruptions at individual observatories. More positively, when operating, Mk3/4 did observe the vast majority of the SEP-associated CMEs also observed by LASCO, so the detection of SEP-associated CMEs by Mk3/4 was not an issue.

\begin{figure}
    \centering
\includegraphics[width=0.7\textwidth]{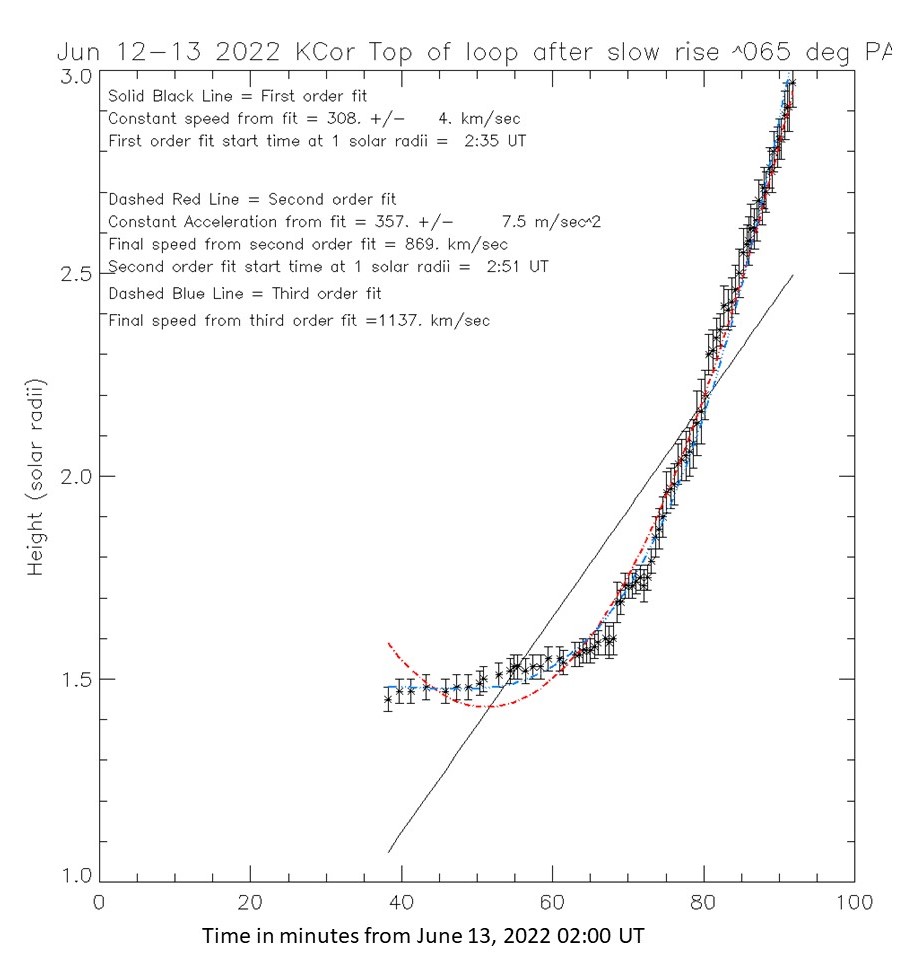}
    \caption{Example of a CME height-time curve derived from 15~s  MLSO K-Cor images, for a north-eastern CME on June 13, 2022. Measurements are made at a position angle of 65$^\circ$. First-, second- and third-order fits are indicated in black, red and blue, respectively.  } 
    \label{KCorht}
\end{figure}

Work is ongoing to add SEP-associated CMEs observed by K-COR since 2013 to our sample of events and to assess whether the higher instrument cadence will allow the CME dynamics to be determined using techniques such as P2P and CSI. Though beyond the scope of this paper, we show in Figure~31 an example of a height-time profile obtained from 15~s cadence K-Cor observations for a CME on June~13, 2022 associated with an M3.4 flare at N15E45 and with a $>20$~MeV proton event observed at spacecraft including STEREO~A and near-earth spacecraft. This clearly shows an initial slow rise followed by a rapid acceleration at $\sim1.6$~R$_s$ to a higher $\sim$terminal speed. First- to third-order fits are shown, with the third-order fit (blue) indicating a final speed of 1137~km/s that is comparable to the speed of 1150~km/s in the mid corona given in the CDAW LASCO CME catalog. The K-Cor height-time profile also shows finer-scale features that might be fitted by other methods. A combined K-Cor-LASCO height-time profile is also shown in \cite{stcyr2017}. Such results indicate that K-Cor observations are able to provide more detailed information on the dynamics of CMEs in the low corona than is possible using Mk3/4.  While solar cycle 25 promises to provide additional events for such analysis, at the time of writing, lava flows from the Mauna Loa volcano eruption on November 28, 2022 have cut power and access to the observatory, so no MLSO observations will be possible at least for the near future.

\begin{acks}
At NASA-GSFC, William Thompson provided expert IDL support, and Jack Ireland and Kevin Schenk assisted with obtaining some spacebased data.  At the Naval Research Laboratory, Russ. Howard and Robin Colanino assisted with obtaining the SOLWIND coronagraph observations, and at the Max Planck Institute for Solar System Research, Borut Podlipnik assisted with obtaining some LASCO C1 data. We thank Hilary Cane for providing a list of SEP events and their solar sources that facilitated this study and Bojan Vr\v{s}nak for help with implementing CSI. Dan Fry of SRAG at NASA-JSC provided valuable guidance early in the study.  This work is supported by funding for the NASA LWS Focused Science Team ``Toward a Systems Approach to Energetic Particle Acceleration and Transport on the Sun and in the Heliosphere", NNH17ZDA001N-LWS, and NASA program NNH19ZDA001N-HSR. The Mauna Loa Solar Observatory is operated by the High Altitude Observatory as part of the National Center for Atmospheric Research (NCAR). NCAR is supported by the National Science Foundation.  The SMM data are courtesy of the HAO/SMM C/P project team and NASA. 
 The Mk3, Mk4 and SMM data sets are identified by DOI:10.5065/D62N5129, DOI: 10.5065/D66972C9 and DOI:10.5065/D64J0CXB, respectively. The energetic particle data used are available from the NASA Space Physics Data Facility (e.g., \url{https://cdaweb.gsfc.nasa.gov/} and \url{https://spdf.gsfc.nasa.gov/research/vepo/}), the ACE Science Center (\url{https://izw1.caltech.edu/ACE/ASC/}) and the neutron monitor database (\url{https://www.nmdb.eu/}) The CDAW LASCO CME catalog (\url{https://cdaw.gsfc.nasa.gov/CME_list/}) is generated and maintained at the CDAW Data Center by NASA and The Catholic University of America in cooperation with the Naval Research Laboratory. SOHO is a project of international cooperation between ESA and NASA.

\end{acks}

\bibliographystyle{spr-mp-sola}
\bibliography{MLSOSEP}  

\begin{thebibliography}{84}
\ifx\bisbn     \undefined \def\bisbn  #1{ISBN #1}\fi
\ifx\binits    \undefined \def\binits#1{#1}\fi
\ifx\bauthor   \undefined \def\bauthor#1{#1}\fi
\ifx\batitle   \undefined \def\batitle#1{#1}\fi
\ifx\bjtitle   \undefined \def\bjtitle#1{\textit{#1}}\fi
\ifx\bvolume   \undefined \def\bvolume#1{\textbf{#1}}\fi
\ifx\byear     \undefined \def\byear#1{#1}\fi
\ifx\bissue    \undefined \def\bissue#1{#1}\fi
\ifx\bfpage    \undefined \def\bfpage#1{#1}\fi
\ifx\blpage    \undefined \def\blpage #1{#1}\fi
\ifx\burl      \undefined \def\burl#1{\textsf{#1}}\fi
\ifx\href      \undefined \def\href#1#2{\textsf{#2}}\fi
\ifx\betal     \undefined \def\betal{\textit{et al.}}\fi
\ifx\bctitle   \undefined \def\bctitle#1{#1}\fi
\ifx\beditor   \undefined \def\beditor#1{#1}\fi
\ifx\bbtitle   \undefined \def\bbtitle#1{\textit{#1}}\fi
\ifx\bedition  \undefined \def\bedition#1{#1}\fi
\ifx\bseriesno \undefined \def\bseriesno#1{\textbf{#1}}\fi
\ifx\blocation \undefined \def\blocation#1{#1}\fi
\ifx\bsertitle \undefined \def\bsertitle#1{\textit{#1}}\fi
\ifx\bsnm      \undefined \def\bsnm#1{#1}\fi
\ifx\bsuffix   \undefined \def\bsuffix#1{#1}\fi
\ifx\bparticle \undefined \def\bparticle#1{#1}\fi
\ifx\barticle  \undefined \def\barticle#1{}\fi
\ifx\binstitute  \undefined \def\binstitute#1{#1}\fi
\ifx\bpublisher  \undefined \def\bpublisher#1{#1}\fi
\ifx\doiurl    \undefined
  \def\doiurl#1{\href{http://dx.doi.org/#1}{\textsf{DOI}}}\fi
\ifx\arxivurl  \undefined
  \def\arxivurl#1{\href{http://arxiv.org/abs/#1}{\textsf{arXiv}}}\fi
\ifx\adsurl    \undefined
  \def\adsurl#1{\href{http://adsabs.harvard.edu/abs/#1}{\textsf{ADS}}}\fi
\ifx\botherref \undefined \def\botherref#1{}\fi
\ifx\url       \undefined \def\url#1{\textsf{#1}}\fi
\ifx\bchapter  \undefined \def\bchapter#1{}\fi
\ifx\bbook     \undefined \def\bbook#1{}\fi
\ifx\bcomment  \undefined \def\bcomment#1{#1}\fi
\ifx\oauthor   \undefined \def\oauthor#1{#1}\fi
\ifx\citeauthoryear \undefined\def \citeauthoryear#1{#1}\fi
\ifx\endbibitem\undefined \def\endbibitem{}\fi
\ifx\bconflocation  \undefined \def\bconflocation#1{#1} \fi

\bibitem[\protect\citeauthoryear{{Bein} \textit{et~al.}}{2011}]{Bein2011}
\begin{barticle}
\bauthor{\bsnm{{Bein}}, \binits{B.M.}},
\bauthor{\bsnm{{Berkebile-Stoiser}}, \binits{S.}},
\bauthor{\bsnm{{Veronig}}, \binits{A.M.}},
\bauthor{\bsnm{{Temmer}}, \binits{M.}},
\bauthor{\bsnm{{Muhr}}, \binits{N.}},
\bauthor{\bsnm{{Kienreich}}, \binits{I.}},
\bauthor{\bsnm{{Utz}}, \binits{D.}},
\bauthor{\bsnm{{Vr{\v{s}}nak}}, \binits{B.}}:
\byear{2011},
\batitle{{Impulsive Acceleration of Coronal Mass Ejections. I. Statistics and
  Coronal Mass Ejection Source Region Characteristics}}.
\bjtitle{\apj}
\bvolume{738}(\bissue{2}),
\bfpage{191}.
\doiurl{10.1088/0004-637X/738/2/191}.
\adsurl{https://ui.adsabs.harvard.edu/abs/2011ApJ...738..191B}.
\end{barticle}
\endbibitem

\bibitem[\protect\citeauthoryear{{Brueckner}
  \textit{et~al.}}{1995}]{brueckner1995}
\begin{barticle}
\bauthor{\bsnm{{Brueckner}}, \binits{G.E.}},
\bauthor{\bsnm{{Howard}}, \binits{R.A.}},
\bauthor{\bsnm{{Koomen}}, \binits{M.J.}},
\bauthor{\bsnm{{Korendyke}}, \binits{C.M.}},
\bauthor{\bsnm{{Michels}}, \binits{D.J.}},
\bauthor{\bsnm{{Moses}}, \binits{J.D.}},
\bauthor{\bsnm{{Socker}}, \binits{D.G.}},
\bauthor{\bsnm{{Dere}}, \binits{K.P.}},
\bauthor{\bsnm{{Lamy}}, \binits{P.L.}},
\bauthor{\bsnm{{Llebaria}}, \binits{A.}},
\bauthor{\bsnm{{Bout}}, \binits{M.V.}},
\bauthor{\bsnm{{Schwenn}}, \binits{R.}},
\bauthor{\bsnm{{Simnett}}, \binits{G.M.}},
\bauthor{\bsnm{{Bedford}}, \binits{D.K.}},
\bauthor{\bsnm{{Eyles}}, \binits{C.J.}}:
\byear{1995},
\batitle{{The Large Angle Spectroscopic Coronagraph (LASCO)}}.
\bjtitle{\solphys}
\bvolume{162}(\bissue{1-2}),
\bfpage{357}.
\doiurl{10.1007/BF00733434}.
\adsurl{https://ui.adsabs.harvard.edu/abs/1995SoPh..162..357B}.
\end{barticle}
\endbibitem

\bibitem[\protect\citeauthoryear{{Bruno} and {Richardson}}{2021}]{bruno2021}
\begin{barticle}
\bauthor{\bsnm{{Bruno}}, \binits{A.}},
\bauthor{\bsnm{{Richardson}}, \binits{I.G.}}:
\byear{2021},
\batitle{{Empirical Model of 10 - 130 MeV Solar Energetic Particle Spectra at 1
  AU Based on Coronal Mass Ejection Speed and Direction}}.
\bjtitle{\solphys}
\bvolume{296}(\bissue{2}),
\bfpage{36}.
\doiurl{10.1007/s11207-021-01779-4}.
\adsurl{https://ui.adsabs.harvard.edu/abs/2021SoPh..296...36B}.
\end{barticle}
\endbibitem

\bibitem[\protect\citeauthoryear{{Bruno} \textit{et~al.}}{2018}]{bruno2018}
\begin{barticle}
\bauthor{\bsnm{{Bruno}}, \binits{A.}},
\bauthor{\bsnm{{Bazilevskaya}}, \binits{G.A.}},
\bauthor{\bsnm{{Boezio}}, \binits{M.}},
\bauthor{\bsnm{{Christian}}, \binits{E.R.}},
\bauthor{\bsnm{{de Nolfo}}, \binits{G.A.}},
\bauthor{\bsnm{{Martucci}}, \binits{M.}},
\bauthor{\bsnm{{Merge}}, \binits{M.}},
\bauthor{\bsnm{{Mikhailov}}, \binits{V.V.}},
\bauthor{\bsnm{{Munini}}, \binits{R.}},
\bauthor{\bsnm{{Richardson}}, \binits{I.G.}},
\bauthor{\bsnm{{Ryan}}, \binits{J.M.}},
\bauthor{\bsnm{{Stochaj}}, \binits{S.}},
\bauthor{\bsnm{{Adriani}}, \binits{O.}},
\bauthor{\bsnm{{Barbarino}}, \binits{G.C.}},
\bauthor{\bsnm{{Bellotti}}, \binits{R.}},
\bauthor{\bsnm{{Bogomolov}}, \binits{E.A.}},
\bauthor{\bsnm{{Bongi}}, \binits{M.}},
\bauthor{\bsnm{{Bonvicini}}, \binits{V.}},
\bauthor{\bsnm{{Bottai}}, \binits{S.}},
\bauthor{\bsnm{{Cafagna}}, \binits{F.}},
\bauthor{\bsnm{{Campana}}, \binits{D.}},
\bauthor{\bsnm{{Carlson}}, \binits{P.}},
\bauthor{\bsnm{{Casolino}}, \binits{M.}},
\bauthor{\bsnm{{Castellini}}, \binits{G.}},
\bauthor{\bsnm{{De Santis}}, \binits{C.}},
\bauthor{\bsnm{{Di Felice}}, \binits{V.}},
\bauthor{\bsnm{{Galper}}, \binits{A.M.}},
\bauthor{\bsnm{{Karelin}}, \binits{A.V.}},
\bauthor{\bsnm{{Koldashov}}, \binits{S.V.}},
\bauthor{\bsnm{{Koldobskiy}}, \binits{S.}},
\bauthor{\bsnm{{Krutkov}}, \binits{S.Y.}},
\bauthor{\bsnm{{Kvashnin}}, \binits{A.N.}},
\bauthor{\bsnm{{Leonov}}, \binits{A.}},
\bauthor{\bsnm{{Malakhov}}, \binits{V.}},
\bauthor{\bsnm{{Marcelli}}, \binits{L.}},
\bauthor{\bsnm{{Mayorov}}, \binits{A.G.}},
\bauthor{\bsnm{{Menn}}, \binits{W.}},
\bauthor{\bsnm{{Mocchiutti}}, \binits{E.}},
\bauthor{\bsnm{{Monaco}}, \binits{A.}},
\bauthor{\bsnm{{Mori}}, \binits{N.}},
\bauthor{\bsnm{{Osteria}}, \binits{G.}},
\bauthor{\bsnm{{Panico}}, \binits{B.}},
\bauthor{\bsnm{{Papini}}, \binits{P.}},
\bauthor{\bsnm{{Pearce}}, \binits{M.}},
\bauthor{\bsnm{{Picozza}}, \binits{P.}},
\bauthor{\bsnm{{Ricci}}, \binits{M.}},
\bauthor{\bsnm{{Ricciarini}}, \binits{S.B.}},
\bauthor{\bsnm{{Simon}}, \binits{M.}},
\bauthor{\bsnm{{Sparvoli}}, \binits{R.}},
\bauthor{\bsnm{{Spillantini}}, \binits{P.}},
\bauthor{\bsnm{{Stozhkov}}, \binits{Y.I.}},
\bauthor{\bsnm{{Vacchi}}, \binits{A.}},
\bauthor{\bsnm{{Vannuccini}}, \binits{E.}},
\bauthor{\bsnm{{Vasilyev}}, \binits{G.I.}},
\bauthor{\bsnm{{Voronov}}, \binits{S.A.}},
\bauthor{\bsnm{{Yurkin}}, \binits{Y.T.}},
\bauthor{\bsnm{{Zampa}}, \binits{G.}},
\bauthor{\bsnm{{Zampa}}, \binits{N.}}:
\byear{2018},
\batitle{{Solar Energetic Particle Events Observed by the PAMELA Mission}}.
\bjtitle{\apj}
\bvolume{862}(\bissue{2}),
\bfpage{97}.
\doiurl{10.3847/1538-4357/aacc26}.
\adsurl{https://ui.adsabs.harvard.edu/abs/2018ApJ...862...97B}.
\end{barticle}
\endbibitem

\bibitem[\protect\citeauthoryear{{Burkepile} and {St.
  Cyr}}{1993}]{Burkepile1993}
\begin{botherref}
\oauthor{\bsnm{{Burkepile}}, \binits{J.T.}},
\oauthor{\bsnm{{St. Cyr}}, \binits{O.C.}}:
1993,
{A Revised and Expanded Catalogue of Mass Ejections Observed by the Solar
  Maximum Mission Coronagraph}.
\textit{NCAR Technical Note; NCAR/TN-369+STR}.
\end{botherref}
\endbibitem

\bibitem[\protect\citeauthoryear{{Cane}, {Erickson}, and
  {Prestage}}{2002}]{cane2002}
\begin{barticle}
\bauthor{\bsnm{{Cane}}, \binits{H.V.}},
\bauthor{\bsnm{{Erickson}}, \binits{W.C.}},
\bauthor{\bsnm{{Prestage}}, \binits{N.P.}}:
\byear{2002},
\batitle{{Solar flares, type III radio bursts, coronal mass ejections, and
  energetic particles}}.
\bjtitle{Journal of Geophysical Research (Space Physics)}
\bvolume{107}(\bissue{A10}),
\bfpage{1315}.
\doiurl{10.1029/2001JA000320}.
\adsurl{https://ui.adsabs.harvard.edu/abs/2002JGRA..107.1315C}.
\end{barticle}
\endbibitem

\bibitem[\protect\citeauthoryear{{Cane}, {Reames}, and {von
  Rosenvinge}}{1988}]{cane1988}
\begin{barticle}
\bauthor{\bsnm{{Cane}}, \binits{H.V.}},
\bauthor{\bsnm{{Reames}}, \binits{D.V.}},
\bauthor{\bsnm{{von Rosenvinge}}, \binits{T.T.}}:
\byear{1988},
\batitle{{The role of interplanetary shocks in the longitude distribution of
  solar energetic particles}}.
\bjtitle{\jgr}
\bvolume{93}(\bissue{A9}),
\bfpage{9555}.
\doiurl{10.1029/JA093iA09p09555}.
\adsurl{https://ui.adsabs.harvard.edu/abs/1988JGR....93.9555C}.
\end{barticle}
\endbibitem

\bibitem[\protect\citeauthoryear{{Cane}, {Richardson}, and {von
  Rosenvinge}}{2010}]{cane2010}
\begin{barticle}
\bauthor{\bsnm{{Cane}}, \binits{H.V.}},
\bauthor{\bsnm{{Richardson}}, \binits{I.G.}},
\bauthor{\bsnm{{von Rosenvinge}}, \binits{T.T.}}:
\byear{2010},
\batitle{{A study of solar energetic particle events of 1997-2006: Their
  composition and associations}}.
\bjtitle{Journal of Geophysical Research (Space Physics)}
\bvolume{115}(\bissue{A8}),
\bfpage{A08101}.
\doiurl{10.1029/2009JA014848}.
\adsurl{https://ui.adsabs.harvard.edu/abs/2010JGRA..115.8101C}.
\end{barticle}
\endbibitem

\bibitem[\protect\citeauthoryear{Cane \textit{et~al.}}{2006}]{cane2006}
\begin{barticle}
\bauthor{\bsnm{Cane}, \binits{H.V.}},
\bauthor{\bsnm{Mewaldt}, \binits{R.A.}},
\bauthor{\bsnm{Cohen}, \binits{C.M.S.}},
\bauthor{\bparticle{von} \bsnm{Rosenvinge}, \binits{T.T.}}:
\byear{2006},
\batitle{Role of flares and shocks in determining solar energetic particle
  abundances}.
\bjtitle{Journal of Geophysical Research: Space Physics}
\bvolume{111}(\bissue{A6}).
\doiurl{https://doi.org/10.1029/2005JA011071}.
\end{barticle}
\endbibitem

\bibitem[\protect\citeauthoryear{{Ciaravella}
  \textit{et~al.}}{2005}]{Ciaravella2005}
\begin{barticle}
\bauthor{\bsnm{{Ciaravella}}, \binits{A.}},
\bauthor{\bsnm{{Raymond}}, \binits{J.C.}},
\bauthor{\bsnm{{Kahler}}, \binits{S.W.}},
\bauthor{\bsnm{{Vourlidas}}, \binits{A.}},
\bauthor{\bsnm{{Li}}, \binits{J.}}:
\byear{2005},
\batitle{{Detection and Diagnostics of a Coronal Shock Wave Driven by a
  Partial-Halo Coronal Mass Ejection on 2000 June 28}}.
\bjtitle{\apj}
\bvolume{621}(\bissue{2}),
\bfpage{1121}.
\doiurl{10.1086/427619}.
\adsurl{https://ui.adsabs.harvard.edu/abs/2005ApJ...621.1121C}.
\end{barticle}
\endbibitem

\bibitem[\protect\citeauthoryear{{de Wijn} \textit{et~al.}}{2012}]{dewijn2012}
\begin{bchapter}
\bauthor{\bsnm{{de Wijn}}, \binits{A.G.}},
\bauthor{\bsnm{{Burkepile}}, \binits{J.T.}},
\bauthor{\bsnm{{Tomczyk}}, \binits{S.}},
\bauthor{\bsnm{{Nelson}}, \binits{P.G.}},
\bauthor{\bsnm{{Huang}}, \binits{P.}},
\bauthor{\bsnm{{Gallagher}}, \binits{D.}}:
\byear{2012},
\bctitle{{Stray light and polarimetry considerations for the COSMO
  K-Coronagraph}}.
In: \beditor{\bsnm{{Stepp}}, \binits{L.M.}},
\beditor{\bsnm{{Gilmozzi}}, \binits{R.}},
\beditor{\bsnm{{Hall}}, \binits{H.J.}} (eds.)
\bbtitle{Ground-based and Airborne Telescopes IV},
\bsertitle{Society of Photo-Optical Instrumentation Engineers (SPIE) Conference
  Series}
\bseriesno{8444},
\bfpage{84443N}.
\doiurl{10.1117/12.926511}.
\adsurl{https://ui.adsabs.harvard.edu/abs/2012SPIE.8444E..3ND}.
\end{bchapter}
\endbibitem

\bibitem[\protect\citeauthoryear{{Elmore} \textit{et~al.}}{2003}]{elmore2003}
\begin{bchapter}
\bauthor{\bsnm{{Elmore}}, \binits{D.F.}},
\bauthor{\bsnm{{Burkepile}}, \binits{J.T.}},
\bauthor{\bsnm{{Darnell}}, \binits{J.A.}},
\bauthor{\bsnm{{Lecinski}}, \binits{A.R.}},
\bauthor{\bsnm{{Stanger}}, \binits{A.L.}}:
\byear{2003},
\bctitle{{Calibration of a ground-based solar coronal polarimeter}}.
In: \beditor{\bsnm{{Fineschi}}, \binits{S.}} (ed.)
\bbtitle{Polarimetry in Astronomy},
\bsertitle{Society of Photo-Optical Instrumentation Engineers (SPIE) Conference
  Series}
\bseriesno{4843},
\bfpage{66}.
\doiurl{10.1117/12.459279}.
\adsurl{https://ui.adsabs.harvard.edu/abs/2003SPIE.4843...66E}.
\end{bchapter}
\endbibitem

\bibitem[\protect\citeauthoryear{{Fisher} \textit{et~al.}}{1981}]{fisher1981}
\begin{barticle}
\bauthor{\bsnm{{Fisher}}, \binits{R.R.}},
\bauthor{\bsnm{{Lee}}, \binits{R.H.}},
\bauthor{\bsnm{{MacQueen}}, \binits{R.M.}},
\bauthor{\bsnm{{Poland}}, \binits{A.I.}}:
\byear{1981},
\batitle{{New Mauna Loa coronagraph systems}}.
\bjtitle{Applied Optics}
\bvolume{20}(\bissue{6}),
\bfpage{1094}.
\doiurl{10.1364/AO.20.001094}.
\adsurl{https://ui.adsabs.harvard.edu/abs/1981ApOpt..20.1094F}.
\end{barticle}
\endbibitem

\bibitem[\protect\citeauthoryear{Fleck and St.~Cyr}{2014}]{Fleck2014}
\begin{bchapter}
\bauthor{\bsnm{Fleck}, \binits{B.}},
\bauthor{\bsnm{St.~Cyr}, \binits{O.C.}}:
\byear{2014},
\bctitle{{Solar and Heliospheric Observatory (SOHO)}}.
In: \beditor{\bsnm{Allahdadi}, \binits{F.}},
\beditor{\bsnm{Pelton}, \binits{J.N.}} (eds.)
\bbtitle{Handbook of Cosmic Hazards and Planetary Defense},
\bpublisher{Springer},
\blocation{Switzerland},
\bfpage{1}.
\doiurl{10.1007/978-3-319-02847-7_14-1}.
\end{bchapter}
\endbibitem

\bibitem[\protect\citeauthoryear{{Gold} \textit{et~al.}}{1998}]{gold1998}
\begin{barticle}
\bauthor{\bsnm{{Gold}}, \binits{R.E.}},
\bauthor{\bsnm{{Krimigis}}, \binits{S.M.}},
\bauthor{\bsnm{{Hawkins}}, \binits{I.} \bsuffix{S.~E.}},
\bauthor{\bsnm{{Haggerty}}, \binits{D.K.}},
\bauthor{\bsnm{{Lohr}}, \binits{D.A.}},
\bauthor{\bsnm{{Fiore}}, \binits{E.}},
\bauthor{\bsnm{{Armstrong}}, \binits{T.P.}},
\bauthor{\bsnm{{Holland}}, \binits{G.}},
\bauthor{\bsnm{{Lanzerotti}}, \binits{L.J.}}:
\byear{1998},
\batitle{{Electron, Proton, and Alpha Monitor on the Advanced Composition
  Explorer spacecraft}}.
\bjtitle{\ssr}
\bvolume{86},
\bfpage{541}.
\doiurl{10.1023/A:1005088115759}.
\adsurl{https://ui.adsabs.harvard.edu/abs/1998SSRv...86..541G}.
\end{barticle}
\endbibitem

\bibitem[\protect\citeauthoryear{{Gopalswamy}
  \textit{et~al.}}{2010}]{gopalswamy2010}
\begin{barticle}
\bauthor{\bsnm{{Gopalswamy}}, \binits{N.}},
\bauthor{\bsnm{{Yashiro}}, \binits{S.}},
\bauthor{\bsnm{{Michalek}}, \binits{G.}},
\bauthor{\bsnm{{Xie}}, \binits{H.}},
\bauthor{\bsnm{{M{\"a}kel{\"a}}}, \binits{P.}},
\bauthor{\bsnm{{Vourlidas}}, \binits{A.}},
\bauthor{\bsnm{{Howard}}, \binits{R.A.}}:
\byear{2010},
\batitle{{A Catalog of Halo Coronal Mass Ejections from SOHO}}.
\bjtitle{Sun and Geosphere}
\bvolume{5}(\bissue{1}),
\bfpage{7}.
\adsurl{https://ui.adsabs.harvard.edu/abs/2010SunGe...5....7G}.
\end{barticle}
\endbibitem

\bibitem[\protect\citeauthoryear{{Gopalswamy}
  \textit{et~al.}}{2012}]{gopalswamy2012}
\begin{barticle}
\bauthor{\bsnm{{Gopalswamy}}, \binits{N.}},
\bauthor{\bsnm{{Xie}}, \binits{H.}},
\bauthor{\bsnm{{Yashiro}}, \binits{S.}},
\bauthor{\bsnm{{Akiyama}}, \binits{S.}},
\bauthor{\bsnm{{M{\"a}kel{\"a}}}, \binits{P.}},
\bauthor{\bsnm{{Usoskin}}, \binits{I.G.}}:
\byear{2012},
\batitle{{Properties of Ground Level Enhancement Events and the Associated
  Solar Eruptions During Solar Cycle 23}}.
\bjtitle{\ssr}
\bvolume{171}(\bissue{1-4}),
\bfpage{23}.
\doiurl{10.1007/s11214-012-9890-4}.
\adsurl{https://ui.adsabs.harvard.edu/abs/2012SSRv..171...23G}.
\end{barticle}
\endbibitem

\bibitem[\protect\citeauthoryear{{Gopalswamy}
  \textit{et~al.}}{2013}]{gopalswamy2013}
\begin{barticle}
\bauthor{\bsnm{{Gopalswamy}}, \binits{N.}},
\bauthor{\bsnm{{Xie}}, \binits{H.}},
\bauthor{\bsnm{{Akiyama}}, \binits{S.}},
\bauthor{\bsnm{{Yashiro}}, \binits{S.}},
\bauthor{\bsnm{{Usoskin}}, \binits{I.G.}},
\bauthor{\bsnm{{Davila}}, \binits{J.M.}}:
\byear{2013},
\batitle{{The First Ground Level Enhancement Event of Solar Cycle 24: Direct
  Observation of Shock Formation and Particle Release Heights}}.
\bjtitle{\apjl}
\bvolume{765}(\bissue{2}),
\bfpage{L30}.
\doiurl{10.1088/2041-8205/765/2/L30}.
\adsurl{https://ui.adsabs.harvard.edu/abs/2013ApJ...765L..30G}.
\end{barticle}
\endbibitem

\bibitem[\protect\citeauthoryear{{Gopalswamy}
  \textit{et~al.}}{2016}]{gopalswamy2016}
\begin{barticle}
\bauthor{\bsnm{{Gopalswamy}}, \binits{N.}},
\bauthor{\bsnm{{Yashiro}}, \binits{S.}},
\bauthor{\bsnm{{Thakur}}, \binits{N.}},
\bauthor{\bsnm{{M{\"a}kel{\"a}}}, \binits{P.}},
\bauthor{\bsnm{{Xie}}, \binits{H.}},
\bauthor{\bsnm{{Akiyama}}, \binits{S.}}:
\byear{2016},
\batitle{{The 2012 July 23 Backside Eruption: An Extreme Energetic Particle
  Event?}}
\bjtitle{\apj}
\bvolume{833}(\bissue{2}),
\bfpage{216}.
\doiurl{10.3847/1538-4357/833/2/216}.
\adsurl{https://ui.adsabs.harvard.edu/abs/2016ApJ...833..216G}.
\end{barticle}
\endbibitem

\bibitem[\protect\citeauthoryear{{Gosling} \textit{et~al.}}{1976}]{Gosling1976}
\begin{barticle}
\bauthor{\bsnm{{Gosling}}, \binits{J.T.}},
\bauthor{\bsnm{{Hildner}}, \binits{E.}},
\bauthor{\bsnm{{MacQueen}}, \binits{R.M.}},
\bauthor{\bsnm{{Munro}}, \binits{R.H.}},
\bauthor{\bsnm{{Poland}}, \binits{A.I.}},
\bauthor{\bsnm{{Ross}}, \binits{C.L.}}:
\byear{1976},
\batitle{{The speeds of coronal mass ejection events.}}
\bjtitle{\solphys}
\bvolume{48}(\bissue{2}),
\bfpage{389}.
\doiurl{10.1007/BF00152004}.
\adsurl{https://ui.adsabs.harvard.edu/abs/1976SoPh...48..389G}.
\end{barticle}
\endbibitem

\bibitem[\protect\citeauthoryear{{Harrison}
  \textit{et~al.}}{1990}]{Harrison1990}
\begin{barticle}
\bauthor{\bsnm{{Harrison}}, \binits{R.A.}},
\bauthor{\bsnm{{Hildner}}, \binits{E.}},
\bauthor{\bsnm{{Hundhausen}}, \binits{A.J.}},
\bauthor{\bsnm{{Sime}}, \binits{D.G.}},
\bauthor{\bsnm{{Simnett}}, \binits{G.M.}}:
\byear{1990},
\batitle{{The launch of solar coronal mass ejections: Results from the coronal
  mass ejection onset program}}.
\bjtitle{\jgr}
\bvolume{95}(\bissue{A2}),
\bfpage{917}.
\doiurl{10.1029/JA095iA02p00917}.
\adsurl{https://ui.adsabs.harvard.edu/abs/1990JGR....95..917H}.
\end{barticle}
\endbibitem

\bibitem[\protect\citeauthoryear{{Howard} \textit{et~al.}}{1985}]{Howard1985}
\begin{barticle}
\bauthor{\bsnm{{Howard}}, \binits{R.A.}},
\bauthor{\bsnm{{Sheeley}}, \binits{J.} \bsuffix{N.~R.}},
\bauthor{\bsnm{{Michels}}, \binits{D.J.}},
\bauthor{\bsnm{{Koomen}}, \binits{M.J.}}:
\byear{1985},
\batitle{{Coronal mass ejections: 1979-1981}}.
\bjtitle{\jgr}
\bvolume{90}(\bissue{A9}),
\bfpage{8173}.
\doiurl{10.1029/JA090iA09p08173}.
\adsurl{https://ui.adsabs.harvard.edu/abs/1985JGR....90.8173H}.
\end{barticle}
\endbibitem

\bibitem[\protect\citeauthoryear{{Howard} \textit{et~al.}}{2008}]{howard2008}
\begin{barticle}
\bauthor{\bsnm{{Howard}}, \binits{R.A.}},
\bauthor{\bsnm{{Moses}}, \binits{J.D.}},
\bauthor{\bsnm{{Vourlidas}}, \binits{A.}},
\bauthor{\bsnm{{Newmark}}, \binits{J.S.}},
\bauthor{\bsnm{{Socker}}, \binits{D.G.}},
\bauthor{\bsnm{{Plunkett}}, \binits{S.P.}},
\bauthor{\bsnm{{Korendyke}}, \binits{C.M.}},
\bauthor{\bsnm{{Cook}}, \binits{J.W.}},
\bauthor{\bsnm{{Hurley}}, \binits{A.}},
\bauthor{\bsnm{{Davila}}, \binits{J.M.}},
\bauthor{\bsnm{{Thompson}}, \binits{W.T.}},
\bauthor{\bsnm{{St Cyr}}, \binits{O.C.}},
\bauthor{\bsnm{{Mentzell}}, \binits{E.}},
\bauthor{\bsnm{{Mehalick}}, \binits{K.}},
\bauthor{\bsnm{{Lemen}}, \binits{J.R.}},
\bauthor{\bsnm{{Wuelser}}, \binits{J.P.}},
\bauthor{\bsnm{{Duncan}}, \binits{D.W.}},
\bauthor{\bsnm{{Tarbell}}, \binits{T.D.}},
\bauthor{\bsnm{{Wolfson}}, \binits{C.J.}},
\bauthor{\bsnm{{Moore}}, \binits{A.}},
\bauthor{\bsnm{{Harrison}}, \binits{R.A.}},
\bauthor{\bsnm{{Waltham}}, \binits{N.R.}},
\bauthor{\bsnm{{Lang}}, \binits{J.}},
\bauthor{\bsnm{{Davis}}, \binits{C.J.}},
\bauthor{\bsnm{{Eyles}}, \binits{C.J.}},
\bauthor{\bsnm{{Mapson-Menard}}, \binits{H.}},
\bauthor{\bsnm{{Simnett}}, \binits{G.M.}},
\bauthor{\bsnm{{Halain}}, \binits{J.P.}},
\bauthor{\bsnm{{Defise}}, \binits{J.M.}},
\bauthor{\bsnm{{Mazy}}, \binits{E.}},
\bauthor{\bsnm{{Rochus}}, \binits{P.}},
\bauthor{\bsnm{{Mercier}}, \binits{R.}},
\bauthor{\bsnm{{Ravet}}, \binits{M.F.}},
\bauthor{\bsnm{{Delmotte}}, \binits{F.}},
\bauthor{\bsnm{{Auchere}}, \binits{F.}},
\bauthor{\bsnm{{Delaboudiniere}}, \binits{J.P.}},
\bauthor{\bsnm{{Bothmer}}, \binits{V.}},
\bauthor{\bsnm{{Deutsch}}, \binits{W.}},
\bauthor{\bsnm{{Wang}}, \binits{D.}},
\bauthor{\bsnm{{Rich}}, \binits{N.}},
\bauthor{\bsnm{{Cooper}}, \binits{S.}},
\bauthor{\bsnm{{Stephens}}, \binits{V.}},
\bauthor{\bsnm{{Maahs}}, \binits{G.}},
\bauthor{\bsnm{{Baugh}}, \binits{R.}},
\bauthor{\bsnm{{McMullin}}, \binits{D.}},
\bauthor{\bsnm{{Carter}}, \binits{T.}}:
\byear{2008},
\batitle{{Sun Earth Connection Coronal and Heliospheric Investigation
  (SECCHI)}}.
\bjtitle{\ssr}
\bvolume{136}(\bissue{1-4}),
\bfpage{67}.
\doiurl{10.1007/s11214-008-9341-4}.
\adsurl{https://ui.adsabs.harvard.edu/abs/2008SSRv..136...67H}.
\end{barticle}
\endbibitem

\bibitem[\protect\citeauthoryear{{Hundhausen}}{1993}]{Hundhausen1993}
\begin{barticle}
\bauthor{\bsnm{{Hundhausen}}, \binits{A.J.}}:
\byear{1993},
\batitle{{Sizes and locations of coronal mass ejections: SMM observations from
  1980 and 1984-1989}}.
\bjtitle{\jgr}
\bvolume{98}(\bissue{A8}),
\bfpage{13177}.
\doiurl{10.1029/93JA00157}.
\adsurl{https://ui.adsabs.harvard.edu/abs/1993JGR....9813177H}.
\end{barticle}
\endbibitem

\bibitem[\protect\citeauthoryear{{Hundhausen}, {Burkepile}, and {St.
  Cyr}}{1994}]{hundhausen1994}
\begin{barticle}
\bauthor{\bsnm{{Hundhausen}}, \binits{A.J.}},
\bauthor{\bsnm{{Burkepile}}, \binits{J.T.}},
\bauthor{\bsnm{{St. Cyr}}, \binits{O.C.}}:
\byear{1994},
\batitle{{Speeds of coronal mass ejections: SMM observations from 1980 and
  1984-1989}}.
\bjtitle{\jgr}
\bvolume{99}(\bissue{A4}),
\bfpage{6543}.
\doiurl{10.1029/93JA03586}.
\adsurl{https://ui.adsabs.harvard.edu/abs/1994JGR....99.6543H}.
\end{barticle}
\endbibitem

\bibitem[\protect\citeauthoryear{{Hurlburt}
  \textit{et~al.}}{2012}]{Hurlburt2012}
\begin{barticle}
\bauthor{\bsnm{{Hurlburt}}, \binits{N.}},
\bauthor{\bsnm{{Cheung}}, \binits{M.}},
\bauthor{\bsnm{{Schrijver}}, \binits{C.}},
\bauthor{\bsnm{{Chang}}, \binits{L.}},
\bauthor{\bsnm{{Freeland}}, \binits{S.}},
\bauthor{\bsnm{{Green}}, \binits{S.}},
\bauthor{\bsnm{{Heck}}, \binits{C.}},
\bauthor{\bsnm{{Jaffey}}, \binits{A.}},
\bauthor{\bsnm{{Kobashi}}, \binits{A.}},
\bauthor{\bsnm{{Schiff}}, \binits{D.}},
\bauthor{\bsnm{{Serafin}}, \binits{J.}},
\bauthor{\bsnm{{Seguin}}, \binits{R.}},
\bauthor{\bsnm{{Slater}}, \binits{G.}},
\bauthor{\bsnm{{Somani}}, \binits{A.}},
\bauthor{\bsnm{{Timmons}}, \binits{R.}}:
\byear{2012},
\batitle{{Heliophysics Event Knowledgebase for the Solar Dynamics Observatory
  (SDO) and Beyond}}.
\bjtitle{\solphys}
\bvolume{275}(\bissue{1-2}),
\bfpage{67}.
\doiurl{10.1007/s11207-010-9624-2}.
\adsurl{https://ui.adsabs.harvard.edu/abs/2012SoPh..275...67H}.
\end{barticle}
\endbibitem

\bibitem[\protect\citeauthoryear{{Kahler}}{1994}]{kahler1994}
\begin{barticle}
\bauthor{\bsnm{{Kahler}}, \binits{S.}}:
\byear{1994},
\batitle{{Injection Profiles of Solar Energetic Particles as Functions of
  Coronal Mass Ejection Heights}}.
\bjtitle{\apj}
\bvolume{428},
\bfpage{837}.
\doiurl{10.1086/174292}.
\adsurl{https://ui.adsabs.harvard.edu/abs/1994ApJ...428..837K}.
\end{barticle}
\endbibitem

\bibitem[\protect\citeauthoryear{{Kahler} and {Vourlidas}}{2005}]{Kahler2005}
\begin{barticle}
\bauthor{\bsnm{{Kahler}}, \binits{S.W.}},
\bauthor{\bsnm{{Vourlidas}}, \binits{A.}}:
\byear{2005},
\batitle{{Fast coronal mass ejection environments and the production of solar
  energetic particle events}}.
\bjtitle{Journal of Geophysical Research (Space Physics)}
\bvolume{110}(\bissue{A12}),
\bfpage{A12S01}.
\doiurl{10.1029/2005JA011073}.
\adsurl{https://ui.adsabs.harvard.edu/abs/2005JGRA..11012S01K}.
\end{barticle}
\endbibitem

\bibitem[\protect\citeauthoryear{{Kahler}, {Hildner}, and {Van
  Hollebeke}}{1978}]{kahler1978}
\begin{barticle}
\bauthor{\bsnm{{Kahler}}, \binits{S.W.}},
\bauthor{\bsnm{{Hildner}}, \binits{E.}},
\bauthor{\bsnm{{Van Hollebeke}}, \binits{M.A.I.}}:
\byear{1978},
\batitle{{Prompt solar proton events and coronal mass ejections.}}
\bjtitle{\solphys}
\bvolume{57}(\bissue{2}),
\bfpage{429}.
\doiurl{10.1007/BF00160116}.
\adsurl{https://ui.adsabs.harvard.edu/abs/1978SoPh...57..429K}.
\end{barticle}
\endbibitem

\bibitem[\protect\citeauthoryear{Kallenrode}{1993}]{kallenrode1993}
\begin{barticle}
\bauthor{\bsnm{Kallenrode}, \binits{M.-B.}}:
\byear{1993},
\batitle{Neutral lines and azimuthal “transport” of solar energetic
  particles}.
\bjtitle{Journal of Geophysical Research: Space Physics}
\bvolume{98}(\bissue{A4}),
\bfpage{5573}.
\doiurl{https://doi.org/10.1029/92JA02778}.
\end{barticle}
\endbibitem

\bibitem[\protect\citeauthoryear{{Lario} and {Pick}}{2008}]{Lario2008}
\begin{bchapter}
\bauthor{\bsnm{{Lario}}, \binits{D.}},
\bauthor{\bsnm{{Pick}}, \binits{M.}}:
\byear{2008},
\bctitle{{Heliospheric energetic particle variations}}.
In: \beditor{\bsnm{{Balogh}}, \binits{A.}},
\beditor{\bsnm{{Lanzerotti}}, \binits{L.J.}},
\beditor{\bsnm{{Suess}}, \binits{S.T.}} (eds.)
\bbtitle{The Heliosphere through the Solar Activity Cycle},
\bfpage{151}.
\doiurl{10.1007/978-3-540-74302-6_5}.
\adsurl{https://ui.adsabs.harvard.edu/abs/2008hsac.book..151L}.
\end{bchapter}
\endbibitem

\bibitem[\protect\citeauthoryear{{Laurenza}
  \textit{et~al.}}{2009}]{laurenza2009}
\begin{barticle}
\bauthor{\bsnm{{Laurenza}}, \binits{M.}},
\bauthor{\bsnm{{Cliver}}, \binits{E.W.}},
\bauthor{\bsnm{{Hewitt}}, \binits{J.}},
\bauthor{\bsnm{{Storini}}, \binits{M.}},
\bauthor{\bsnm{{Ling}}, \binits{A.G.}},
\bauthor{\bsnm{{Balch}}, \binits{C.C.}},
\bauthor{\bsnm{{Kaiser}}, \binits{M.L.}}:
\byear{2009},
\batitle{{A technique for short-term warning of solar energetic particle events
  based on flare location, flare size, and evidence of particle escape}}.
\bjtitle{Space Weather}
\bvolume{7}(\bissue{4}),
\bfpage{S04008}.
\doiurl{10.1029/2007SW000379}.
\adsurl{https://ui.adsabs.harvard.edu/abs/2009SpWea...7.4008L}.
\end{barticle}
\endbibitem

\bibitem[\protect\citeauthoryear{{Ling} \textit{et~al.}}{2014}]{ling2014}
\begin{barticle}
\bauthor{\bsnm{{Ling}}, \binits{A.G.}},
\bauthor{\bsnm{{Webb}}, \binits{D.F.}},
\bauthor{\bsnm{{Burkepile}}, \binits{J.T.}},
\bauthor{\bsnm{{Cliver}}, \binits{E.W.}}:
\byear{2014},
\batitle{{Development of a Current Sheet in the Wake of a Fast Coronal Mass
  Ejection}}.
\bjtitle{\apj}
\bvolume{784}(\bissue{2}),
\bfpage{91}.
\doiurl{10.1088/0004-637X/784/2/91}.
\adsurl{https://ui.adsabs.harvard.edu/abs/2014ApJ...784...91L}.
\end{barticle}
\endbibitem

\bibitem[\protect\citeauthoryear{{MacQueen}}{1985}]{macqueen1985}
\begin{barticle}
\bauthor{\bsnm{{MacQueen}}, \binits{R.M.}}:
\byear{1985},
\batitle{{Coronal Mass Ejections - Acceleration and Surface Associations}}.
\bjtitle{\solphys}
\bvolume{95}(\bissue{2}),
\bfpage{359}.
\doiurl{10.1007/BF00152412}.
\adsurl{https://ui.adsabs.harvard.edu/abs/1985SoPh...95..359M}.
\end{barticle}
\endbibitem

\bibitem[\protect\citeauthoryear{{MacQueen} and {St. Cyr}}{1991}]{macqueen1991}
\begin{barticle}
\bauthor{\bsnm{{MacQueen}}, \binits{R.M.}},
\bauthor{\bsnm{{St. Cyr}}, \binits{O.C.}}:
\byear{1991},
\batitle{{Sungrazing comets observed by the solar maximum mission
  coronagraph}}.
\bjtitle{Icarus}
\bvolume{90}(\bissue{1}),
\bfpage{96}.
\doiurl{10.1016/0019-1035(91)90071-Z}.
\adsurl{https://ui.adsabs.harvard.edu/abs/1991Icar...90...96M}.
\end{barticle}
\endbibitem

\bibitem[\protect\citeauthoryear{{MacQueen}
  \textit{et~al.}}{1980}]{macqueen1980}
\begin{barticle}
\bauthor{\bsnm{{MacQueen}}, \binits{R.M.}},
\bauthor{\bsnm{{Csoeke-Poeckh}}, \binits{A.}},
\bauthor{\bsnm{{Hildner}}, \binits{E.}},
\bauthor{\bsnm{{House}}, \binits{L.}},
\bauthor{\bsnm{{Reynolds}}, \binits{R.}},
\bauthor{\bsnm{{Stanger}}, \binits{A.}},
\bauthor{\bsnm{{Tepoel}}, \binits{H.}},
\bauthor{\bsnm{{Wagner}}, \binits{W.}}:
\byear{1980},
\batitle{{The High Altitude Observatory coronagraph/polarimeter on the Solar
  Maximum Mission.}}
\bjtitle{\solphys}
\bvolume{65}(\bissue{1}),
\bfpage{91}.
\doiurl{10.1007/BF00151386}.
\adsurl{https://ui.adsabs.harvard.edu/abs/1980SoPh...65...91M}.
\end{barticle}
\endbibitem

\bibitem[\protect\citeauthoryear{{MacQueen}
  \textit{et~al.}}{2001}]{macqueen2001}
\begin{barticle}
\bauthor{\bsnm{{MacQueen}}, \binits{R.M.}},
\bauthor{\bsnm{{Burkepile}}, \binits{J.T.}},
\bauthor{\bsnm{{Holzer}}, \binits{T.E.}},
\bauthor{\bsnm{{Stanger}}, \binits{A.L.}},
\bauthor{\bsnm{{Spence}}, \binits{K.E.}}:
\byear{2001},
\batitle{{Solar Coronal Brightness Changes and Mass Ejections during Solar
  Cycle 22}}.
\bjtitle{\apj}
\bvolume{549}(\bissue{2}),
\bfpage{1175}.
\doiurl{10.1086/319464}.
\adsurl{https://ui.adsabs.harvard.edu/abs/2001ApJ...549.1175M}.
\end{barticle}
\endbibitem

\bibitem[\protect\citeauthoryear{Majumdar \textit{et~al.}}{2020}]{Majumdar2020}
\begin{barticle}
\bauthor{\bsnm{Majumdar}, \binits{S.}},
\bauthor{\bsnm{Pant}, \binits{V.}},
\bauthor{\bsnm{Patel}, \binits{R.}},
\bauthor{\bsnm{Banerjee}, \binits{D.}}:
\byear{2020},
\batitle{{Connecting 3D Evolution of Coronal Mass Ejections to Their Source
  Regions}}.
\bjtitle{The Astrophysical Journal}
\bvolume{899}(\bissue{1}),
\bfpage{6}.
\doiurl{10.3847/1538-4357/aba1f2}.
\end{barticle}
\endbibitem

\bibitem[\protect\citeauthoryear{{Mari{\v{c}}i{\'c}}
  \textit{et~al.}}{2004}]{Maricic2004}
\begin{barticle}
\bauthor{\bsnm{{Mari{\v{c}}i{\'c}}}, \binits{D.}},
\bauthor{\bsnm{{Vr{\v{s}}nak}}, \binits{B.}},
\bauthor{\bsnm{{Stanger}}, \binits{A.L.}},
\bauthor{\bsnm{{Veronig}}, \binits{A.}}:
\byear{2004},
\batitle{{Coronal Mass Ejection of 15 May 2001: I. Evolution of Morphological
  Features of the Eruption}}.
\bjtitle{\solphys}
\bvolume{225}(\bissue{2}),
\bfpage{337}.
\doiurl{10.1007/s11207-004-3748-1}.
\adsurl{https://ui.adsabs.harvard.edu/abs/2004SoPh..225..337M}.
\end{barticle}
\endbibitem

\bibitem[\protect\citeauthoryear{{McGuire}, {von Rosenvinge}, and
  {McDonald}}{1986}]{mcguire1986}
\begin{barticle}
\bauthor{\bsnm{{McGuire}}, \binits{R.E.}},
\bauthor{\bsnm{{von Rosenvinge}}, \binits{T.T.}},
\bauthor{\bsnm{{McDonald}}, \binits{F.B.}}:
\byear{1986},
\batitle{{The Composition of Solar Energetic Particles}}.
\bjtitle{\apj}
\bvolume{301},
\bfpage{938}.
\doiurl{10.1086/163958}.
\adsurl{https://ui.adsabs.harvard.edu/abs/1986ApJ...301..938M}.
\end{barticle}
\endbibitem

\bibitem[\protect\citeauthoryear{{Mishev} \textit{et~al.}}{2021}]{mishev2021}
\begin{barticle}
\bauthor{\bsnm{{Mishev}}, \binits{A.L.}},
\bauthor{\bsnm{{Koldobskiy}}, \binits{S.A.}},
\bauthor{\bsnm{{Kocharov}}, \binits{L.G.}},
\bauthor{\bsnm{{Usoskin}}, \binits{I.G.}}:
\byear{2021},
\batitle{{GLE \#67 Event on 2 November 2003: An Analysis of the Spectral and
  Anisotropy Characteristics Using Verified Yield Function and Detrended
  Neutron Monitor Data}}.
\bjtitle{\solphys}
\bvolume{296}(\bissue{5}),
\bfpage{79}.
\doiurl{10.1007/s11207-021-01832-2}.
\adsurl{https://ui.adsabs.harvard.edu/abs/2021SoPh..296...79M}.
\end{barticle}
\endbibitem

\bibitem[\protect\citeauthoryear{{Miteva}, {Samwel}, and
  {Costa-Duarte}}{2018}]{miteva2018}
\begin{barticle}
\bauthor{\bsnm{{Miteva}}, \binits{R.}},
\bauthor{\bsnm{{Samwel}}, \binits{S.W.}},
\bauthor{\bsnm{{Costa-Duarte}}, \binits{M.V.}}:
\byear{2018},
\batitle{{The Wind/EPACT Proton Event Catalog (1996 - 2016)}}.
\bjtitle{\solphys}
\bvolume{293}(\bissue{2}),
\bfpage{27}.
\doiurl{10.1007/s11207-018-1241-5}.
\adsurl{https://ui.adsabs.harvard.edu/abs/2018SoPh..293...27M}.
\end{barticle}
\endbibitem

\bibitem[\protect\citeauthoryear{{M{\"u}ller-Mellin}
  \textit{et~al.}}{1995}]{muller1995}
\begin{barticle}
\bauthor{\bsnm{{M{\"u}ller-Mellin}}, \binits{R.}},
\bauthor{\bsnm{{Kunow}}, \binits{H.}},
\bauthor{\bsnm{{Flei{\ss}ner}}, \binits{V.}},
\bauthor{\bsnm{{Pehlke}}, \binits{E.}},
\bauthor{\bsnm{{Rode}}, \binits{E.}},
\bauthor{\bsnm{{R{\"o}schmann}}, \binits{N.}},
\bauthor{\bsnm{{Scharmberg}}, \binits{C.}},
\bauthor{\bsnm{{Sierks}}, \binits{H.}},
\bauthor{\bsnm{{Rusznyak}}, \binits{P.}},
\bauthor{\bsnm{{McKenna-Lawlor}}, \binits{S.}},
\bauthor{\bsnm{{Elendt}}, \binits{I.}},
\bauthor{\bsnm{{Sequeiros}}, \binits{J.}},
\bauthor{\bsnm{{Meziat}}, \binits{D.}},
\bauthor{\bsnm{{Sanchez}}, \binits{S.}},
\bauthor{\bsnm{{Medina}}, \binits{J.}},
\bauthor{\bsnm{{Del Peral}}, \binits{L.}},
\bauthor{\bsnm{{Witte}}, \binits{M.}},
\bauthor{\bsnm{{Marsden}}, \binits{R.}},
\bauthor{\bsnm{{Henrion}}, \binits{J.}}:
\byear{1995},
\batitle{{COSTEP - Comprehensive Suprathermal and Energetic Particle
  Analyser}}.
\bjtitle{\solphys}
\bvolume{162}(\bissue{1-2}),
\bfpage{483}.
\doiurl{10.1007/BF00733437}.
\adsurl{https://ui.adsabs.harvard.edu/abs/1995SoPh..162..483M}.
\end{barticle}
\endbibitem

\bibitem[\protect\citeauthoryear{{Papaioannou}
  \textit{et~al.}}{2016}]{papaioannou2016}
\begin{barticle}
\bauthor{\bsnm{{Papaioannou}}, \binits{A.}},
\bauthor{\bsnm{{Sandberg}}, \binits{I.}},
\bauthor{\bsnm{{Anastasiadis}}, \binits{A.}},
\bauthor{\bsnm{{Kouloumvakos}}, \binits{A.}},
\bauthor{\bsnm{{Georgoulis}}, \binits{M.K.}},
\bauthor{\bsnm{{Tziotziou}}, \binits{K.}},
\bauthor{\bsnm{{Tsiropoula}}, \binits{G.}},
\bauthor{\bsnm{{Jiggens}}, \binits{P.}},
\bauthor{\bsnm{{Hilgers}}, \binits{A.}}:
\byear{2016},
\batitle{{Solar flares, coronal mass ejections and solar energetic particle
  event characteristics}}.
\bjtitle{Journal of Space Weather and Space Climate}
\bvolume{6},
\bfpage{A42}.
\doiurl{10.1051/swsc/2016035}.
\adsurl{https://ui.adsabs.harvard.edu/abs/2016JSWSC...6A..42P}.
\end{barticle}
\endbibitem

\bibitem[\protect\citeauthoryear{Posner}{2007}]{posner2007}
\begin{barticle}
\bauthor{\bsnm{Posner}, \binits{A.}}:
\byear{2007},
\batitle{{Up to 1-hour forecasting of radiation hazards from solar energetic
  ion events with relativistic electrons}}.
\bjtitle{Space Weather}
\bvolume{5}(\bissue{5}).
\doiurl{https://doi.org/10.1029/2006SW000268}.
\end{barticle}
\endbibitem

\bibitem[\protect\citeauthoryear{{Raymond} \textit{et~al.}}{2000}]{Raymond2000}
\begin{barticle}
\bauthor{\bsnm{{Raymond}}, \binits{J.C.}},
\bauthor{\bsnm{{Thompson}}, \binits{B.J.}},
\bauthor{\bsnm{{St. Cyr}}, \binits{O.C.}},
\bauthor{\bsnm{{Gopalswamy}}, \binits{N.}},
\bauthor{\bsnm{{Kahler}}, \binits{S.}},
\bauthor{\bsnm{{Kaiser}}, \binits{M.}},
\bauthor{\bsnm{{Lara}}, \binits{A.}},
\bauthor{\bsnm{{Ciaravella}}, \binits{A.}},
\bauthor{\bsnm{{Romoli}}, \binits{M.}},
\bauthor{\bsnm{{O'Neal}}, \binits{R.}}:
\byear{2000},
\batitle{{SOHO and radio observations of a CME shock wave}}.
\bjtitle{\grl}
\bvolume{27}(\bissue{10}),
\bfpage{1439}.
\doiurl{10.1029/1999GL003669}.
\adsurl{https://ui.adsabs.harvard.edu/abs/2000GeoRL..27.1439R}.
\end{barticle}
\endbibitem

\bibitem[\protect\citeauthoryear{{Reames}}{1999}]{reames1999}
\begin{barticle}
\bauthor{\bsnm{{Reames}}, \binits{D.V.}}:
\byear{1999},
\batitle{{Particle acceleration at the Sun and in the heliosphere}}.
\bjtitle{\ssr}
\bvolume{90},
\bfpage{413}.
\doiurl{10.1023/A:1005105831781}.
\adsurl{https://ui.adsabs.harvard.edu/abs/1999SSRv...90..413R}.
\end{barticle}
\endbibitem

\bibitem[\protect\citeauthoryear{{Reames}}{2009}]{reames2009}
\begin{barticle}
\bauthor{\bsnm{{Reames}}, \binits{D.V.}}:
\byear{2009},
\batitle{{Solar Release Times of Energetic Particles in Ground-Level Events}}.
\bjtitle{\apj}
\bvolume{693}(\bissue{1}),
\bfpage{812}.
\doiurl{10.1088/0004-637X/693/1/812}.
\adsurl{https://ui.adsabs.harvard.edu/abs/2009ApJ...693..812R}.
\end{barticle}
\endbibitem

\bibitem[\protect\citeauthoryear{{Richardson} and
  {Cane}}{1996}]{richardson1996}
\begin{barticle}
\bauthor{\bsnm{{Richardson}}, \binits{I.G.}},
\bauthor{\bsnm{{Cane}}, \binits{H.V.}}:
\byear{1996},
\batitle{{Particle flows observed in ejecta during solar event onsets and their
  implication for the magnetic field topology}}.
\bjtitle{\jgr}
\bvolume{101}(\bissue{A12}),
\bfpage{27521}.
\doiurl{10.1029/96JA02643}.
\adsurl{https://ui.adsabs.harvard.edu/abs/1996JGR...10127521R}.
\end{barticle}
\endbibitem

\bibitem[\protect\citeauthoryear{{Richardson} and
  {Cane}}{2010}]{richardson2010}
\begin{barticle}
\bauthor{\bsnm{{Richardson}}, \binits{I.G.}},
\bauthor{\bsnm{{Cane}}, \binits{H.V.}}:
\byear{2010},
\batitle{{Near-Earth Interplanetary Coronal Mass Ejections During Solar Cycle
  23 (1996 - 2009): Catalog and Summary of Properties}}.
\bjtitle{\solphys}
\bvolume{264}(\bissue{1}),
\bfpage{189}.
\doiurl{10.1007/s11207-010-9568-6}.
\adsurl{https://ui.adsabs.harvard.edu/abs/2010SoPh..264..189R}.
\end{barticle}
\endbibitem

\bibitem[\protect\citeauthoryear{Richardson, Mays, and
  Thompson}{2018}]{richardson2018}
\begin{barticle}
\bauthor{\bsnm{Richardson}, \binits{I.G.}},
\bauthor{\bsnm{Mays}, \binits{M.L.}},
\bauthor{\bsnm{Thompson}, \binits{B.J.}}:
\byear{2018},
\batitle{{Prediction of Solar Energetic Particle Event Peak Proton Intensity
  Using a Simple Algorithm Based on CME Speed and Direction and Observations of
  Associated Solar Phenomena}}.
\bjtitle{Space Weather}
\bvolume{16}(\bissue{11}),
\bfpage{1862}.
\doiurl{https://doi.org/10.1029/2018SW002032}.
\end{barticle}
\endbibitem

\bibitem[\protect\citeauthoryear{{Richardson}, {von Rosenvinge}, and
  {Cane}}{2015}]{richardson2015}
\begin{barticle}
\bauthor{\bsnm{{Richardson}}, \binits{I.G.}},
\bauthor{\bsnm{{von Rosenvinge}}, \binits{T.T.}},
\bauthor{\bsnm{{Cane}}, \binits{H.V.}}:
\byear{2015},
\batitle{{The Properties of Solar Energetic Particle Event-Associated Coronal
  Mass Ejections Reported in Different CME Catalogs}}.
\bjtitle{\solphys}
\bvolume{290}(\bissue{6}),
\bfpage{1741}.
\doiurl{10.1007/s11207-015-0701-4}.
\adsurl{https://ui.adsabs.harvard.edu/abs/2015SoPh..290.1741R}.
\end{barticle}
\endbibitem

\bibitem[\protect\citeauthoryear{{Richardson}, {von Rosenvinge}, and
  {Cane}}{2016}]{richardson2016}
\begin{barticle}
\bauthor{\bsnm{{Richardson}}, \binits{I.G.}},
\bauthor{\bsnm{{von Rosenvinge}}, \binits{T.T.}},
\bauthor{\bsnm{{Cane}}, \binits{H.V.}}:
\byear{2016},
\batitle{{North/South Hemispheric Periodicities in the $>25$ MeV Solar Proton
  Event Rate During the Rising and Peak Phases of Solar Cycle 24}}.
\bjtitle{\solphys}
\bvolume{291}(\bissue{7}),
\bfpage{2117}.
\doiurl{10.1007/s11207-016-0948-4}.
\adsurl{https://ui.adsabs.harvard.edu/abs/2016SoPh..291.2117R}.
\end{barticle}
\endbibitem

\bibitem[\protect\citeauthoryear{{Richardson}, {von Rosenvinge}, and
  {Cane}}{2017}]{richardson2017}
\begin{barticle}
\bauthor{\bsnm{{Richardson}}, \binits{I.G.}},
\bauthor{\bsnm{{von Rosenvinge}}, \binits{T.T.}},
\bauthor{\bsnm{{Cane}}, \binits{H.V.}}:
\byear{2017},
\batitle{{25~MeV solar proton events in Cycle 24 and previous cycles}}.
\bjtitle{Advances in Space Research}
\bvolume{60}(\bissue{4}),
\bfpage{755}.
\doiurl{10.1016/j.asr.2016.07.035}.
\adsurl{https://ui.adsabs.harvard.edu/abs/2017AdSpR..60..755R}.
\end{barticle}
\endbibitem

\bibitem[\protect\citeauthoryear{{Richardson}
  \textit{et~al.}}{2000}]{richardson2000}
\begin{barticle}
\bauthor{\bsnm{{Richardson}}, \binits{I.G.}},
\bauthor{\bsnm{{Dvornikov}}, \binits{V.M.}},
\bauthor{\bsnm{{Sdobnov}}, \binits{V.E.}},
\bauthor{\bsnm{{Cane}}, \binits{H.V.}}:
\byear{2000},
\batitle{{Bidirectional particle flows at cosmic ray and lower ($\sim1$ MeV)
  energies and their association with interplanetary coronal mass
  ejections/ejecta}}.
\bjtitle{\jgr}
\bvolume{105}(\bissue{A6}),
\bfpage{12579}.
\doiurl{10.1029/1999JA000331}.
\adsurl{https://ui.adsabs.harvard.edu/abs/2000JGR...10512579R}.
\end{barticle}
\endbibitem

\bibitem[\protect\citeauthoryear{{Richardson}
  \textit{et~al.}}{2014}]{richardson2014}
\begin{barticle}
\bauthor{\bsnm{{Richardson}}, \binits{I.G.}},
\bauthor{\bsnm{{von Rosenvinge}}, \binits{T.T.}},
\bauthor{\bsnm{{Cane}}, \binits{H.V.}},
\bauthor{\bsnm{{Christian}}, \binits{E.R.}},
\bauthor{\bsnm{{Cohen}}, \binits{C.M.S.}},
\bauthor{\bsnm{{Labrador}}, \binits{A.W.}},
\bauthor{\bsnm{{Leske}}, \binits{R.A.}},
\bauthor{\bsnm{{Mewaldt}}, \binits{R.A.}},
\bauthor{\bsnm{{Wiedenbeck}}, \binits{M.E.}},
\bauthor{\bsnm{{Stone}}, \binits{E.C.}}:
\byear{2014},
\batitle{{$> 25$ MeV Proton Events Observed by the High Energy Telescopes on
  the STEREO A and B Spacecraft and/or at Earth During the First
  {\ensuremath{\sim}} Seven Years of the STEREO Mission}}.
\bjtitle{\solphys}
\bvolume{289}(\bissue{8}),
\bfpage{3059}.
\doiurl{10.1007/s11207-014-0524-8}.
\adsurl{https://ui.adsabs.harvard.edu/abs/2014SoPh..289.3059R}.
\end{barticle}
\endbibitem

\bibitem[\protect\citeauthoryear{{Shanmugaraju}
  \textit{et~al.}}{2003}]{Shanmugaraju2003}
\begin{barticle}
\bauthor{\bsnm{{Shanmugaraju}}, \binits{A.}},
\bauthor{\bsnm{{Moon}}, \binits{Y.-J.}},
\bauthor{\bsnm{{Dryer}}, \binits{M.}},
\bauthor{\bsnm{{Umapathy}}, \binits{S.}}:
\byear{2003},
\batitle{{On the kinematic evolution of flare-associated CMEs}}.
\bjtitle{\solphys}
\bvolume{215}(\bissue{1}),
\bfpage{185}.
\doiurl{10.1023/A:1024808819850}.
\adsurl{https://ui.adsabs.harvard.edu/abs/2003SoPh..215..185S}.
\end{barticle}
\endbibitem

\bibitem[\protect\citeauthoryear{{Sheeley} \textit{et~al.}}{1980}]{sheeley1980}
\begin{barticle}
\bauthor{\bsnm{{Sheeley}}, \binits{J.} \bsuffix{N.~R.}},
\bauthor{\bsnm{{Michels}}, \binits{D.J.}},
\bauthor{\bsnm{{Howard}}, \binits{R.A.}},
\bauthor{\bsnm{{Koomen}}, \binits{M.J.}}:
\byear{1980},
\batitle{{Initial observations with the SOLWIND coronagraph}}.
\bjtitle{\apjl}
\bvolume{237},
\bfpage{L99}.
\doiurl{10.1086/183243}.
\adsurl{https://ui.adsabs.harvard.edu/abs/1980ApJ...237L..99S}.
\end{barticle}
\endbibitem

\bibitem[\protect\citeauthoryear{St.~Cyr, Fleck, and Davila}{2014}]{stcyr2014}
\begin{barticle}
\bauthor{\bsnm{St.~Cyr}, \binits{O.C.}},
\bauthor{\bsnm{Fleck}, \binits{B.}},
\bauthor{\bsnm{Davila}, \binits{J.M.}}:
\byear{2014},
\batitle{The impact of coronagraphs}.
\bjtitle{Eos, Transactions American Geophysical Union}
\bvolume{95}(\bissue{41}),
\bfpage{369}.
\doiurl{https://doi.org/10.1002/2014EO410001}.
\end{barticle}
\endbibitem

\bibitem[\protect\citeauthoryear{St.~Cyr, Posner, and
  Burkepile}{2017}]{stcyr2017}
\begin{barticle}
\bauthor{\bsnm{St.~Cyr}, \binits{O.C.}},
\bauthor{\bsnm{Posner}, \binits{A.}},
\bauthor{\bsnm{Burkepile}, \binits{J.T.}}:
\byear{2017},
\batitle{{Solar energetic particle warnings from a coronagraph}}.
\bjtitle{Space Weather}
\bvolume{15}(\bissue{1}),
\bfpage{240}.
\doiurl{https://doi.org/10.1002/2016SW001545}.
\end{barticle}
\endbibitem

\bibitem[\protect\citeauthoryear{{St. Cyr} \textit{et~al.}}{1999}]{stcyr1999}
\begin{barticle}
\bauthor{\bsnm{{St. Cyr}}, \binits{O.C.}},
\bauthor{\bsnm{{Burkepile}}, \binits{J.T.}},
\bauthor{\bsnm{{Hundhausen}}, \binits{A.J.}},
\bauthor{\bsnm{{Lecinski}}, \binits{A.R.}}:
\byear{1999},
\batitle{{A comparison of ground-based and spacecraft observations of coronal
  mass ejections from 1980-1989}}.
\bjtitle{\jgr}
\bvolume{104}(\bissue{A6}),
\bfpage{12493}.
\doiurl{10.1029/1999JA900045}.
\adsurl{https://ui.adsabs.harvard.edu/abs/1999JGR...10412493S}.
\end{barticle}
\endbibitem

\bibitem[\protect\citeauthoryear{{St. Cyr} \textit{et~al.}}{2000}]{stcyr2000}
\begin{barticle}
\bauthor{\bsnm{{St. Cyr}}, \binits{O.C.}},
\bauthor{\bsnm{{Plunkett}}, \binits{S.P.}},
\bauthor{\bsnm{{Michels}}, \binits{D.J.}},
\bauthor{\bsnm{{Paswaters}}, \binits{S.E.}},
\bauthor{\bsnm{{Koomen}}, \binits{M.J.}},
\bauthor{\bsnm{{Simnett}}, \binits{G.M.}},
\bauthor{\bsnm{{Thompson}}, \binits{B.J.}},
\bauthor{\bsnm{{Gurman}}, \binits{J.B.}},
\bauthor{\bsnm{{Schwenn}}, \binits{R.}},
\bauthor{\bsnm{{Webb}}, \binits{D.F.}},
\bauthor{\bsnm{{Hildner}}, \binits{E.}},
\bauthor{\bsnm{{Lamy}}, \binits{P.L.}}:
\byear{2000},
\batitle{{Properties of coronal mass ejections: SOHO LASCO observations from
  January 1996 to June 1998}}.
\bjtitle{\jgr}
\bvolume{105}(\bissue{A8}),
\bfpage{18169}.
\doiurl{10.1029/1999JA000381}.
\adsurl{https://ui.adsabs.harvard.edu/abs/2000JGR...10518169S}.
\end{barticle}
\endbibitem

\bibitem[\protect\citeauthoryear{{St. Cyr} \textit{et~al.}}{2015}]{stcyr2015}
\begin{barticle}
\bauthor{\bsnm{{St. Cyr}}, \binits{O.C.}},
\bauthor{\bsnm{{Flint}}, \binits{Q.A.}},
\bauthor{\bsnm{{Xie}}, \binits{H.}},
\bauthor{\bsnm{{Webb}}, \binits{D.F.}},
\bauthor{\bsnm{{Burkepile}}, \binits{J.T.}},
\bauthor{\bsnm{{Lecinski}}, \binits{A.R.}},
\bauthor{\bsnm{{Quirk}}, \binits{C.}},
\bauthor{\bsnm{{Stanger}}, \binits{A.L.}}:
\byear{2015},
\batitle{{MLSO Mark III K-Coronameter Observations of the CME Rate from 1989 -
  1996}}.
\bjtitle{\solphys}
\bvolume{290}(\bissue{10}),
\bfpage{2951}.
\doiurl{10.1007/s11207-015-0780-2}.
\adsurl{https://ui.adsabs.harvard.edu/abs/2015SoPh..290.2951S}.
\end{barticle}
\endbibitem

\bibitem[\protect\citeauthoryear{{Stone} \textit{et~al.}}{1998}]{stone1998}
\begin{barticle}
\bauthor{\bsnm{{Stone}}, \binits{E.C.}},
\bauthor{\bsnm{{Cohen}}, \binits{C.M.S.}},
\bauthor{\bsnm{{Cook}}, \binits{W.R.}},
\bauthor{\bsnm{{Cummings}}, \binits{A.C.}},
\bauthor{\bsnm{{Gauld}}, \binits{B.}},
\bauthor{\bsnm{{Kecman}}, \binits{B.}},
\bauthor{\bsnm{{Leske}}, \binits{R.A.}},
\bauthor{\bsnm{{Mewaldt}}, \binits{R.A.}},
\bauthor{\bsnm{{Thayer}}, \binits{M.R.}},
\bauthor{\bsnm{{Dougherty}}, \binits{B.L.}},
\bauthor{\bsnm{{Grumm}}, \binits{R.L.}},
\bauthor{\bsnm{{Milliken}}, \binits{B.D.}},
\bauthor{\bsnm{{Radocinski}}, \binits{R.G.}},
\bauthor{\bsnm{{Wiedenbeck}}, \binits{M.E.}},
\bauthor{\bsnm{{Christian}}, \binits{E.R.}},
\bauthor{\bsnm{{Shuman}}, \binits{S.}},
\bauthor{\bsnm{{von Rosenvinge}}, \binits{T.T.}}:
\byear{1998},
\batitle{{The Solar Isotope Spectrometer for the Advanced Composition
  Explorer}}.
\bjtitle{\ssr}
\bvolume{86},
\bfpage{357}.
\doiurl{10.1023/A:1005027929871}.
\adsurl{https://ui.adsabs.harvard.edu/abs/1998SSRv...86..357S}.
\end{barticle}
\endbibitem

\bibitem[\protect\citeauthoryear{Thompson \textit{et~al.}}{2017}]{thompson2017}
\begin{barticle}
\bauthor{\bsnm{Thompson}, \binits{W.T.}},
\bauthor{\bsnm{St.~Cyr}, \binits{O.C.}},
\bauthor{\bsnm{Burkepile}, \binits{J.T.}},
\bauthor{\bsnm{Posner}, \binits{A.}}:
\byear{2017},
\batitle{{Automatic Near-Real-Time Detection of CMEs in Mauna Loa K-Cor
  Coronagraph Images}}.
\bjtitle{Space Weather}
\bvolume{15}(\bissue{10}),
\bfpage{1288}.
\doiurl{https://doi.org/10.1002/2017SW001694}.
\burl{https://agupubs.onlinelibrary.wiley.com/doi/abs/10.1002/2017SW001694}.
\end{barticle}
\endbibitem

\bibitem[\protect\citeauthoryear{{Torsti} \textit{et~al.}}{1995}]{torsti1995}
\begin{barticle}
\bauthor{\bsnm{{Torsti}}, \binits{J.}},
\bauthor{\bsnm{{Valtonen}}, \binits{E.}},
\bauthor{\bsnm{{Lumme}}, \binits{M.}},
\bauthor{\bsnm{{Peltonen}}, \binits{P.}},
\bauthor{\bsnm{{Eronen}}, \binits{T.}},
\bauthor{\bsnm{{Louhola}}, \binits{M.}},
\bauthor{\bsnm{{Riihonen}}, \binits{E.}},
\bauthor{\bsnm{{Schultz}}, \binits{G.}},
\bauthor{\bsnm{{Teittinen}}, \binits{M.}},
\bauthor{\bsnm{{Ahola}}, \binits{K.}},
\bauthor{\bsnm{{Holmlund}}, \binits{C.}},
\bauthor{\bsnm{{Kelh{\"a}}}, \binits{V.}},
\bauthor{\bsnm{{Lepp{\"a}l{\"a}}}, \binits{K.}},
\bauthor{\bsnm{{Ruuska}}, \binits{P.}},
\bauthor{\bsnm{{Str{\"o}mmer}}, \binits{E.}}:
\byear{1995},
\batitle{{Energetic Particle Experiment ERNE}}.
\bjtitle{\solphys}
\bvolume{162}(\bissue{1-2}),
\bfpage{505}.
\doiurl{10.1007/BF00733438}.
\adsurl{https://ui.adsabs.harvard.edu/abs/1995SoPh..162..505T}.
\end{barticle}
\endbibitem

\bibitem[\protect\citeauthoryear{Tsurutani and von
  Rosenvinge}{1984}]{tsurutani1984}
\begin{barticle}
\bauthor{\bsnm{Tsurutani}, \binits{B.T.}},
\bauthor{\bparticle{von} \bsnm{Rosenvinge}, \binits{T.T.}}:
\byear{1984},
\batitle{{ISEE-3 distant Geotail results}}.
\bjtitle{Geophysical Research Letters}
\bvolume{11}(\bissue{10}),
\bfpage{1027}.
\doiurl{https://doi.org/10.1029/GL011i010p01027}.
\end{barticle}
\endbibitem

\bibitem[\protect\citeauthoryear{{Van Hollebeke}, {Ma Sung}, and
  {McDonald}}{1975}]{VanHollebeke1975}
\begin{barticle}
\bauthor{\bsnm{{Van Hollebeke}}, \binits{M.A.I.}},
\bauthor{\bsnm{{Ma Sung}}, \binits{L.S.}},
\bauthor{\bsnm{{McDonald}}, \binits{F.B.}}:
\byear{1975},
\batitle{{The Variation of Solar Proton Energy Spectra and Size Distribution
  with Heliolongitude}}.
\bjtitle{\solphys}
\bvolume{41}(\bissue{1}),
\bfpage{189}.
\doiurl{10.1007/BF00152967}.
\adsurl{https://ui.adsabs.harvard.edu/abs/1975SoPh...41..189V}.
\end{barticle}
\endbibitem

\bibitem[\protect\citeauthoryear{{von Rosenvinge}
  \textit{et~al.}}{1978}]{vonrosen1978}
\begin{barticle}
\bauthor{\bsnm{{von Rosenvinge}}, \binits{T.T.}},
\bauthor{\bsnm{{McDonald}}, \binits{F.B.}},
\bauthor{\bsnm{{Trainor}}, \binits{J.H.}},
\bauthor{\bsnm{{Van Hollebeke}}, \binits{M.A.I.}},
\bauthor{\bsnm{{Fisk}}, \binits{L.A.}}:
\byear{1978},
\batitle{{The medium energy cosmic ray experiment for ISEE-C.}}
\bjtitle{IEEE Transactions on Geoscience Electronics}
\bvolume{16},
\bfpage{208}.
\doiurl{10.1109/TGE.1978.294548}.
\adsurl{https://ui.adsabs.harvard.edu/abs/1978ITGE...16..208V}.
\end{barticle}
\endbibitem

\bibitem[\protect\citeauthoryear{{von Rosenvinge}
  \textit{et~al.}}{1995}]{vonrosen1995}
\begin{barticle}
\bauthor{\bsnm{{von Rosenvinge}}, \binits{T.T.}},
\bauthor{\bsnm{{Barbier}}, \binits{L.M.}},
\bauthor{\bsnm{{Karsch}}, \binits{J.}},
\bauthor{\bsnm{{Liberman}}, \binits{R.}},
\bauthor{\bsnm{{Madden}}, \binits{M.P.}},
\bauthor{\bsnm{{Nolan}}, \binits{T.}},
\bauthor{\bsnm{{Reames}}, \binits{D.V.}},
\bauthor{\bsnm{{Ryan}}, \binits{L.}},
\bauthor{\bsnm{{Singh}}, \binits{S.}},
\bauthor{\bsnm{{Trexel}}, \binits{H.}},
\bauthor{\bsnm{{Winkert}}, \binits{G.}},
\bauthor{\bsnm{{Mason}}, \binits{G.M.}},
\bauthor{\bsnm{{Hamilton}}, \binits{D.C.}},
\bauthor{\bsnm{{Walpole}}, \binits{P.}}:
\byear{1995},
\batitle{{The Energetic Particles: Acceleration, Composition, and Transport
  (EPACT) investigation on the WIND spacecraft}}.
\bjtitle{\ssr}
\bvolume{71}(\bissue{1-4}),
\bfpage{155}.
\doiurl{10.1007/BF00751329}.
\adsurl{https://ui.adsabs.harvard.edu/abs/1995SSRv...71..155V}.
\end{barticle}
\endbibitem

\bibitem[\protect\citeauthoryear{{Vourlidas}
  \textit{et~al.}}{2010}]{Vourlidas2010}
\begin{barticle}
\bauthor{\bsnm{{Vourlidas}}, \binits{A.}},
\bauthor{\bsnm{{Howard}}, \binits{R.A.}},
\bauthor{\bsnm{{Esfandiari}}, \binits{E.}},
\bauthor{\bsnm{{Patsourakos}}, \binits{S.}},
\bauthor{\bsnm{{Yashiro}}, \binits{S.}},
\bauthor{\bsnm{{Michalek}}, \binits{G.}}:
\byear{2010},
\batitle{{Comprehensive Analysis of Coronal Mass Ejection Mass and Energy
  Properties Over a Full Solar Cycle}}.
\bjtitle{\apj}
\bvolume{722}(\bissue{2}),
\bfpage{1522}.
\doiurl{10.1088/0004-637X/722/2/1522}.
\adsurl{https://ui.adsabs.harvard.edu/abs/2010ApJ...722.1522V}.
\end{barticle}
\endbibitem

\bibitem[\protect\citeauthoryear{{Vourlidas}
  \textit{et~al.}}{2020}]{Vourlidas2020}
\begin{barticle}
\bauthor{\bsnm{{Vourlidas}}, \binits{A.}},
\bauthor{\bsnm{{Balmaceda}}, \binits{L.A.}},
\bauthor{\bsnm{{Xie}}, \binits{H.}},
\bauthor{\bsnm{{St. Cyr}}, \binits{O.C.}}:
\byear{2020},
\batitle{{The Coronal Mass Ejection Visibility Function of Modern
  Coronagraphs}}.
\bjtitle{\apj}
\bvolume{900}(\bissue{2}),
\bfpage{161}.
\doiurl{10.3847/1538-4357/abada5}.
\adsurl{https://ui.adsabs.harvard.edu/abs/2020ApJ...900..161V}.
\end{barticle}
\endbibitem

\bibitem[\protect\citeauthoryear{{Vr{\v{s}}nak}}{2001}]{Vrsnak2001}
\begin{barticle}
\bauthor{\bsnm{{Vr{\v{s}}nak}}, \binits{B.}}:
\byear{2001},
\batitle{{Dynamics of solar coronal eruptions}}.
\bjtitle{\jgr}
\bvolume{106}(\bissue{A11}),
\bfpage{25249}.
\doiurl{10.1029/2000JA004007}.
\adsurl{https://ui.adsabs.harvard.edu/abs/2001JGR...10625249V}.
\end{barticle}
\endbibitem

\bibitem[\protect\citeauthoryear{{Vr{\v{s}}nak}
  \textit{et~al.}}{2007}]{vrsnak2007}
\begin{barticle}
\bauthor{\bsnm{{Vr{\v{s}}nak}}, \binits{B.}},
\bauthor{\bsnm{{Mari{\v{c}}i{\'c}}}, \binits{D.}},
\bauthor{\bsnm{{Stanger}}, \binits{A.L.}},
\bauthor{\bsnm{{Veronig}}, \binits{A.M.}},
\bauthor{\bsnm{{Temmer}}, \binits{M.}},
\bauthor{\bsnm{{Ro{\v{s}}a}}, \binits{D.}}:
\byear{2007},
\batitle{{Acceleration Phase of Coronal Mass Ejections: I. Temporal and Spatial
  Scales}}.
\bjtitle{\solphys}
\bvolume{241}(\bissue{1}),
\bfpage{85}.
\doiurl{10.1007/s11207-006-0290-3}.
\adsurl{https://ui.adsabs.harvard.edu/abs/2007SoPh..241...85V}.
\end{barticle}
\endbibitem

\bibitem[\protect\citeauthoryear{{Webb} and {Howard}}{1994}]{Webb1994}
\begin{barticle}
\bauthor{\bsnm{{Webb}}, \binits{D.F.}},
\bauthor{\bsnm{{Howard}}, \binits{R.A.}}:
\byear{1994},
\batitle{{The solar cycle variation of coronal mass ejections and the solar
  wind mass flux}}.
\bjtitle{\jgr}
\bvolume{99}(\bissue{A3}),
\bfpage{4201}.
\doiurl{10.1029/93JA02742}.
\adsurl{https://ui.adsabs.harvard.edu/abs/1994JGR....99.4201W}.
\end{barticle}
\endbibitem

\bibitem[\protect\citeauthoryear{{Webb} and {Vourlidas}}{2016}]{webb2016}
\begin{barticle}
\bauthor{\bsnm{{Webb}}, \binits{D.F.}},
\bauthor{\bsnm{{Vourlidas}}, \binits{A.}}:
\byear{2016},
\batitle{{LASCO White-Light Observations of Eruptive Current Sheets Trailing
  CMEs}}.
\bjtitle{\solphys}
\bvolume{291}(\bissue{12}),
\bfpage{3725}.
\doiurl{10.1007/s11207-016-0988-9}.
\adsurl{https://ui.adsabs.harvard.edu/abs/2016SoPh..291.3725W}.
\end{barticle}
\endbibitem

\bibitem[\protect\citeauthoryear{Whitman \textit{et~al.}}{2022}]{Whitman2022}
\begin{botherref}
\oauthor{\bsnm{Whitman}, \binits{K.}},
\oauthor{\bsnm{Egeland}, \binits{R.}},
\oauthor{\bsnm{Richardson}, \binits{I.G.}},
\oauthor{\bsnm{Allison}, \binits{C.}},
\oauthor{\bsnm{Quinn}, \binits{P.}},
\oauthor{\bsnm{Barzilla}, \binits{J.}},
\oauthor{\bsnm{Kitiashvili}, \binits{I.}},
\oauthor{\bsnm{Sadykov}, \binits{V.}},
\oauthor{\bsnm{Bain}, \binits{H.M.}},
\oauthor{\bsnm{Dierckxsens}, \binits{M.}},
\oauthor{\bsnm{Mays}, \binits{M.L.}},
\oauthor{\bsnm{Tadesse}, \binits{T.}},
\oauthor{\bsnm{Lee}, \binits{K.T.}},
\oauthor{\bsnm{Semones}, \binits{E.}},
\oauthor{\bsnm{Luhmann}, \binits{J.G.}},
\oauthor{\bsnm{Nunez}, \binits{M.}},
\oauthor{\bsnm{White}, \binits{S.M.}},
\oauthor{\bsnm{Kahler}, \binits{S.W.}},
\oauthor{\bsnm{Ling}, \binits{A.G.}},
\oauthor{\bsnm{Smart}, \binits{D.F.}},
\oauthor{\bsnm{Shea}, \binits{M.A.}},
\oauthor{\bsnm{Tenishev}, \binits{V.}},
\oauthor{\bsnm{Boubrahimi}, \binits{S.F.}},
\oauthor{\bsnm{Aydin}, \binits{B.}},
\oauthor{\bsnm{Martens}, \binits{P.}},
\oauthor{\bsnm{Angryk}, \binits{R.}},
\oauthor{\bsnm{Marsh}, \binits{M.S.}},
\oauthor{\bsnm{Dalla}, \binits{S.}},
\oauthor{\bsnm{Crosby}, \binits{N.}},
\oauthor{\bsnm{Schwadron}, \binits{N.A.}},
\oauthor{\bsnm{Kozarev}, \binits{K.}},
\oauthor{\bsnm{Gorby}, \binits{M.}},
\oauthor{\bsnm{Young}, \binits{M.A.}},
\oauthor{\bsnm{Laurenza}, \binits{M.}},
\oauthor{\bsnm{Cliver}, \binits{E.W.}},
\oauthor{\bsnm{Alberti}, \binits{T.}},
\oauthor{\bsnm{Stumpo}, \binits{M.}},
\oauthor{\bsnm{Benella}, \binits{S.}},
\oauthor{\bsnm{Papaioannou}, \binits{A.}},
\oauthor{\bsnm{Anastasiadis}, \binits{A.}},
\oauthor{\bsnm{Sandberg}, \binits{I.}},
\oauthor{\bsnm{Georgoulis}, \binits{M.K.}},
\oauthor{\bsnm{Ji}, \binits{A.}},
\oauthor{\bsnm{Kempton}, \binits{D.}},
\oauthor{\bsnm{Pandey}, \binits{C.}},
\oauthor{\bsnm{Li}, \binits{G.}},
\oauthor{\bsnm{Hu}, \binits{J.}},
\oauthor{\bsnm{Zank}, \binits{G.P.}},
\oauthor{\bsnm{Lavasa}, \binits{E.}},
\oauthor{\bsnm{Giannopoulos}, \binits{G.}},
\oauthor{\bsnm{Falconer}, \binits{D.}},
\oauthor{\bsnm{Kadadi}, \binits{Y.}},
\oauthor{\bsnm{Fernandes}, \binits{I.}},
\oauthor{\bsnm{Dayeh}, \binits{M.A.}},
\oauthor{\bsnm{Munoz-Jaramillo}, \binits{A.}},
\oauthor{\bsnm{Chatterjee}, \binits{S.}},
\oauthor{\bsnm{Moreland}, \binits{K.D.}},
\oauthor{\bsnm{Sokolov}, \binits{I.V.}},
\oauthor{\bsnm{Roussev}, \binits{I.I.}},
\oauthor{\bsnm{Taktakishvili}, \binits{A.}},
\oauthor{\bsnm{Effenberger}, \binits{F.}},
\oauthor{\bsnm{Gombosi}, \binits{T.}},
\oauthor{\bsnm{Huang}, \binits{Z.}},
\oauthor{\bsnm{Zhao}, \binits{L.}},
\oauthor{\bsnm{Wijsen}, \binits{N.}},
\oauthor{\bsnm{Aran}, \binits{A.}},
\oauthor{\bsnm{Poedts}, \binits{S.}},
\oauthor{\bsnm{Kouloumvakos}, \binits{A.}},
\oauthor{\bsnm{Paassilta}, \binits{M.}},
\oauthor{\bsnm{Vainio}, \binits{R.}},
\oauthor{\bsnm{Belov}, \binits{A.}},
\oauthor{\bsnm{Eroshenko}, \binits{E.A.}},
\oauthor{\bsnm{Abunina}, \binits{M.A.}},
\oauthor{\bsnm{Abunin}, \binits{A.A.}},
\oauthor{\bsnm{Balch}, \binits{C.C.}},
\oauthor{\bsnm{Malandraki}, \binits{O.}},
\oauthor{\bsnm{Karavolos}, \binits{M.}},
\oauthor{\bsnm{Heber}, \binits{B.}},
\oauthor{\bsnm{Labrenz}, \binits{J.}},
\oauthor{\bsnm{Kuhl}, \binits{P.}},
\oauthor{\bsnm{Kosovichev}, \binits{A.G.}},
\oauthor{\bsnm{Oria}, \binits{V.}},
\oauthor{\bsnm{Nita}, \binits{G.M.}},
\oauthor{\bsnm{Illarionov}, \binits{E.}},
\oauthor{\bsnm{O'Keefe}, \binits{P.M.}},
\oauthor{\bsnm{Jiang}, \binits{Y.}},
\oauthor{\bsnm{Fereira}, \binits{S.H.}},
\oauthor{\bsnm{Ali}, \binits{A.}},
\oauthor{\bsnm{Paouris}, \binits{E.}},
\oauthor{\bsnm{Aminalragia-Giamini}, \binits{S.}},
\oauthor{\bsnm{Jiggens}, \binits{P.}},
\oauthor{\bsnm{Jin}, \binits{M.}},
\oauthor{\bsnm{Lee}, \binits{C.O.}},
\oauthor{\bsnm{Palmerio}, \binits{E.}},
\oauthor{\bsnm{Bruno}, \binits{A.}},
\oauthor{\bsnm{Kasapis}, \binits{S.}},
\oauthor{\bsnm{Wang}, \binits{X.}},
\oauthor{\bsnm{Chen}, \binits{Y.}},
\oauthor{\bsnm{Sanahuja}, \binits{B.}},
\oauthor{\bsnm{Lario}, \binits{D.}},
\oauthor{\bsnm{Jacobs}, \binits{C.}},
\oauthor{\bsnm{Strauss}, \binits{D.T.}},
\oauthor{\bsnm{Steyn}, \binits{R.}},
\oauthor{\bsnm{{van den Berg}}, \binits{J.}},
\oauthor{\bsnm{Swalwell}, \binits{B.}},
\oauthor{\bsnm{Waterfall}, \binits{C.}},
\oauthor{\bsnm{Nedal}, \binits{M.}},
\oauthor{\bsnm{Miteva}, \binits{R.}},
\oauthor{\bsnm{Dechev}, \binits{M.}},
\oauthor{\bsnm{Zucca}, \binits{P.}},
\oauthor{\bsnm{Engell}, \binits{A.}},
\oauthor{\bsnm{Maze}, \binits{B.}},
\oauthor{\bsnm{Farmer}, \binits{H.}},
\oauthor{\bsnm{Kerber}, \binits{T.}},
\oauthor{\bsnm{Barnett}, \binits{B.}},
\oauthor{\bsnm{Loomis}, \binits{J.}},
\oauthor{\bsnm{Grey}, \binits{N.}},
\oauthor{\bsnm{Thompson}, \binits{B.J.}},
\oauthor{\bsnm{Linker}, \binits{J.A.}},
\oauthor{\bsnm{Caplan}, \binits{R.M.}},
\oauthor{\bsnm{Downs}, \binits{C.}},
\oauthor{\bsnm{T\"or\"ok}, \binits{T.}},
\oauthor{\bsnm{Lionello}, \binits{R.}},
\oauthor{\bsnm{Titov}, \binits{V.}},
\oauthor{\bsnm{Zhang}, \binits{M.}},
\oauthor{\bsnm{Hosseinzadeh}, \binits{P.}}:
2022,
Review of solar energetic particle models.
\textit{Advances in Space Research}.
\doiurl{https://doi.org/10.1016/j.asr.2022.08.006}.
\end{botherref}
\endbibitem

\bibitem[\protect\citeauthoryear{{Wild}}{1950}]{wild1950}
\begin{barticle}
\bauthor{\bsnm{{Wild}}, \binits{J.P.}}:
\byear{1950},
\batitle{{Observations of the Spectrum of High-Intensity Solar Radiation at
  Metre Wavelengths. II. Outbursts}}.
\bjtitle{Australian Journal of Scientific Research A Physical Sciences}
\bvolume{3},
\bfpage{399}.
\doiurl{10.1071/CH9500399}.
\adsurl{https://ui.adsabs.harvard.edu/abs/1950AuSRA...3..399W}.
\end{barticle}
\endbibitem

\bibitem[\protect\citeauthoryear{{Wilson III}
  \textit{et~al.}}{2021}]{wilson2021}
\begin{barticle}
\bauthor{\bsnm{{Wilson III}}, \binits{L.B.}},
\bauthor{\bsnm{{Brosius}}, \binits{A.L.}},
\bauthor{\bsnm{{Gopalswamy}}, \binits{N.}},
\bauthor{\bsnm{{Nieves-Chinchilla}}, \binits{T.}},
\bauthor{\bsnm{{Szabo}}, \binits{A.}},
\bauthor{\bsnm{{Hurley}}, \binits{K.}},
\bauthor{\bsnm{{Phan}}, \binits{T.}},
\bauthor{\bsnm{{Kasper}}, \binits{J.C.}},
\bauthor{\bsnm{{Lugaz}}, \binits{N.}},
\bauthor{\bsnm{{Richardson}}, \binits{I.G.}},
\bauthor{\bsnm{{Chen}}, \binits{C.H.K.}},
\bauthor{\bsnm{{Verscharen}}, \binits{D.}},
\bauthor{\bsnm{{Wicks}}, \binits{R.T.}},
\bauthor{\bsnm{{TenBarge}}, \binits{J.M.}}:
\byear{2021},
\batitle{{A Quarter Century of Wind Spacecraft Discoveries}}.
\bjtitle{Reviews of Geophysics}
\bvolume{59}(\bissue{2}),
\bfpage{e2020RG000714}.
\doiurl{10.1029/2020RG000714}.
\adsurl{https://ui.adsabs.harvard.edu/abs/2021RvGeo..5900714W}.
\end{barticle}
\endbibitem

\bibitem[\protect\citeauthoryear{{Winter} and {Ledbetter}}{2015}]{winter2015}
\begin{barticle}
\bauthor{\bsnm{{Winter}}, \binits{L.M.}},
\bauthor{\bsnm{{Ledbetter}}, \binits{K.}}:
\byear{2015},
\batitle{{Type II and Type III Radio Bursts and their Correlation with Solar
  Energetic Proton Events}}.
\bjtitle{\apj}
\bvolume{809}(\bissue{1}),
\bfpage{105}.
\doiurl{10.1088/0004-637X/809/1/105}.
\adsurl{https://ui.adsabs.harvard.edu/abs/2015ApJ...809..105W}.
\end{barticle}
\endbibitem

\bibitem[\protect\citeauthoryear{{Yashiro} \textit{et~al.}}{2004}]{Yashiro2004}
\begin{barticle}
\bauthor{\bsnm{{Yashiro}}, \binits{S.}},
\bauthor{\bsnm{{Gopalswamy}}, \binits{N.}},
\bauthor{\bsnm{{Michalek}}, \binits{G.}},
\bauthor{\bsnm{{St. Cyr}}, \binits{O.C.}},
\bauthor{\bsnm{{Plunkett}}, \binits{S.P.}},
\bauthor{\bsnm{{Rich}}, \binits{N.B.}},
\bauthor{\bsnm{{Howard}}, \binits{R.A.}}:
\byear{2004},
\batitle{{A catalog of white light coronal mass ejections observed by the SOHO
  spacecraft}}.
\bjtitle{Journal of Geophysical Research (Space Physics)}
\bvolume{109}(\bissue{A7}),
\bfpage{A07105}.
\doiurl{10.1029/2003JA010282}.
\adsurl{https://ui.adsabs.harvard.edu/abs/2004JGRA..109.7105Y}.
\end{barticle}
\endbibitem

\bibitem[\protect\citeauthoryear{{Yashiro} \textit{et~al.}}{2008}]{yashiro2008}
\begin{barticle}
\bauthor{\bsnm{{Yashiro}}, \binits{S.}},
\bauthor{\bsnm{{Michalek}}, \binits{G.}},
\bauthor{\bsnm{{Akiyama}}, \binits{S.}},
\bauthor{\bsnm{{Gopalswamy}}, \binits{N.}},
\bauthor{\bsnm{{Howard}}, \binits{R.A.}}:
\byear{2008},
\batitle{{Spatial Relationship between Solar Flares and Coronal Mass
  Ejections}}.
\bjtitle{\apj}
\bvolume{673}(\bissue{2}),
\bfpage{1174}.
\doiurl{10.1086/524927}.
\adsurl{https://ui.adsabs.harvard.edu/abs/2008ApJ...673.1174Y}.
\end{barticle}
\endbibitem

\bibitem[\protect\citeauthoryear{{Zhang} and {Dere}}{2006}]{zhang2006}
\begin{barticle}
\bauthor{\bsnm{{Zhang}}, \binits{J.}},
\bauthor{\bsnm{{Dere}}, \binits{K.P.}}:
\byear{2006},
\batitle{{A Statistical Study of Main and Residual Accelerations of Coronal
  Mass Ejections}}.
\bjtitle{\apj}
\bvolume{649}(\bissue{2}),
\bfpage{1100}.
\doiurl{10.1086/506903}.
\adsurl{https://ui.adsabs.harvard.edu/abs/2006ApJ...649.1100Z}.
\end{barticle}
\endbibitem

\bibitem[\protect\citeauthoryear{{Zhang} \textit{et~al.}}{2001}]{Zhang2001}
\begin{barticle}
\bauthor{\bsnm{{Zhang}}, \binits{J.}},
\bauthor{\bsnm{{Dere}}, \binits{K.P.}},
\bauthor{\bsnm{{Howard}}, \binits{R.A.}},
\bauthor{\bsnm{{Kundu}}, \binits{M.R.}},
\bauthor{\bsnm{{White}}, \binits{S.M.}}:
\byear{2001},
\batitle{{On the Temporal Relationship between Coronal Mass Ejections and
  Flares}}.
\bjtitle{\apj}
\bvolume{559}(\bissue{1}),
\bfpage{452}.
\doiurl{10.1086/322405}.
\adsurl{https://ui.adsabs.harvard.edu/abs/2001ApJ...559..452Z}.
\end{barticle}
\endbibitem

\end{thebibliography}

\IfFileExists{\jobname.bbl}{} {\typeout{}
\typeout{****************************************************}
\typeout{****************************************************}
\typeout{** Please run "bibtex \jobname" to obtain} \typeout{**
the bibliography and then re-run LaTeX} \typeout{** twice to fix
the references !}
\typeout{****************************************************}
\typeout{****************************************************}
\typeout{}}

\end{article} 

\end{document}